\documentclass{aa}  
\usepackage{graphicx}
\usepackage{txfonts}
\usepackage{lscape}
\usepackage{multirow}
\usepackage{footnote}
\usepackage{hyperref}  

\usepackage[usenames,dvipsnames]{xcolor}

\usepackage{amsmath}
\usepackage{float}

\def\teff{$T_{\rm eff}$}
\def\logg{$\log g$}
\def\vsini{$v \sin i$}
\def\kms{km\,s$^{-1}$}

\begin{document}

   \title{The ESO UVES/FEROS Large Programs of TESS OB pulsators\thanks{Based on observations collected at the European Southern Observatory, Chile under programs 1104.D-0230 and 0104.A-9001}}

   \subtitle{I. Global stellar parameters from high-resolution spectroscopy}

\author{Nadya Serebriakova\inst{\ref{KUL}}
\and Andrew Tkachenko\inst{\ref{KUL}}
\and Sarah Gebruers\inst{\ref{KUL},\ref{MPIA}}
\and Dominic M. Bowman\inst{\ref{KUL}} 
\and Timothy Van Reeth\inst{\ref{KUL}}
\and Laurent Mahy\inst{\ref{ROB}}
\and Siemen Burssens\inst{\ref{KUL}} 
\and Luc IJspeert\inst{\ref{KUL}}
\and Hugues Sana\inst{\ref{KUL}}
\and Conny Aerts\inst{\ref{KUL},\ref{MPIA},\ref{Radboud}}
}
\institute{Institute of Astronomy, KU Leuven, Celestijnenlaan 200D, B-3001 Leuven, Belgium \\ \email{nadya.serebriakova@kuleuven.be} \label{KUL} 
\and Max Planck Institute for Astronomy, K\"onigstuhl 17, 69117 Heidelberg, Germany \label{MPIA}
\and Royal Observatory of Belgium, Ringlaan 3, B-1180 Brussels, Belgium \label{ROB} 
\and Department of Astrophysics, IMAPP, Radboud University Nijmegen, PO Box 9010, 6500 GL Nijmegen, The Netherlands\label{Radboud}
}

   \date{Received ...; accepted ...}


  \abstract
   {Modern stellar structure and evolution theory experiences a lack of observational calibrations for the interior physics of intermediate- and high-mass stars. This leads to discrepancies between theoretical predictions and observed phenomena mostly related to angular momentum and element transport. Analyses of large samples of massive stars connecting state-of-the-art spectroscopy to asteroseismology may provide clues on how to improve our understanding of their interior structure.}
   {We aim to deliver a sample of O- and B-type stars at metallicity regimes of the Milky Way and the Large Magellanic Cloud (LMC) galaxies with accurate atmospheric parameters from high-resolution spectroscopy, along with a detailed investigation of line-profile broadening, for future asteroseismic studies.}
  {After describing the general aims of our two Large Programs, we develop dedicated methodology to fit spectral lines and deduce accurate global stellar parameters from high-resolution multi-epoch {\sc uves} and {\sc feros} spectroscopy. We use the best available atmosphere models for three regimes covered by our global sample, given its breadth in terms of mass, effective temperature, and evolutionary stage.}
   {Aside from accurate atmospheric parameters and locations in the Hertzsprung-Russell diagram, we deliver detailed analyses of macroturbulent line broadening, including estimation of the radial and tangential components. We find that these two components are difficult to disentangle from spectra with signal-to-noise ratios below 250. }
   {Future asteroseismic modelling of the deep interior physics of the most promising stars in our sample will improve the existing dearth of such knowledge for large samples of OB stars, including those of low metallicity in the LMC.}

   \keywords{stars: fundamental parameters -- stars: massive -- stars: early-type -- stars: rotation -- asteroseismology -- stars: oscillations 
               }

   \maketitle


\section{Introduction}\label{sec:intoduction}

The field of stellar astrophysics has been revolutionized with the advent of space missions such as CoRoT \citep{Auvergne2009}, {\it Kepler}/K2 \citep{kepler}, and TESS \citep{tess}. This revolution became possible thanks to the high duty-cycle, nearly uninterrupted, and unprecedentedly long (from about a month in a single TESS sector to four years for the entire {\it Kepler\/} nominal mission duration) photometric time series at the sub-mmag level of precision that these missions delivered. This high precision data is currently available for tens of thousands of stars.  These ensembles provide stringent tests for the theories of stellar structure, evolution, atmospheres, tidal interactions, etc. In particular, numerous recent studies point to both severe and less critical deficiencies in the above-mentioned theories, notably on angular momentum transport and on chemical mixing. Some of the deficiencies remain unresolved, whereas others are currently being tackled in detail from theoretical improvements \citep[e.g.,][]{Mosser2012,Beck2012,Fuller2012,Cantiello2014,Deheuvels2016,Fuller2017a,Fuller2017b,Hekker2017,Aerts2019,Bowman2020rev,Deheuvels2020,Aerts2021,Southworth2021}.

The field of binary stars is among those that received a major boost in recent years. This is in part associated with the superior quality of the space-based photometry when compared to its ground-based counterpart. It is also due to synergies enabled by various types of intrinsic variability often observed in one or both binary components. This way, solid observational detections of tidally-induced pulsations were reported in the HD\,174884 system observed by CoRoT \citep{Maceroni2009}, followed by  the discovery of high-amplitude tidally-induced oscillations in the {\it Kepler\/} object of interest KOI-54 \citep{Welsh2011}. KOI-54's observed properties were subsequently  modelled by \citet{Fuller2012}. Numerous binary systems with stellar pulsations that are either induced or perturbed by tides have been meanwhile found and modelled \citep[e.g.,][]{Bowman2019b,Guo2019,Handler2020,Kurtz2020,Fuller2020,Fuller2021,Rappaport2021,Lee2021,VanReeth2022,Jayaraman2022}. Applications occur most notably for the class of highly eccentric systems dubbed ``heart-beat stars'' \citep[e.g.,][]{Thompson2012,Hambleton2013,Beck2012,Hambleton2016,Guo2020,Cheng2020,Kolaczek2021}. Furthermore, a sample of low- and intermediate-mass binaries observed by {\it Kepler} has enabled empirical tests of state-of-the-art tidal evolution theory \citep{Zahn2013} with respect to its predictions for the circularisation of binary orbits \citep{VanEylen2016}. Ultimately, (eclipsing) binary stars allow us to constrain the amount of core-boundary mixing and the convective core masses in intermediate- and high-mass stars \citep[e.g.,][and references therein]{claret,Claret2019,binTkachenko,Johnston2021}, thanks to the unique property of eclipsing, spectroscopic double-lined binaries delivering masses and radii with precision and accuracy reaching 1\% level, and in some cases even better \citep[e.g.,][]{Maxted2018,Maxted2020,Mahy2020b,Mahy2020a,Serenelli2021,Pavlovski2022}.

Yet another field in stellar astrophysics that advances enormously with the launch and operations of space-based missions is asteroseismology. This field exploits the fact 
that, as stars evolve and cross different regions in the Hertzsprung-Russell (HR) diagram, they become unstable to pulsations. Two principle types of stellar pulsations are distinguished following their dominant restoring force: pressure (p) modes which are acoustic waves dominantly restored by pressure gradients, and gravity (g) modes 
restored by buoyancy. While p~modes are typically confined to the envelope of stars, g~modes have their largest amplitudes in the near-core regions and are therefore most sensitive to the physical conditions in the deep stellar interior. One of the outstanding observational findings of space-based missions is that the vast majority of stars are intrinsically variable, and that pulsating stars are found everywhere across the HR diagram \citep{Aerts2021,Kurtz2022}. 

Asteroseismology probing the near-core region in intermediate- to high-mass stars (spectral types early-F to late-O) had to await space-based observations, owing to the long periods of their g~modes (of the order of a day). Theoretically predicted informative series of (quasi-)uniformly spaced g-mode periods dubbed period spacing patterns \citep{Miglio2008,Bouabid2013} were only recently 
detected from space-based photometric observations of B- and F-type stars \citep[e.g.,][]{VanReeth2015a,VanReeth2015b,Papics2017,Szewczuk2021}. Detailed asteroseismic modelling of the period spacing patterns enables probes of the level and functional form of convective core boundary and envelope mixing \citep[e.g.,][]{Moravveji2015,Moravveji2016,Szewczuk2018,Michielsen2019,Michielsen2021,Mombarg2019,Mombarg2021,Pedersen2018,Pedersen2021,Pedersen2022,Bowman2021c,Szewczuk2022}, internal rotation \citep[e.g.,][]{Triana2015,VanReeth2016,VanReeth2018,Ouazzani2017,Ouazzani2020,Li2020}, the physics of radiative levitation \citep{Deal2016,Deal2017,Mombarg2020,Mombarg2022}, and impact of magnetism \citep{Buysschaert2018c,Lecoanet2022a} in these stars. In particular, these and other studies lead to the conclusion that standard stellar structure and evolution models of intermediate and high-mass stars incorrectly predict their convective core masses \citep[e.g.,][and references therein]{Johnston2021} and lack mechanism(s) of efficient angular momentum transport already in their longest and most stable main-sequence stage of evolution \citep[e.g.,][and references therein]{Aerts2017a,Aerts2019,Aerts2021}. These conclusions for the main sequence are in agreement with both observational and numerical studies of intermediate-mass stars in post-main sequence stages \citep[e.g.,][]{Mosser2012,Beck2012,Cantiello2014,Hermes2017,Aerts2019}.

Typically space-based photometry in tandem with ground-based spectroscopy reveal pulsators in regions in the HR diagram where they were not expected. On the other hand, no signatures of pulsations are sometimes found in stars that are expected to pulsate according to their location in the HR diagram. For example, recent studies of A- and F-type stars based on {\it Kepler} and Gaia photometric data reveal a considerable fraction of pulsators outside of their respective theoretical instability regions \citep[e.g.,][]{Uytterhoeven2011,Bowman2018,Murphy2019,DeRidder2022}. On the other hand, \citet{Balona2014,Balona2015} report that at least 50\% of A-type stars do not pulsate while being located within the borders of the theoretical instability strips. Furthermore, \citet{Balona2020} find a group of B-type stars pulsating in high-frequencies while being too cool to be explained as rapidly rotating $\beta$~Cep-type p-mode pulsators whose effective temperature is substantially affected by gravity darkening. Such stars are found in both ground-based and space-based high-cadence photometry and are rather interpreted as Slowly Pulsating B (SPB) stars whose observed (prograde) g-mode frequencies are shifted to high-frequencies, sometimes above $\sim 3\,$d$^{-1}$, by the influence of the Coriolis acceleration on the modes and by the Doppler effect
\citep{Aerts2005,Saesen2010,Saesen2013,Mowlavi2013,Mowlavi2016,Mozdzierski2014,Mozdzierski2019,Sharma2022,Gebruers2022,DeRidder2022}. This interpretation is in agreement with the theoretical computations by \citet{Bouabid2013,Salmon2014}.

Stars of spectral type O and B were under-represented in the nominal {\it Kepler\/} field as its selection was dictated by the mission's core science of detecting exoplanets orbiting solar-type stars. Moreover, models of these intermediate- to high-mass stars are most uncertain among those in stellar evolution computations \citep[e.g.,][]{Martins2013}. This current state of affairs motivates us to look closely at photometric data delivered by the all-sky TESS mission as its observational strategy does not bias against O and B stars, and allows us to target these objects both in the Milky Way and Large Magellanic Cloud (LMC) galaxies. The LMC is a challenging laboratory for the theory of heat-driven stellar pulsations 
as it depends strongly on the metallicity. The lower value of metallicity of the LMC and its implied low number of predicted pulsators is therefore of particular interest \citep[cf.,][]{Salmon2014}. Preliminary investigations of TESS data of massive stars in the LMC indeed indicate few heat-driven pulsators \citep{Bowman2019a}.

The current paper is the first one in a series aiming at a full application of asteroseismology to O and B stars, from a combined approach using space photometry and astrometry together with high-resolution spectroscopy. As is necessarily the case in massive stars asteroseismology, by nature of those stars' oscillation characteristics, it concerns a long-term project and the overall study involves several (PhD) sub-projects. Here, we first present the overall project aims and subsequently focus on 
the derivation of the global spectroscopic parameters of the target stars, which were selected from Cycle\,1 TESS space-based photometry.
In Sect.~\ref{sec:LP}, we present the concrete scientific motivation, the role of TESS space-based photometry, the asteroseismic quantities we are after, the role of the ESO high-resolution ground-based spectroscopy in our Large Programs, and the sample selection. The spectroscopic observations and custom data processing are described in Sect.~\ref{sec:spec_data_reduction}, while spectroscopic classification, inference of atmospheric parameters, and investigation of line broadening mechanisms in the spectra of the sample stars are detailed in Sect.~\ref{sec:spectroscopic_analysis}. 

We provide a detailed discussion of the obtained results, keeping in mind the photometric classification of stars from Cycle\,1 TESS photometry, and conclude the paper in Sect.~\ref{sec:discussion_conclusions}. Subsequent papers will tackle abundance determinations, frequency analyses from the full Cycle 1 to Cycle 4 TESS photometric light curves, and stellar modelling for all O and B pulsators with sufficient identified modes based on all observational input, including all the Gaia data that will be available at that stage of our project.
   
\section{Two ESO Large Programs with the {\sc feros} and {\sc uves} spectrographs}\label{sec:LP}

\subsection{The overall project goals}

The ultimate goal of our two ESO large programs ({\sc uves}: 1104.D-0230; {\sc feros}: 0104.A-9001) is to remedy the current lack of a high-precision observational calibration for the theoretical description of stellar interiors in stellar structure and evolution models of intermediate- to high-mass stars. Our concrete goals can only be achieved by exploiting complementary (ESO) high-resolution spectroscopy along with TESS space-based photometry and Gaia astrometry for a wide range of stellar masses covered by OB(AF)-type stars. 

With our two large programs, we aim to (i) resolve the current two-orders-of-magnitude discrepancies between theory and observations of angular momentum transport; (ii) deliver asteroseismically calibrated mixing profiles for stellar interiors, which currently do not exist in a quantitative sense for high-mass stars and are scarce for intermediate-mass stars; (iii) challenge state-of-the-art models of stellar structure and pulsations by searching for coherent g-mode pulsators in the low-metallicity regime of the LMC; and (iv) investigate possible connections between coherently and/or stochastically excited gravity waves and turbulent velocity fields in the atmospheres of OB stars. Furthermore, our program is designed in such a way as to enable an observational check for binarity, which also allows us to deliver an unprecedented legacy for future studies of pulsators in (close) binaries. Tidal asteroseismology is capable of quantifying the effect of tides by investigating the internal rotation and mixing in close binaries and comparing the outcome to the single-star case \citep[e.g.,][for reviews]{Guo2021,Lampens2021}.

\subsection{Asteroseismology of massive stars: the role of TESS}\label{sec:LP_seismology}

Unlike ``classical'' observational methods (for example, snapshot spectroscopy and multi-colour photometry) that largely concern surface properties of stars, asteroseismology offers insight into the internal physics of stars through the interpretation of detected and identified stellar oscillations. Hence it is an optimal tool to evaluate and calibrate models of stellar interiors, provided that we can detect and identify sufficient modes in terms of their $l$, $m$, and $n$ numbers \citep{asteroseismology}.


In terms of angular momentum transport in intermediate-mass dwarfs, the asteroseismic ensemble studies by 
\citet{VanReeth2016,VanReeth2018,Aerts2017b,Ouazzani2019}
demonstrate that stars born with a convective core tend to show almost rigid rotation in their radiative envelopes independent of their near-core rotation rate and evolutionary stage. This implies that strong coupling between the convective core and radiative envelope occurs on the main sequence. This conclusion is based on studies of stars with mass between 1.3~M$_{\odot}$ and 9~M$_{\odot}$, with the mass distribution being largely skewed towards late-B and F-type stars ($M\lesssim3.0$~M$_{\odot}$). Only for this mass regime, the later evolutionary stages have been mapped in great detail \citep[e.g.,][for a review]{Aerts2019}.
The goal of our ESO {\sc uves+feros} large programs is to expand the parameter space towards the higher stellar mass regime ($M\gtrsim3.0$~M$_{\odot}$) by focusing on stars of spectral type O and B on the main sequence, as well as in the hydrogen-shell and core-helium burning phases covered by blue supergiants. These phases constitute $\sim$90\% of the total lifetime of massive stars, excluding the remaining short period prior to the remnant formation when considerable mass loss occurs due to a radiation- or dust-driven wind. During this final phase, unambiguous detections of modes is challenging due to the strong wind.

Previously, analyses of the CoRoT and {\it Kepler}/K2 photometric data revealed the presence of non-radial oscillation modes in core-hydrogen burning stars with masses between 3 and $\sim\!25\,$M$_{\odot}$ \citep[e.g.,][]{Neiner2009,Huat2009,Diago2009,Gutierrez2009,Degroote2010a,Briquet2011,Mahy2011,Thoul2013,Buysschaert2015,Papics2017,Szewczuk2021,Pedersen2021} 
and in blue supergiants \citep{Aerts2010,Aerts2017a,Aerts2018,SSD2018}. Forward asteroseismic modelling is only available for about 30 of them, with quite different levels of depth
\citep{Aerts2011,Briquet2011,Neiner2012b,Moravveji2015,Moravveji2016,Michielsen2021,Pedersen2021}. The majority of the other intermediate-to-high mass pulsators lack proper mode identification, particularly for those with short light curves (less than half a year). Prior to space asteroseismology, long-term ground-based campaigns did already lead to such modelling for a few bright $\beta\,$Cep stars, yet based on only a few identified modes
\citep{Aerts2003,Handler2006,Dziembowski2008,Handler2009,Briquet2012}.

CoRoT data gave rise to the detection of stochastic high-frequency oscillations in a few OB-type stars \citep{Belkacem2009,Degroote2010c}, driven by either turbulent envelope convection \citep{Belkacem2010} or nonlinear resonant mode coupling \citep{Degroote2009}. Moreover, \citet{Blomme2011} found stochastic low-frequency (SLF) variability in the CoRoT data of several rapidly-rotating O-type stars.
Such SLF was later found to be present, often in addition to coherent oscillation modes, for 
the vast majority of O and B stars observed with CoRoT or K2 \citep{Bowman2019c}.  

A similar discovery was made for a sample of some 50 LMC O- and B-type stars observed with TESS \citep{Bowman2019a}. The lack of dependence of the properties of SLF variability on the metallicity of the star and its apparent scaling with the stellar (convective core) luminosity led to the interpretation that 
internal gravity waves (IGWs) excited stochastically at the core boundary 
are the cause of the observed photospheric variability 
\citep[see, e.g.,][]{Rogers2013,Edelmann2019,Ratnasingam2020,Horst2020}, although alternative interpretations have also been suggested \citep{Lecoanet2019}. In any case, SLF variability is ubiquitious in OB stars and is related to  their mass and age  \citep{Bowman2020b}. 

In conclusion, too few O- and B-type pulsators have been modelled in sufficient asteroseismic depth to improve the physics of stellar interiors for stars with masses above $\sim\,$5\,M$_\odot$ \citep{Bowman2020rev}. This is why we resort to the TESS mission, offering hundreds of potential massive candidate pulsators with high-cadence photometry covering at least 6 months of uninterrupted light curves, following our approved TESS guest observer programs covering cycles 1 to 5 of the extended mission (PI Bowman\footnote{\url{https://heasarc.gsfc.nasa.gov/docs/tess/approved-programs.html}}).

\subsection{The quest for mixing and rotation profiles\label{sec:LP_profiles}}

Because coherent g-modes and stochastically excited IGWs are both confined to the low-frequency regime (typically below some 3~d$^{-1}$ or $\sim35\ \mu$Hz), while asteroseismic analyses require individual pulsation modes to be resolved, our goals can only be achieved by exploiting TESS light curves of longer duration than the CoRoT light curves. Therefore, we focus on OB stars in the Southern Continuous Viewing Zone (CVZ-S) of TESS and its vicinity – the regions of the sky that have between some 200~d and 1-yr long, high duty cycle photometric data, which is long enough to enable both p- and g-mode asteroseismology. We take an observationally driven approach with the goal to estimate the following three quantities for a representative sample of single intermediate- and high-mass stars from their asteroseismic analysis: (i) $\Omega(r,t)$ - the interior rotation frequency and its evolution in time, (ii) $D_{\rm ov}(r,t)$ - the convective core overshooting profile and its evolution in time, and (iii) $D_{\rm mix}(r,t)$ - the chemical mixing profile in the radiative envelope and its evolution in time. Here, $t$ is the age of the star and $r\in[0, R(t)]$ is the depth inside the star from the centre $r=0$ to the surface at $r=R(t)$. The quantity $D_{\rm mix}(r,t)$ captures the cumulative effect of all active mixing phenomena in the radiatively stratified envelope from the star's birth until $t$. 

With an addition of the four minimum free parameters that are required to compute evolutionary models, namely the stellar mass $M$, the initial hydrogen and metal mass fractions $(X, Z)$, and the age $t$, asteroseismology turns out to be at least a 7D problem, even in the simplest case where the three above mentioned internal profiles would only have one free parameter that does not change in time. However, as shown by \citet{VanReeth2016} this 7D estimation problem can be broken down into a multi-step approach. In particular, interpretation of the period spacing patterns of g~modes (cf. Sect.~\ref{sec:intoduction}) allows us to estimate the near-core rotation frequency of the star, $\Omega_{\rm core}$ \citep{VanReeth2016,Ouazzani2017,Li2020}.
Furthermore, the inclusion of envelope mixing $D_{\rm mix}(r)$ appreciably affects high-order g~modes and it determines the surface abundances of CNO-produced chemical elements. In addition, g~modes are capable of differentiating between different $D_{\rm ov}(r)$ profiles for main-sequence stars \citep{Pedersen2018,Pedersen2021,Pedersen2022,Michielsen2019,Michielsen2021}. This way, combined asteroseismology and surface abundances information can be used to constrain the amount and functional form of mixing in both the near-core region and radiative envelope of intermediate- to high-mass stars. \citet{Aerts2018} provided methodology for forward asteroseismic modelling encapsulating all the above-described analysis steps, relying on state-of-the-art statistical principles to account for theoretical model prediction uncertainties as well as strong correlations among the numerous free parameters to estimate.

\subsection{The role of ESO high-resolution spectroscopy}\label{sec:LP_spectroscopy_sample}

High-quality spectroscopic observations are essential to address the above science questions and to meet our goals. We therefore applied for high-resolution (R$\gtrsim$40\,000), high signal-to-noise ratio (S/N$\sim$100) two-epoch spectroscopy for a sample of TESS O- and B-type variable stars in the Milky Way and LMC galaxies. These data were obtained with the Ultraviolet and Visual \'Echelle Spectrograph \citep[{\sc uves};][]{Dekker2000} and the Fiber-fed Extended Range Optical Spectrograph \citep[{\sc feros};][]{Kaufer1999} instruments attached to the UT2@VLT and MPG/ESO 2.2-m telescopes, respectively. 

{\sc uves} is a cross-dispersed \'echelle spectrograph that covers the entire optical range with its two arms (300-500~nm and 420-1100~nm in the blue and red arms, respectively) and offers a variety of spectral resolving power possibilities (up to 80\,000 and 110\,000 in the blue and red arm, respectively) regulated by the slit width of the instrument. We have opted for the {\sc uves}\footnote{\url{https://www.eso.org/public/teles-instr/paranal-observatory/vlt/vlt-instr/uves/}} DIC-2 blue arm standard setting (CD2, 437+760~nm) as it covers a number of Balmer (including H$_{\alpha}$) and helium lines – the main diagnostic lines for the determination of fundamental atmospheric parameters – as well as important diagnostic lines of metals for the determination of \vsini, micro- and macro-turbulent velocities, and surface abundances of OB stars. Because massive stars are typically moderate to fast rotators, we chose a slit width of 1 arcsecond, corresponding to a resolving power of $R\sim40\,000$, which is sufficient to properly resolve individual lines in the spectra of these stars. 

{\sc feros}\footnote{\url{https://www.eso.org/public/teles-instr/lasilla/mpg22/feros/}} is a fiber-fed \'echelle spectrograph covering the wavelength range from some 350~nm to 920~nm. It operates at a resolving power of $R\sim48\,000$. This combination of a wide wavelength range and high spectral resolution makes the instrument ideal to feed the stellar astrophysics community with high-quality spectroscopic observations of relatively bright stars, including intermediate- to high-mass stars of spectral type O and B.

\subsection{Sample selection}\label{sec:LP_sampleselection}

Our sample selection was largely based on the studies of massive stars 
by \citet{Bowman2019a} and \citet{Pedersen2019} with respect to their photometric variability as detected in the first sector(s) of TESS data. Their 
samples have been complemented with a global search for stars of spectral type O and B in the Southern CVZ of TESS, using the infrastructure and functionality of the Simbad astronomical database\footnote{\url{http://simbad.u-strasbg.fr/simbad/}} \citep{Wenger2000}. This obtained list of OB targets has been cross-matched with the catalogues of \citet{Bowman2019a,Bowman2019b,Bowman2020b,Pedersen2019} and the targets that did not have a match were kept in the sample for further consideration. For these extra targets, we have performed custom extraction of their light curves from the TESS Full Frame Images (FFIs) and performed their visual photometric variability classification based on the dominant signal detected in the TESS data. All eclipsing binaries have been excluded, making our sample biased against this type of objects --- TESS (pulsating) eclipsing binaries constitute a separate (PhD) study \citep[cf.,][]{IJspeert2021}. This way, we ended up with slightly over 350 targets that were prioritised according to their potential for future asteroseismic modelling. All stars were subjected to a check for existing spectroscopic data of required quality ($R\gtrsim40\,000$ and S/N$\sim$100) and cadence (minimum two-epochs) in public (including ESO) archives. 

We were left with a total of 272 stars that we ultimately targeted with the {\sc uves} and {\sc feros} instruments. The sample was divided in two parts according to the $V$-band magnitude of the target. Fainter stars with $V_{\rm mag}>10$ were observed with the {\sc uves} instrument while brighter objects with $V_{\rm mag}\leq10$ were targeted with the {\sc feros} instrument. This strategy allowed us to achieve maximal observing efficiency. In particular, this implied that none of the {\sc uves} science exposure 
were dominated by the instrument's system overhead. 

The choice to take two-epoch spectroscopic observations was driven by the need to (i) uncover potential binaries in our sample, and (ii) detect possible variations in spectral lines (He, Si, Mg, etc.) due to large macroturbulent velocity fields. High values of macroturbulence were previously reported for many high-mass stars \citep[e.g.,][]{Simon-Diaz2014,Simon-Diaz2017}. Moreover, a link between the physics underlying this phenomenon and non-radial gravity modes and/or IGWs has been suggested \citep{Aerts2015,Bowman2020b}. To avoid 
misinterpretation of detected variability due to strong tidal interactions between two stars in a (close) binary system rather than being intrinsic, uncovering such systems is of great importance. The two spectroscopic observations per object have been separated by at least 2-3 weeks to ensure detection capability for both short- and intermediate-period binaries where tidal forces are expected to be strongest.
   
 \subsection{Further content and goals of this paper}\label{sec:LP_thispaper}

   As of now, we focus on the global spectroscopic analysis of a sub-part of our overall stellar sample of 272 OB stars in the Galaxy and LMC. In particular, we perform the analysis of the entire {\sc uves} sample of 105 stars, composed of mostly blue supergiants in the LMC and a few dwarfs in the Milky Way. We extend this sub-sample with 43 LMC blue supergiant stars observed with the {\sc feros} instrument, bringing the total of treated stars in this paper to 148. The remaining 124 galactic objects (98 single stars and 26 spectroscopic binaries) in the {\sc feros} sample are analysed and presented in the accompanying paper by \citet{Gebruers2022}. 
   
   Our goal for the 148 OB stars (of which 24 are duplicates that are stars observed with both instruments) treated here is twofold: (i) perform a check for binarity and derive stellar atmospheric parameters such as the effective temperature \teff, surface gravity \logg, and projected rotational velocity \vsini\ and (ii) should the data quality allow for it, investigate the effect of macroturbulent broadening on spectral lines across the sample and its possible connection with stellar oscillations. In the final section, we reconsider the TESS variability classification of the stars according to their spectroscopic parameters. 
   
\section{Custom spectroscopic data reduction}\label{sec:spec_data_reduction}

\begin{figure*}
   \begin{centering}
            {\includegraphics[clip,width=450pt]{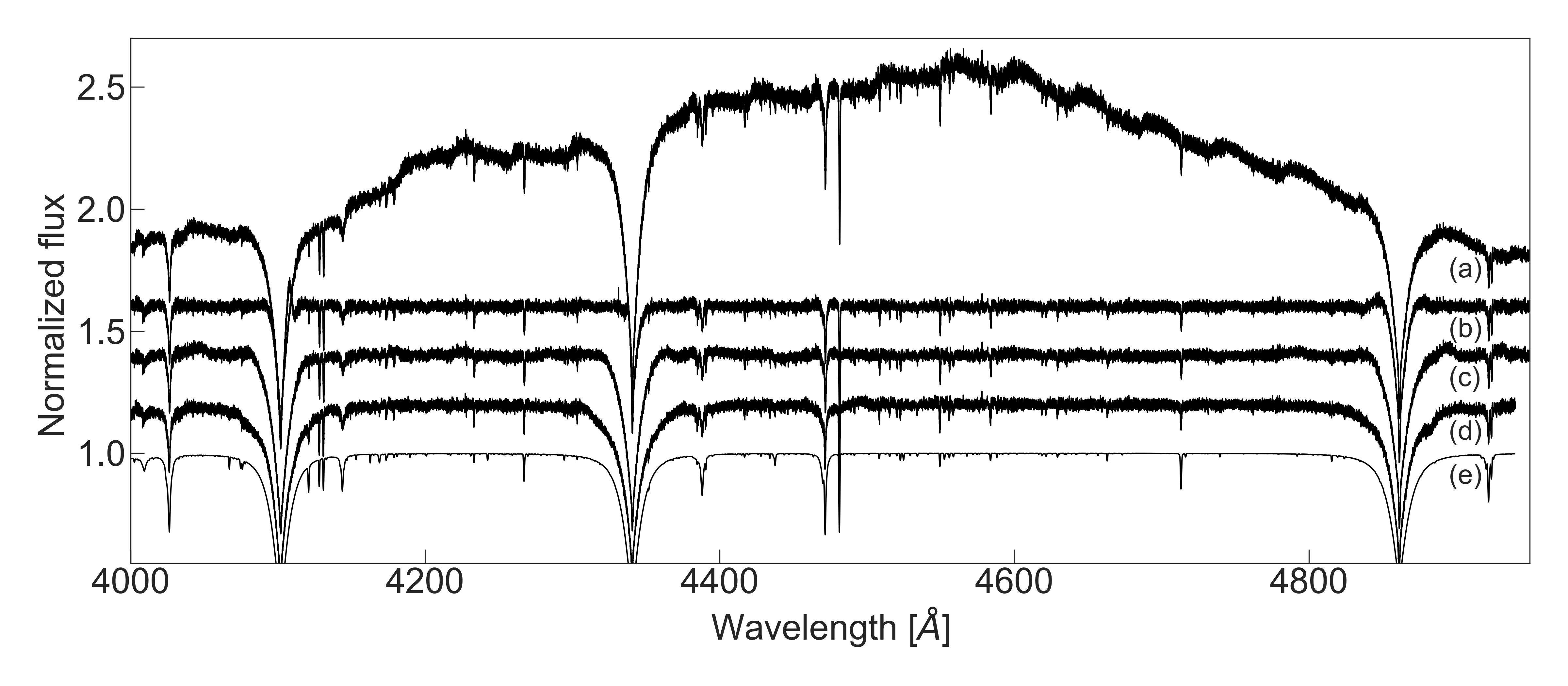}}
      \caption{The {\sc uves} spectrum of galactic main-sequence star HD 41297. Panel (a) shows the order-merged non-normalised spectrum extracted with the ESO data reduction pipeline. Panels (b), (c), and (d) demonstrate the spectra normalised by fitting a low-degree polynomial to the individual \'echelle orders, fitting a 2D smoothing spline to all 30 \'echelle orders in the {\sc uves} blue arm, and iterative fitting of synthetic spectra, respectively. The best fit synthetic spectrum obtained for this star is shown in panel (e) for comparison. See text for details.}
         \label{renorm_broad}
   \end{centering}
   \end{figure*}
   
 \begin{figure*}
   \begin{centering}
            {\includegraphics[clip,width=450pt]{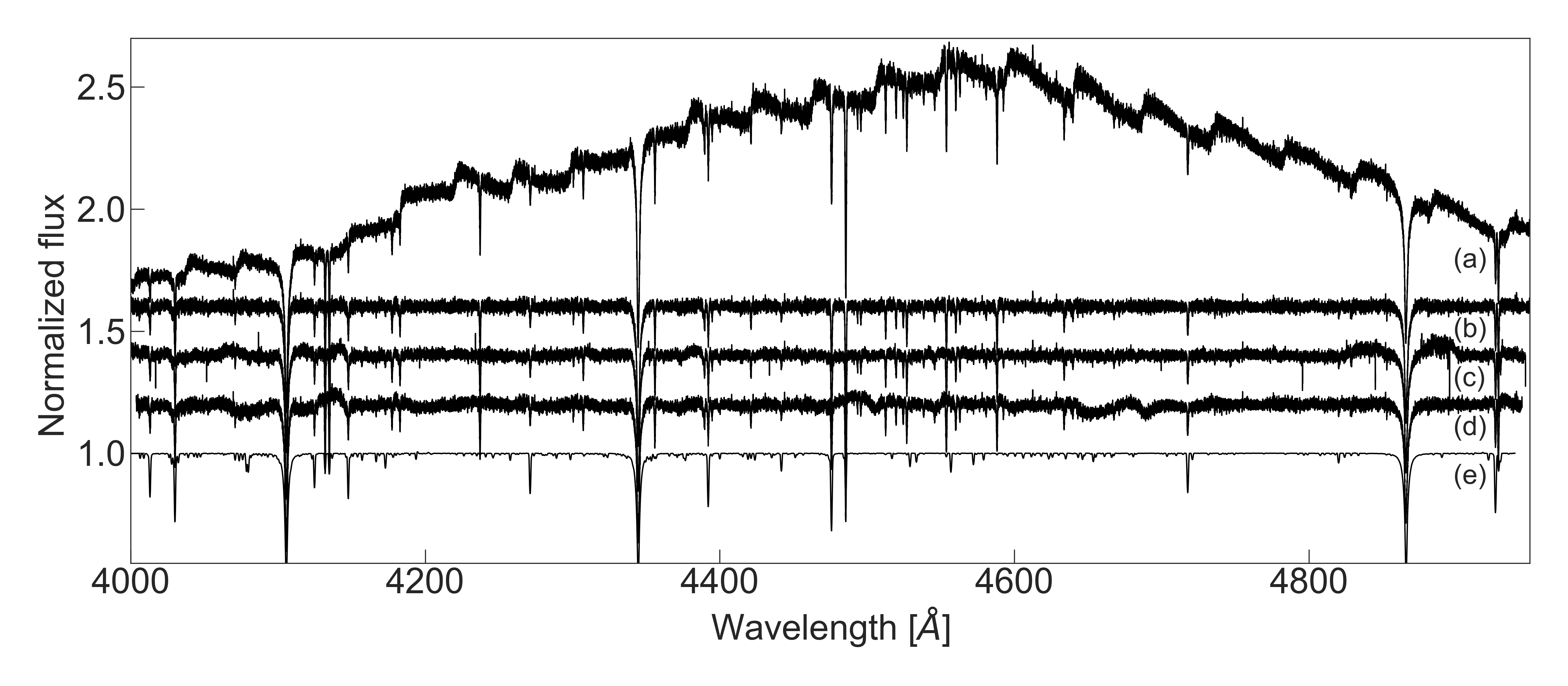}}
      \caption{Same as Fig.~\ref{renorm_broad} but for the {\sc uves} spectrum of supergiant HD 270123. See text for details. }
         \label{renorm_narrow}
   \end{centering}
   \end{figure*}

The power of asteroseismology for probing the effects of mixing in the radiative envelopes and atomic diffusion in intermediate- to high-mass stars is greatly increased when high-precision inference of atmospheric chemical composition is available for these stars (for example, \citealt{Mombarg2022}).
The required high precision in the determination of atmospheric parameters and chemical composition of stars can only be achieved when the raw spectra are extracted and normalised to the local continuum in an optimal way. As mentioned in Sect.~\ref{sec:LP}, a sub-sample of 124 Galactic objects from our program targeted with the {\sc feros} instrument are analysed in the accompanying paper by \citet{Gebruers2022}, where a modified version of the {\sc ceres} data reduction pipeline\footnote{\url{https://github.com/rabrahm/ceres}} \citep{Brahm2017} was used to achieve the required precision in the processed spectra. 

In this work, we employ the {\sc EsoReflex} data reduction pipeline\footnote{\url{https://www.eso.org/sci/software/pipelines/}} \citep{Freudling2013} to process the data obtained with the {\sc uves} instrument. The data reduction steps include bias subtraction, flat fielding, wavelength calibration, order merging, and removal of cosmic hits. Figures~\ref{renorm_broad}(a) and \ref{renorm_narrow}(a) show the extracted, wavelength-calibrated, order-merged {\sc uves} spectra of an example main-sequence star, HD~41297, and an example blue supergiant, HD~270123, respectively. Both spectra contain a prominent wave-like pattern reminiscent of the individual \'echelle orders, and this pattern is detected in the reduced spectra of most of our targets. The pattern cannot be properly accounted for during the normalisation process and strongly affects the end results, especially in the wavelength regions of broad hydrogen lines in the spectra of main-sequence stars. 

Given that the inferior quality of the spectrum extraction and normalisation is in direct conflict with our needs for future high-precision asteroseismic modelling, we opt to develop a custom reduction pipeline instead. We start with an intermediate product of the ESO pipeline in the form of the blaze-corrected, non-merged \'echelle orders. These non-merged spectra are subjected to normalisation to the local continuum where we consider the following two approaches: (i) an `order-by-order' method where each individual \'echelle order is normalised independently by fitting a low-degree polynomial to it (dubbed `method 1'), and (ii) a `2D method' where all 30 \'echelle orders in the blue arm are considered as an ensemble and approximated by a 2D surface of cubic smoothing splines (dubbed `method 2'). The individual normalised orders are ultimately merged together with the flux being represented by the weighted mean in the overlapping regions of the consecutive orders. The weights are assumed to be increasing with the distance from the edge of the order such that the order edges characterised by high levels of noise have negligible contribution to the mean flux value.

We find that both of these two new methods give comparable results. Typical examples of the normalised spectra are shown in Figs.~\ref{renorm_broad}(b,c) and \ref{renorm_narrow}(b,c) for a galactic main-sequence star with broad hydrogen lines and a blue supergiant in the LMC, respectively. One can see that, in the case of blue supergiants that do not display broad hydrogen lines spanning the whole \'echelle order or beyond (cf. Fig.~\ref{renorm_narrow}(c)), a simple order-by-order-based method leads to normalised spectra of high quality. In particular, these spectra are free of the pattern reminiscent of the individual \'echelle orders that is otherwise present when the normalisation is performed on the order-merged spectra delivered by the ESO data reduction pipeline.

In the case of main-sequence stars whose broad hydrogen lines often span the whole or even extend beyond a single \'echelle order, neither of the above-described normalisation methods delivers satisfactory results. We find that the outer wings of hydrogen lines appear to be most affected by the normalisation process (cf. Fig.~\ref{renorm_broad}(c)) which in turn leads to erroneous inference of \teff\ and \logg\ of the star. In the accompanying paper by \citet{Gebruers2022}, a method that assumes simultaneous optimisation of stellar atmospheric parameters and properties of the normalisation function is discussed and employed for the analysis of B- and A-type main-sequence stars. Here, we follow a similar approach. The general idea is to use information on the local continuum placement from a grid of synthetic spectra that spans the parameter range of O- and B-type stars. We employ the Grid Search in Stellar Parameters \citep[GSSP;][]{gssp} software package\footnote{\url{https://fys.kuleuven.be/ster/meetings/binary-2015/gssp-software-package}} to compute a grid of synthetic spectra in the region of the parameter space corresponding to late B-type stars, while O- and early B-type stars are represented by the {\sc ostar2002} \citep{Lanz2003} and {\sc bstar2006} \citep{Lanz2007} {\sc tlusty} grids.

The normalisation procedure (dubbed `method 3') consists of the following steps: (i) the blaze-corrected non-merged observed {\sc uves} spectrum is taken as input and all individual \'echelle orders are normalised by dividing with low-degree ($n = 3$) Chebyshev polynomial functions; (ii) the coefficients of the individual Chebyshev polynomials are optimised by minimising the difference between the normalised \'echelle order and synthetic spectrum in the wavelength range covered by that order; (iii) all individual normalised \'echelle orders are ultimately merged together in a weighted scheme similar to the one described above; and (iv) the best fit synthetic spectrum and normalisation function are obtained by minimising the $\chi^2$ merit function across the entire grid of synthetic spectra in the wavelength range between  4000 and 5000~\AA. Figure~\ref{renorm_broad}(d) shows the spectrum of an example main-sequence star, HD~41297, normalised using the procedure described above. The difference in quality of normalisation with respect to all other methods considered in this work is most noticeable around the Balmer lines though small-scale improvements in normalisation are also seen throughout the entire wavelength interval.

For completeness, we apply `method 3' of normalisation to the spectrum of blue supergiant stars. As can be seen in the typical example in Fig.~\ref{renorm_narrow}(d), there is no improvement in the quality of normalisation compared to the less complex recipe of the order-by-order normalisation method. On the contrary, the quality gets worse in wavelength regions associated with the edges of those \'echelle orders that are rich with rather narrow absoprtion features. Therefore, in the rest of this work, we adopt the `method 1' spectrum normalisation for all blue supergiants in our sample.

In order to validate `method 3' of spectrum normalisation, we pick two main-sequence stars, HD~41297 and HD~45527, that were also observed with the {\sc feros} instrument and analysed in \citet{Gebruers2022}. In the latter study, the {\sc feros} spectra were processed with the modified version of the {\sc ceres} data reduction pipeline and subsequently analysed with the {\sc zeta-Payne} method \citep{Straumit2022}. The {\sc zeta-Payne} method employs a pre-trained neural network to predict normalised synthetic spectrum for an arbitrary set of stellar labels such as \teff, \logg, \vsini, and [M/H] (and optionally microturbulence $\xi$). The normalisation function is represented by a series of Chebyshev polynomials and is optimised in the method along with the above-mentioned atmospheric parameters of the star. Application of our `method 3' for normalisation to the {\sc uves} spectra of HD~41297 and HD~45527 delivers atmospheric parameters that are consistent within the 1$\sigma$ confidence interval with those reported in \citet{Gebruers2022}.

Finally, we quantify the effect of spectrum normalisation on the inferred atmospheric parameters of three main-sequence stars 
and three blue supergiants. 
In this exercise, we consider three different approaches for spectrum normalisation: (i) fitting and dividing the blaze-corrected order-merged observed spectrum (c.f. Figs~\ref{renorm_broad}(a) and \ref{renorm_narrow}(a)) with a low-degree polynomial (that is the procedure commonly used in the literature); (ii) our `method 1'; and (iii) our `method 3'. All three methods deliver atmospheric parameters for the blue supergiants that are consistent within the 1$\sigma$ confidence interval. For the main-sequence stars, however, we find that our `method 3' provides a substantially better representation of the observed spectrum by the best fit model and delivers \teff\ and \logg\ that differ by some 1\,000~K and 0.9~dex, respectively, from the parameters inferred with the other two methods. Therefore, in the rest of this work, we adopt`method 3' for the spectrum normalisation of all main-sequence stars in our sample according to initial classification of targets' type (see section \ref{sec:spectroscopic_analysis} for details on classification). 

\section{Analysis of the {\sc uves} and {\sc feros} spectroscopy}\label{sec:spectroscopic_analysis}

In order to provide an empirical estimate for the effective temperature and luminosity of the star and to inform the above-described normalisation process, we first perform an initial classification of all the spectra in our sample. The classification is done by means of visual comparison of the selected spectral line intensities and their ratios to the digital spectral classification atlas of Gray\footnote{\url{https://ned.ipac.caltech.edu/level5/Gray/frames.html}} \citep{Gray2009}. We use the pair of \ion{Mg}{ii}~4481~\AA\ and \ion{He}{i}~4471~\AA\ spectral lines as a main diagnostic to assign a spectral type to all stars in the sample. 
Luminosity classes were decided upon from the strengths of spectral lines of Si, N, O and Fe, depending on the \teff\ regime: (i) \ion{Si}{iv}~4116~\AA\ to \ion{He}{i}~4121~\AA\ and \ion{N}{iii}~4379~\AA\ to \ion{He}{i}~4387~\AA\ line strength ratios for late O-type stars; (ii) \ion{O}{ii}~4070~\AA, 4348~\AA, 4416~\AA\ and \ion{Si}{iii}~4553~\AA\ spectral line strengths for early B-type stars; and (iii) line strength of the \ion{Si}{ii}~4128/4130~\AA\ doublet for late B-type stars. In this way, we identify  5, 49, and 63 stars of late O-, early-B, and late-B spectral classes, respectively. Moreover, not unexpectedly, we confirm our sample to be dominated with evolved blue supergiants (see Table~\ref{tab:results}).

\subsection{Identification of line profile variations and binary candidates}

The strategy of multi-epoch spectroscopic observations allows us to check all our targets for possible binary nature and for particularly strong line profile variations (LPV) due to intrinsic variability of the star. Both of these phenomena are readily revealed in the analysis of radial velocity (RV) variations that occur due to either global line shifts in the spectra of spectroscopic binaries or line distortions (asymmetries, skewness, etc.) due to variability intrinsic to the star (abundance spots, pulsations, etc.). 

We employ the Least-Squares Deconvolution \citep[LSD;][]{Donati1997} method as implemented in \citet{lsd} to compute high S/N average profiles from the absorption lines for all our targets.  These average profiles are subsequently analysed to deduce RV values by fitting them with an asymmetric Gaussian profile of the following form:
\begin{equation}
      A(x,x_0,\sigma,r) = \frac{1}{\sqrt{\pi \sigma^2} (r+1) } 
      \begin{cases}  
      \exp\,\displaystyle{\left\{- \frac{(x-x_0)^2}{2 \sigma^2 }\right\}}\hspace{-0.2cm}& \text{if $x > x_{0}$},\\[0.3cm]
      \exp\,\displaystyle{\left\{- \frac{(x-x_{0})^2}{2 r^2 \sigma^2 }\right\}}\hspace{-0.2cm}&\text{otherwise}. \\
      \end{cases}
   \label{asymgaussian}
\end{equation}
Here $r$ is the asymmetry coefficient, $\sigma$ controls the width of the profile, and $x_0$ represents the position of the minimum of the profile.

Spectroscopic classification of a star that presumably shows line profile variations due to its intrinsic variability requires a joint interpretation of the inferred RV and coefficient of asymmetry \citep[c.f.][]{Aerts1999,DeCatAerts2002,Telting2006}. We classify a star as a spectroscopic binary (dubbed either SB1 or SB2 in Table~\ref{tab:results} for the case of single- and double-lined binary, respectively) when (i) the RV difference between the two epochs is larger than 3$\sigma$, or (ii) either a global shift of spectral lines and/or double-lined nature are confirmed visually. We take a conservative 3$\sigma$ interval to account for the fact that the obtained uncertainties are purely statistical resulting from the least-squares fitting method. They are hence an underestimation ignoring any systematic uncertainties. The same two criteria, that is visual inspection and a relative RV difference below 3$\sigma$, are used to classify a star as ``LPV'' in Table~\ref{tab:results}, where the visual inspection focuses on LPV in the form of (local) spectral line distortions instead of global shifts. Furthermore, stars that show large global asymmetries in their observed line profiles are put into a separate class, dubbed ``asymm'' in Table~\ref{tab:results}. Ultimately, stars that do not show significant RV variability nor appreciable line asymmetries in the two observed epochs are classified as apparently constant (dubbed ``const'' in Table~\ref{tab:results}). Ambiguous classifications are supplemented with ``?''. The results of our classification are summarised in Table~\ref{tab:results} (column designated ``variability'').

\begin{figure*}
   \centering
   \includegraphics[trim={0.2cm 0 1cm 1cm},clip,width=230px]{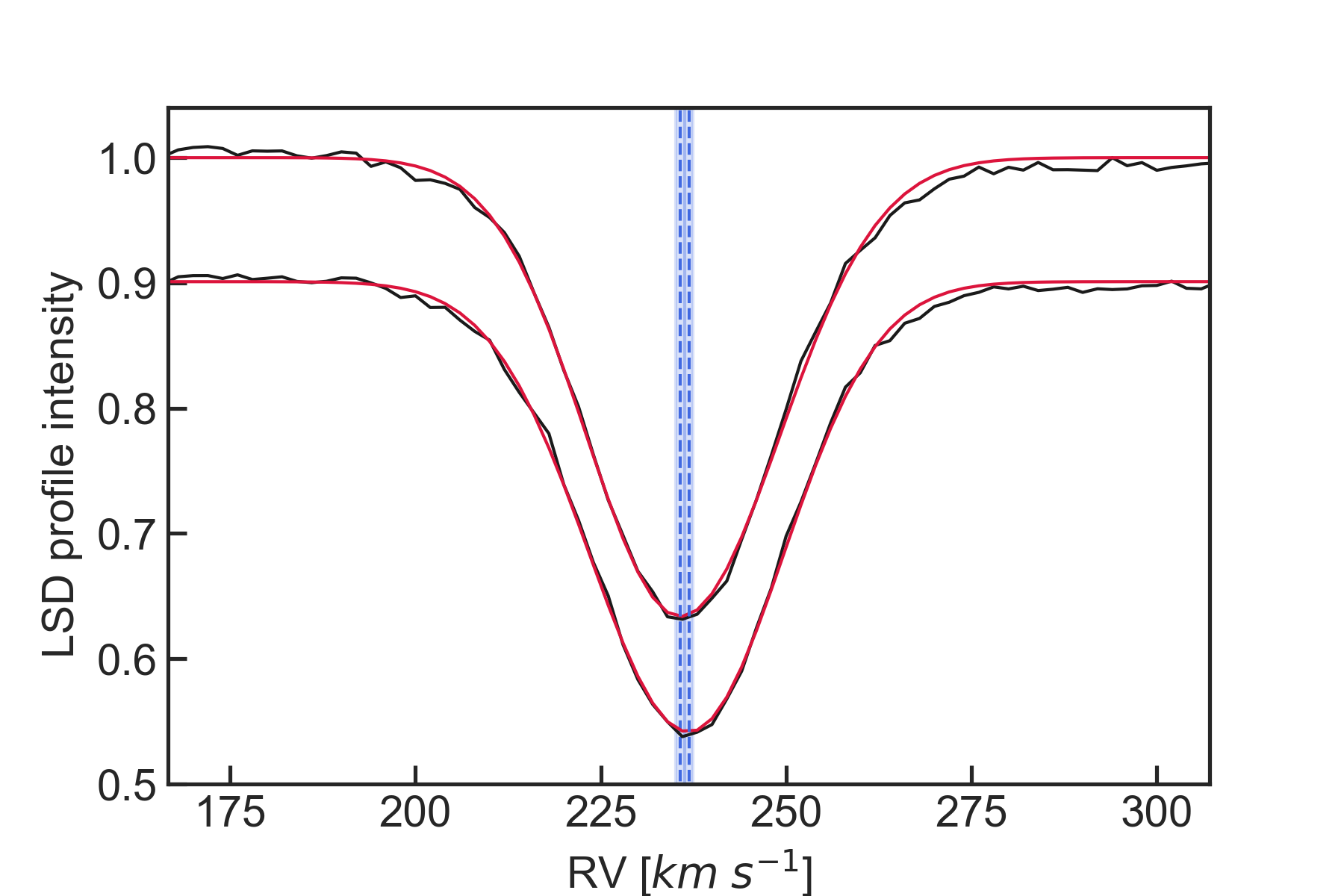}
   \includegraphics[trim={0.2cm 0 1cm 1cm},clip,width=230px]{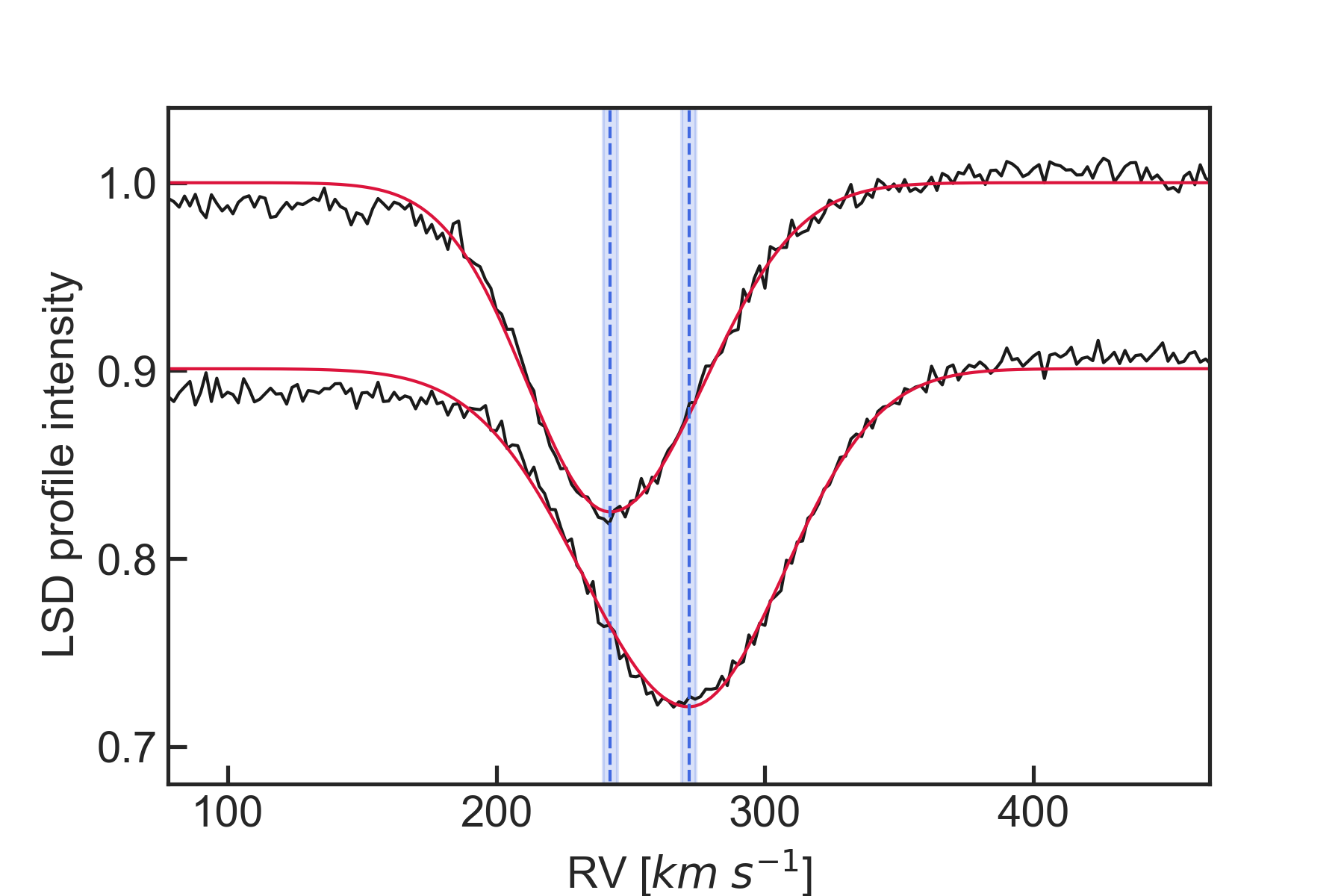}
   \includegraphics[trim={0.2cm 0 1cm 1cm},clip,width=230px]{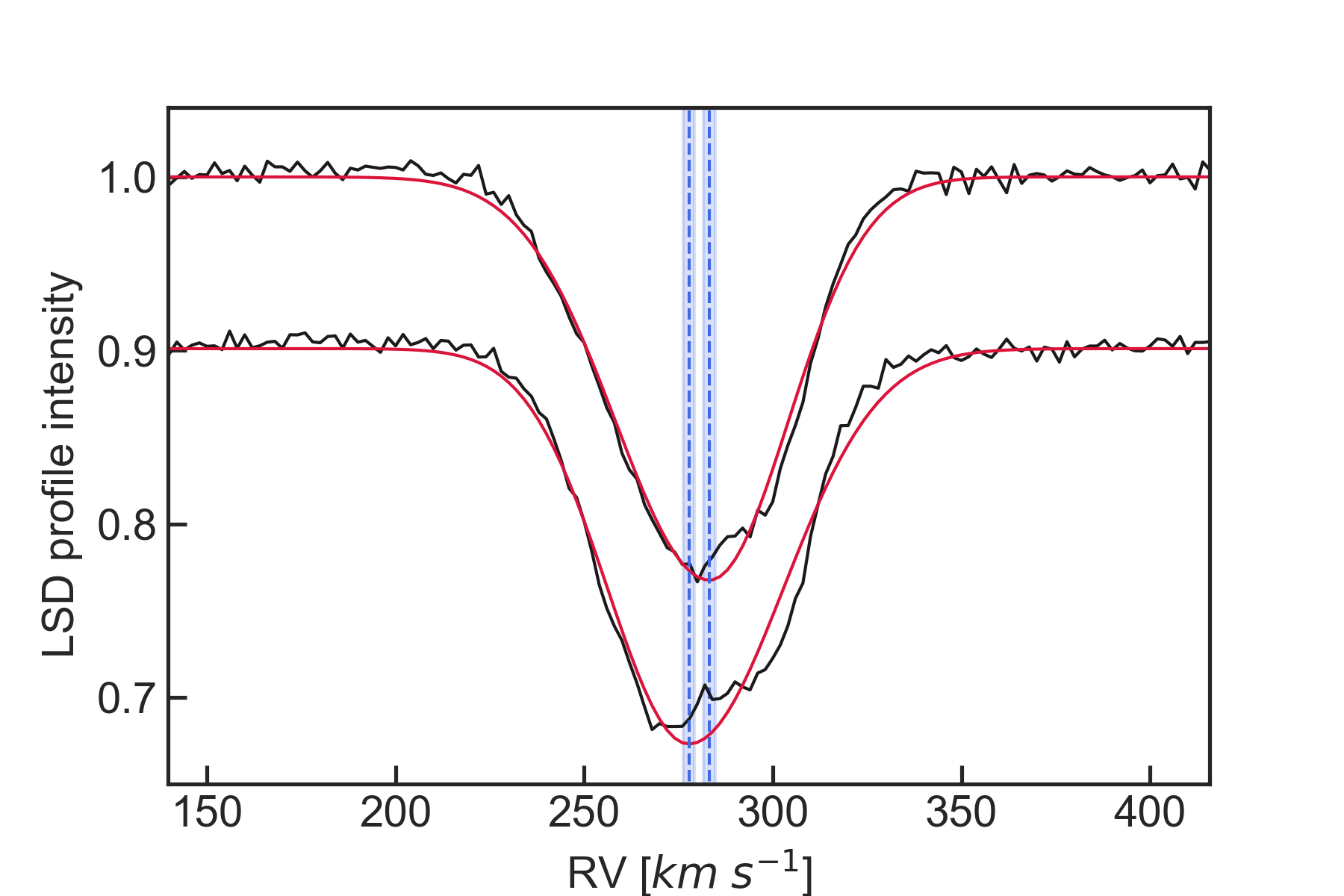}
   \includegraphics[trim={0.2cm 0 1cm 1cm},clip,width=230px]{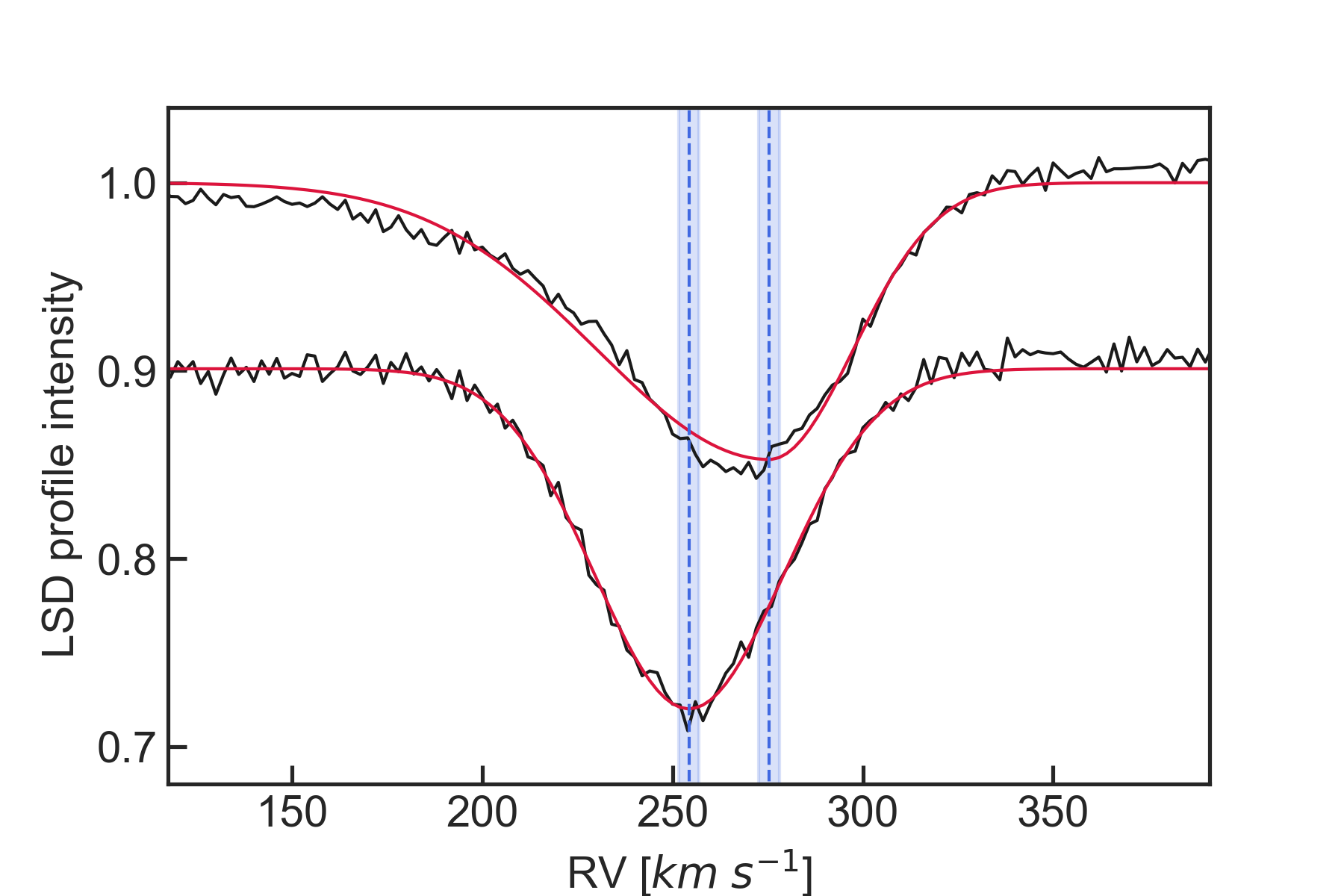}
   \caption{Comparison of the observed LSD profiles with the best fit asymmetric Gaussian profile models. From top left to bottom right: SK -70 77, HD 268729, SOI 404, and HD~268654 (all were classified as blue supergiants). Blue lines indicate the inferred RVs while the shaded areas reflect the corresponding 1$\sigma$ uncertainty intervals.}
              \label{LSDRV}
\end{figure*}

We stress that no distinction is made at this stage between intrinsic variability of the star originating from pulsations, rotational modulation, or any other physical mechanism, due to the limited amount of information (only two spectroscopic epochs) and simplicity of the employed classification method. Figure~\ref{LSDRV} shows examples of the observed LSD profiles with the best fit asymmetric Gaussian profile models overlaid for SK~-70~77 (top left), HD~268729 (top right), SOI~404 (bottom left), and HD~268654 (bottom right), which represent typical behaviour for different variability groups. These stars are classified as apparently constant (``const''), spectroscopic binary candidate (``SB1''), line profile variable (``LPV''), and one showing appreciable line asymmetries (``asymm''), respectively. 

In this way we identify 65 (21), 6 (1), 15 (13), and 22 (15) apparently constant stars, stars with LPV, spectroscopic binary candidates, and stars with appreciable line asymmetries, respectively, in the {\sc uves} ({\sc feros}) sample. Of those detections, 0 (0), 3 (1), 12 (11), and 9 (12) are tentative classifications meaning that more observational epochs are required to draw definitive conclusions as to the exact type of spectroscopic variability.

\subsection{Inference of the projected rotational velocity \vsini}\label{sect:VsiniInference}

Measuring the projected rotational velocity \vsini\ from the spectrum of an early-type star is not a straightforward task. Lines are typically broadened by several physical mechanisms acting simultaneously in the atmospheres of these stars, ultimately leading to an appreciable overestimation of the \vsini\ parameter when other mechanisms are ignored in the analysis. \citet{Simon-Diaz2014} and \citet{Simon-Diaz2017} stress that the contribution of a macroturbulent velocity, $\Theta$, to the total broadening of spectral lines in O- and B-type stars can be as large as the contribution associated with \vsini, and in many cases even exceed it. \citet{Aerts2014} demonstrate that the macroturbulent and projected rotational velocities cannot be reliably inferred from single snapshot spectroscopic observations of intrinsically variable (spotted and/or pulsating) early-type stars whose spectral lines show pronounced LPV. In particular, the authors stress that \vsini\ and $\Theta$ vary appreciably and in anti-phase with each other during the pulsation cycle, where $\Theta$ can vary by as much as a few tens of \kms during the pulsation cycles.

\citet{Cantiello2009} report a strong anti-correlation between the microturbulent velocity $\xi$ and surface gravity \logg\ in the LMC sample of B-type stars analysed in \citet{Hunter2008}. The authors suggest a connection between microturbulence and velocity fields triggered by gravity waves induced in the iron sub-surface convection zone. It is also worth mentioning that the most evolved among B-type stars studied in \citet{Hunter2008} show microturbulence as large as some 20~\kms\ which is in line with \citet[][Chapter 17]{Gray2008} who also suggests vertical stratification of microturbulence in photospheres of evolved hot stars. In summary, the total line broadening observed in spectra of intermediate- to high-mass stars is typically a combination of several line broadening mechanisms such as stellar rotation, micro- and macroturbulence, and stellar oscillations. In the most extreme cases, all of these mechanisms will have comparable magnitudes which unavoidably leads to strong degeneracy between the corresponding parametric descriptions in the modelling of individual spectral lines or their ensembles.

\begin{figure*}
            {\includegraphics[trim={0.2cm 0 0 0},clip,scale=0.6]{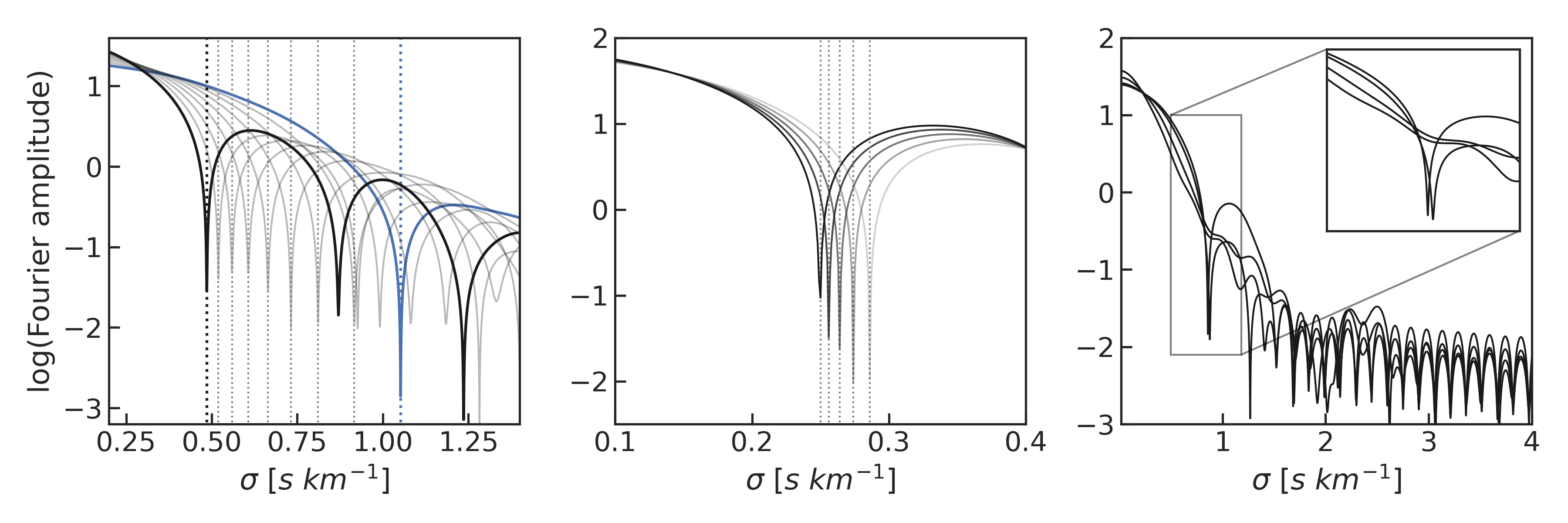}}
      \caption{Fourier transforms of synthetic line profiles with positions of the first zero indicated with vertical dotted lines.  Left panel:  the \vsini\ parameter is varied from 42~\kms\  (blue) to 90~\kms\ (black) with a step of 6~\kms\ (grey). Middle panel: the linear limb darkening coefficient $\epsilon$ is varied from 0.2 (black) to 1.0 (light grey), while \vsini\ is kept fixed at 50~\kms. Right panel: a different contribution of the macroturbulence $\Theta$ to the total line broadening is assumed: $\Theta/v \sin i = 0.3, 0.6, 1.2$ and $1.8$, with \vsini\ fixed to 50~km\,s$^{-1}$. The inset is a close up into the region of the position of the first zero. }
         \label{fourier_rp}
   \end{figure*}

In order to provide as reliable estimate of \vsini\ as possible for all stars in our sample, we employ a method that is based on the Fourier transform (FT) of a spectral line profile \citep[][Chapter 2]{Carroll1933,Gray2008}. This method relies on the fact that the FT of a rotationally broadened profile consists of multiple lobes separated by zeros (see the right panel of Fig. \ref{fourier_rp}). The position of the first zero and the associated frequency in the FT translates directly into the projected rotational velocity \vsini\ of the star (denoted as \vsini$_{\rm FT}$), provided that the dominant line broadening contribution delivers a time-independent profile \citep[see Eq.\,(1) in][]{Simon-Diaz2014}. An application is shown in the left panel of Fig. \ref{fourier_rp}. A convolution of the rotational profile with profiles due to other broadening mechanisms transforms into multiplication in Fourier space, which preserves the positions of the zeros. Therefore, identification of the first zero in the FT allows us to isolate the effect of \vsini\ from other broadening mechanisms affecting the line profile, provided that: (i) additional line broadening mechanisms can be represented by a convolution, that is they are independent of each other; and (ii) these mechanisms do not share the same frequency regime with rotational broadening. Such conditions are typically not satisfied when \vsini$\lesssim50$~\kms and/or when time-dependent phenomena are dominant in the line-forming region \citep{Aerts2014}. Note that the FT method also becomes less reliable if the macroturbulent velocity parameter $\Theta$ exceeds \vsini\ as lobes and zeroes smear out, as shown in the right panel of Fig.~\ref{fourier_rp}.

The position of the first zero in the FT of the line profile is also affected by the limb darkening effect. Following \cite{limb_darkening}, the linear limb darkening coefficient $\epsilon$ is found to be around 0.35 in the case of B-type stars, which is significantly lower than the value of $\epsilon=0.6$ often used as reference in studies of solar-type stars. The uncertainty in $\epsilon$  leads to a noticeable shift of the first zero in the FT of the line profile (see the middle panel of Fig. \ref{fourier_rp}) which translates into a small yet systematic difference of up to 10\,$\%$ in the inferred value of  \vsini. Therefore, we fix $\epsilon=0.35$ for the determination of \vsini$_{\rm FT}$ and neglect any small object-to-object variations of the coefficient within the sample. 

Because the FT method is most suitable for isolated and undistorted line profiles, a careful selection of the latter is required. We rely on the \ion{Si}{ii}~4128~\AA\ and \ion{Si}{iii}~4552~\AA\ lines for the late and early B-type stars in our sample, respectively. The \ion{Mg}{ii}~4481~\AA\ line is used in all other cases unless its visual inspection reveals either poor quality in terms of S/N or a high degree of asymmetry. In those cases, we choose \ion{He}{i}~4471~\AA\ as an alternative, even though helium lines are generally not the best choice because of their large intrinsic pressure-dominated broadening. However, because of a general lack of metal lines in the spectra of hot stars and our sample being dominated by blue supergiants characterised by low \logg\ values, the \ion{He}{i}~4471~\AA\ line is a reasonable alternative in the absence of isolated undistorted metal lines. Furthermore, given that LPV due to stellar pulsations may lead to appreciable variations in \vsini$_{\rm FT}$ for each star \citep{Aerts2014}, we infer the stellar projected rotational velocities from individual epoch spectra instead of their co-added version. The finally accepted value of \vsini$_{\rm FT}$ is the average between the two epochs of observations while its temporal variability (if detected) is interpreted as systematic uncertainty due to temporal variability of the line profiles. The respective \vsini\ values and their uncertainties are reported in Table~\ref{tab:results_lp} (column designated \vsini$_{\rm FT}$).

\begin{figure*}
   \begin{center}
      {\includegraphics[trim={1cm 0 0 0},clip, width=0.58\linewidth]{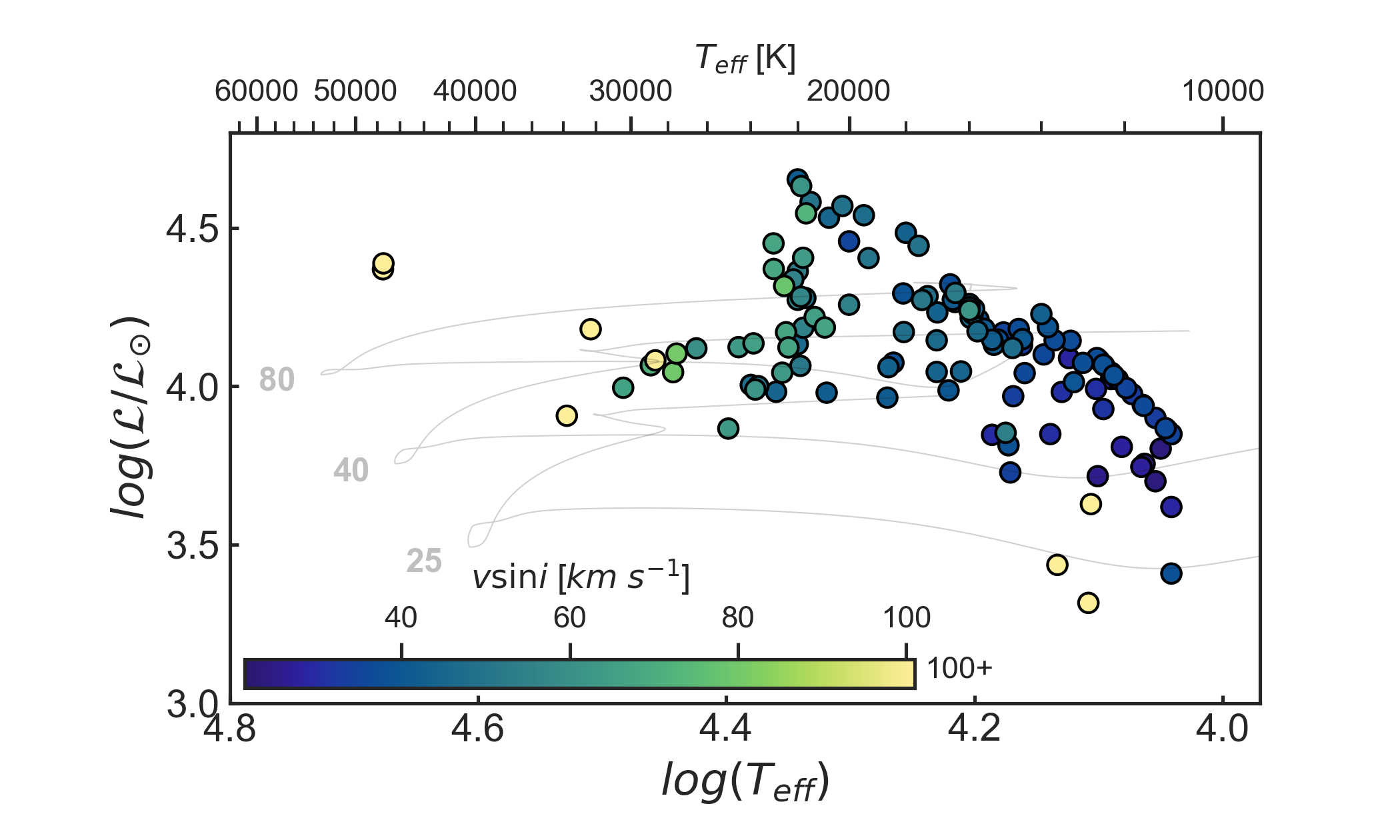}}
      {\includegraphics[trim={0cm 0 0cm 0cm}, clip, width=0.33\linewidth]{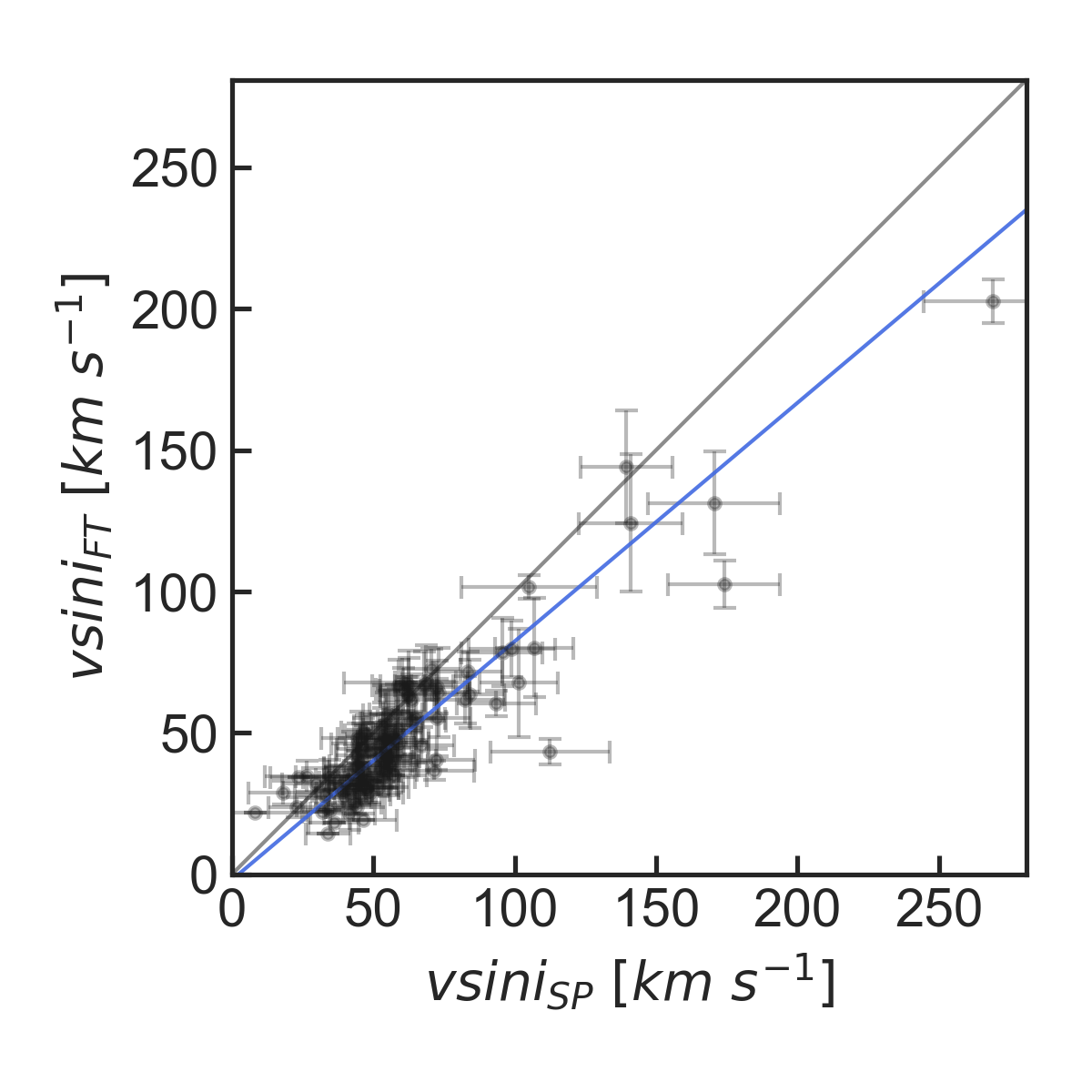}}
      \caption{Results of \vsini\ inferences. Left panel: positions of the sample stars in the spectroscopic HR diagram. Individual objects are colour-coded by their \vsini\ value inferred from the Fourier method. Right panel: comparison of \vsini$_{\rm SP}$  with \vsini$_{\rm FT}$. The grey line shows the case where \vsini$_{\rm FT}$ and \vsini$_{\rm SP}$ are equal, while the blue line is an error-weighted fit to the observed values.  }
         \label{rotation}
    \end{center}
   \end{figure*}

 Another commonly used method to infer \vsini\ is through fitting either individual line profiles or the entire spectrum of the star by a set of synthetic models. As discussed above, we expect this method to lead to an over-estimation of \vsini\ if other broadening mechanisms are ignored in the fitting procedure while they may contribute to the total line broadening, as is often the case in the spectra of early-type stars. To quantify the effect, we estimate \vsini\ for all stars in the sample by performing the fit to the full spectra with synthetic models and by assuming no macroturbulent velocity. The respective parameter values are reported in Table~\ref{tab:results_lp} and are compared to the \vsini\ values inferred with the FT method in Fig.~\ref{rotation} (right panel). One can see that the \vsini\ values inferred from the method of synthetic spectra are up to 30\,\% higher compared to those inferred with the FT method. Indeed, the former method gives an indication of the total broadening when other broadening mechanisms are not directly included in the fitting, while the latter method to a large extent treats the rotational broadening separately from the other mechanisms.

\subsection{Inference of the effective temperature \teff\ and surface gravity \logg\ of the sample stars} \label{sect:TeffLoggInference}

\begin{figure*}
   \begin{center}
            {\includegraphics[clip,scale=0.5]{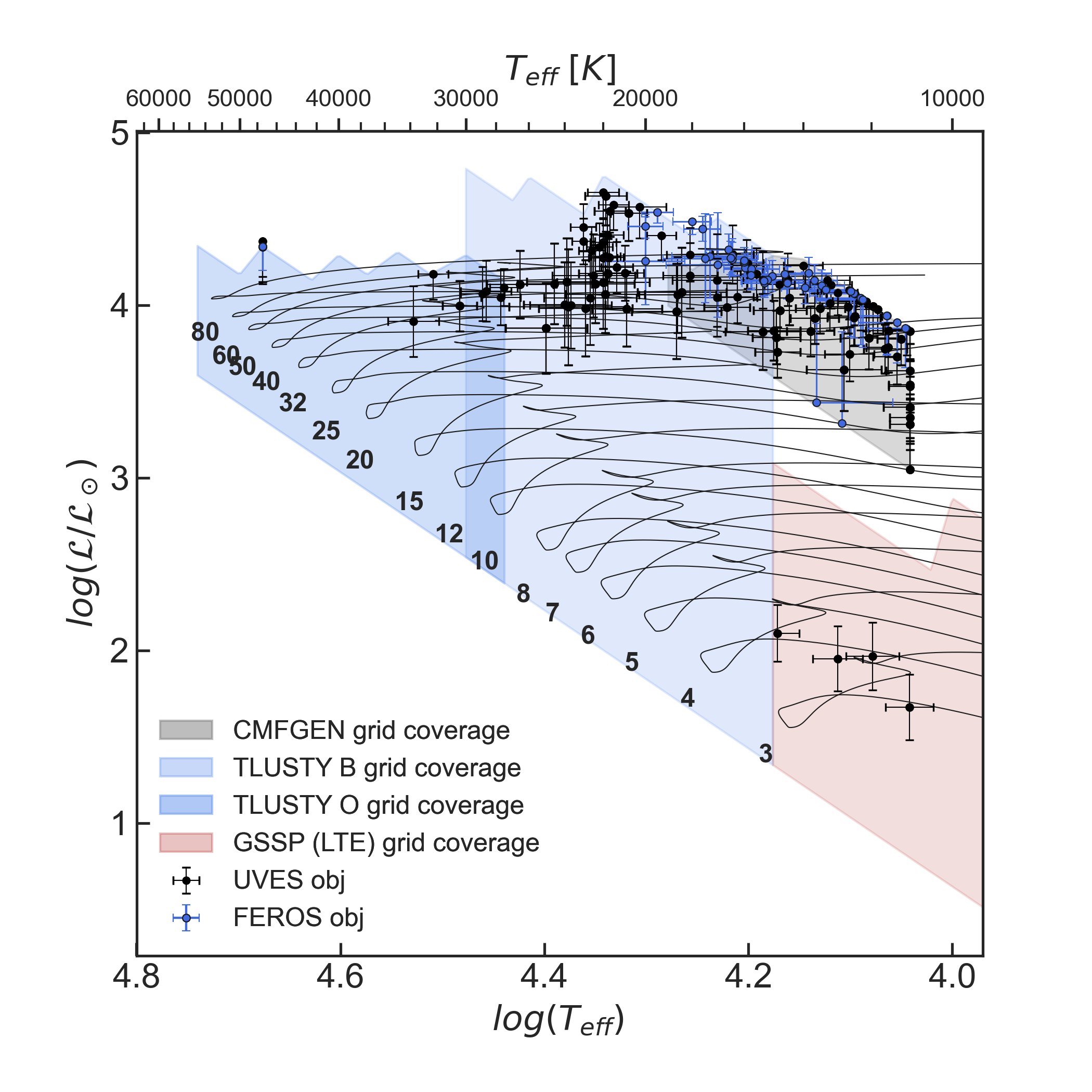}}
            {\includegraphics[clip,scale=0.5]{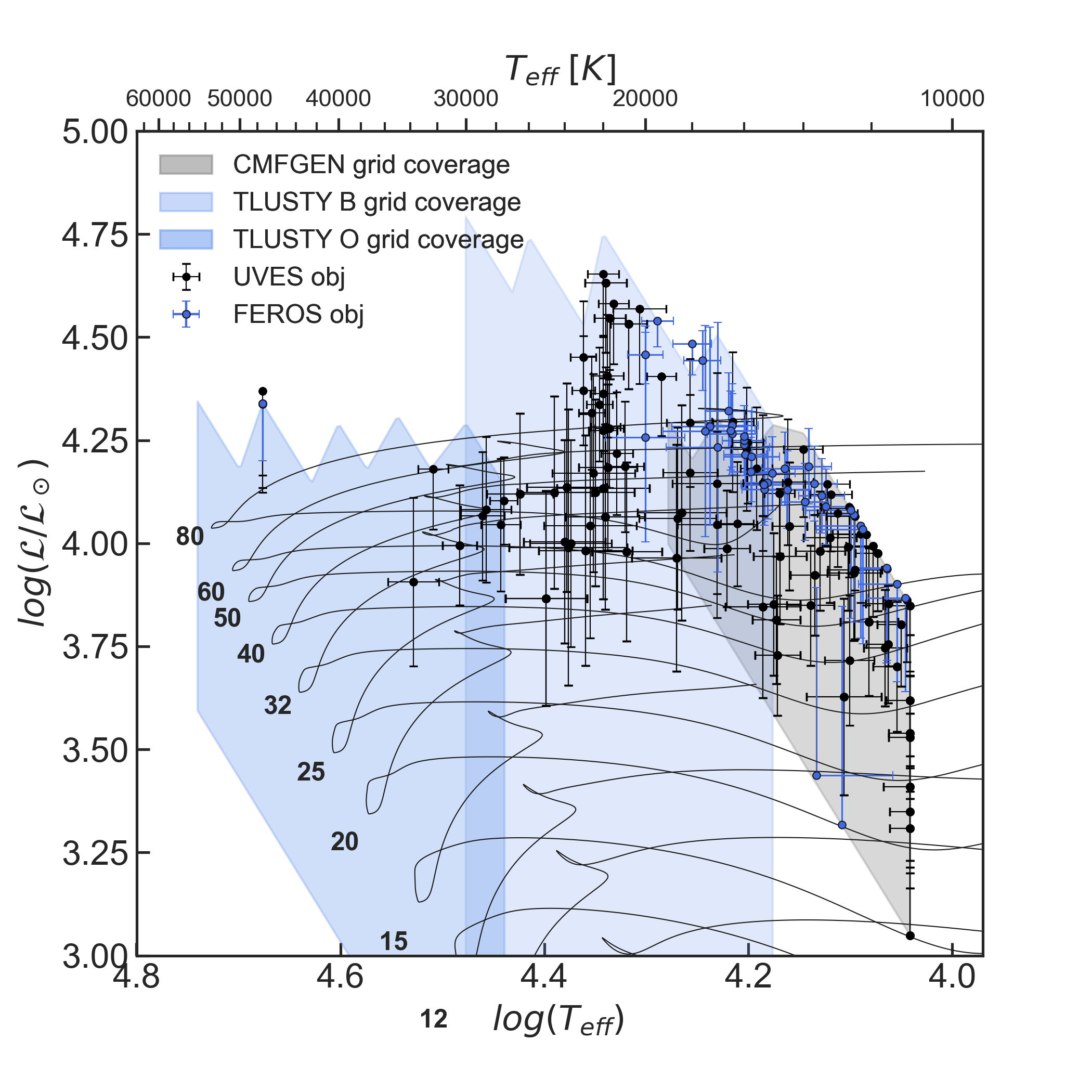}}
      \caption{Positions of the sample stars in the spectroscopic
      $\log(T_{\text{eff}})$--$\log (\mathcal{L}/\mathcal{L}_{\odot})$ HR diagram. The {\sc uves} and {\sc feros} samples are shown with black and blue symbols with error bars, respectively. 
      Regions of the parameter space covered by the {\sc gssp}, {\sc tlusty}, and {\sc cmfgen} models are marked with red, blue, and grey shaded areas, respectively. Nonrotating {\sc mesa} stellar evolution tracks computed for $Z=0.006$ are shown with black solid lines and are taken from \citet{Burssens2020}. Left panel: full parameter range. Right panel: zoom into the high-luminosity region where the bulk of the sample is situated.  } 
         \label{spec_hrd}
    \end{center}
   \end{figure*}

As mentioned in Sect.~\ref{sec:spec_data_reduction},  we rely on three grids of synthetic spectra for our analyses. We fit the normalised observed spectra with those predicted by models in the grids by means of a $\chi^2$-minimisation to find the best-fit parameters. The fitting is performed in the wavelength range of 4000-5000~\AA, which includes a plethora of metal and helium lines, as well as the $H_\beta$, $H_\gamma$, and $H_\delta$ lines. The minimisation is performed with the Trust Region Reflective algorithm \citep{TRF}  applied to the \teff -- \logg -- \vsini\ parameter space, where the model spectra for the parameter values are interpolated from those in the grids at each step of the iteration. The {\sc ostar2002} \citep{Lanz2003} and {\sc bstar2006} \citep{Lanz2007} grids of {\sc tlusty} models are employed for the analysis of a few predominantly upper main-sequence stars in the sample. Furthermore, owing to the fact that our sample is dominated by evolved blue supergiants in the LMC with \teff$\gtrsim$10\,000~K and that the {\sc ostar2002} and {\sc bstar2006} grids do not extend towards low \logg\ values characteristic for supergiant stars, we compute a dedicated grid of {\sc cmfgen} \citep{Hillier1998} models covering \logg\ values as low as 1.7~dex. The parameter space of the grid is dictated by the lowest \logg\ possible for a given \teff\ such that the star remains in hydrostatic equilibrium. This way, the grid extends from 11\,000~K to 19\,000~K in \teff\ while \logg\ ranges from 1.7-2.5~dex at 11\,000~K and from 2.2-2.5~dex at 19\,000~K. Owing to the large computation time, a step of 1\,000~K and 0.1~dex is chosen for \teff\ and \logg, respectively. The parameter space covered with the {\sc cmfgen} grid is indicated with the grey shaded area in Fig.~\ref{spec_hrd}.

A grid of the LTE-based {\sc gssp} \citep{gssp} models is used for the analysis of a few unevolved stars with \teff$\lesssim$15\,000~K, a range in the parameter space not covered by the {\sc bstar} grid of the {\sc tlusty} models and where the use of {\sc cmfgen} is not justified. The {\sc gssp} grid is designed such as to cover a \teff\ range between 9\,000~K and 15\,000~K. By analogy with the {\sc cmfgen} grid and for exactly the same reason (i.e. the star needs to remain in the hydrostatic equilibrium), the \logg\ range is set to 2.5-4.5~dex for \teff$<$10\,000~K and to 3.0-4.5~dex for \teff$\geq$10\,000~K. The step in \teff\ is 100~K and 250~K for \teff$<$10\,000~K and \teff$>$10\,000~K, respectively, while the step in \logg\ is fixed to 0.1~dex. We assume  the solar chemical composition as derived in \citet{Grevesse2007}.  This way, we adopt adequate tools for each of the three regimes of stars in our sample. The regions of the parameter space covered by the grids of {\sc gssp}, {\sc tlusty}, and {\sc cmfgen} models used in this study are illustrated with shaded areas in Fig.~\ref{spec_hrd}. 

\begin{figure*}
   \begin{center}
            {\includegraphics[trim={1.4cm 0.3cm 0cm 0.3cm}, clip, width=0.5\linewidth]{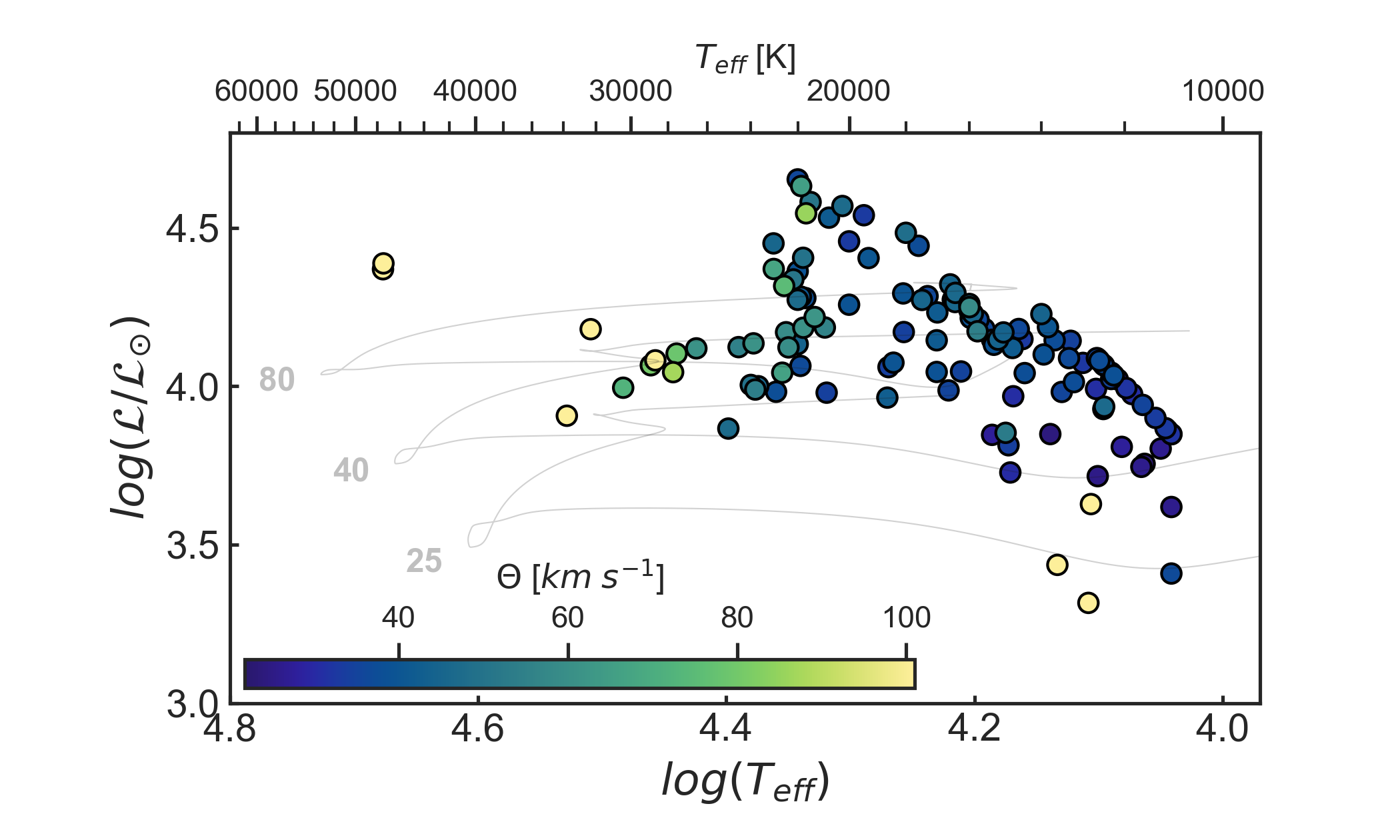}}
            {\includegraphics[trim={0cm 0cm 0cm 0cm},clip, width=0.42\linewidth]{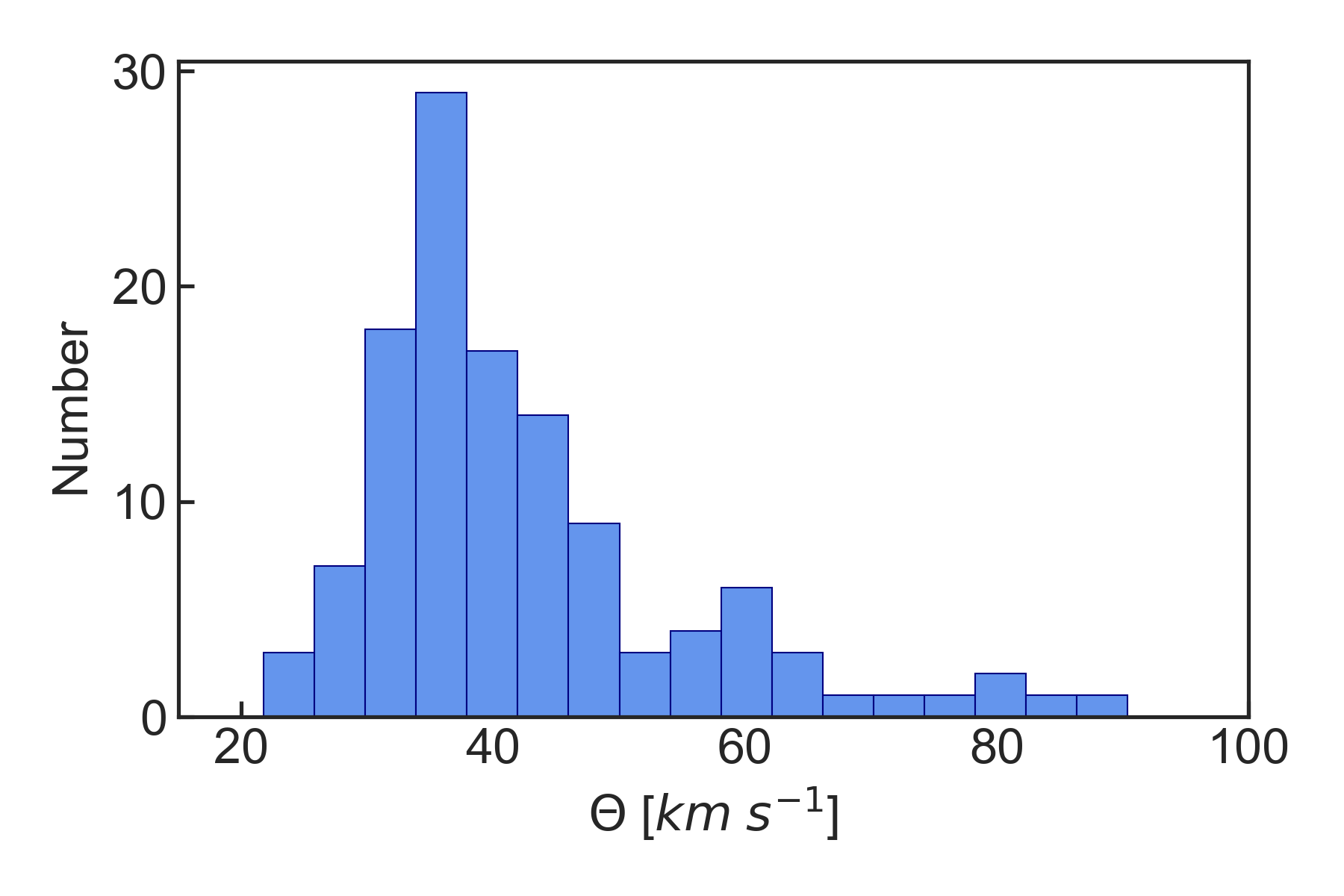}}
      \caption{Results of the inference of the macroturbulence parameter $\Theta$. Left panel: stellar positions in the spectroscopic HR diagram colour coded by the inferred value of the macroturbulent velocity parameter $\Theta$. Value of the macroturbulent velocity is an error-weighted of two epochs. Right panel: distribution of the macroturbulent velocity parameter $\Theta$ across the sample.}
         \label{macroturbulence}
    \end{center}
   \end{figure*}

Spectroscopic inference of surface abundances for the sample stars is the subject of a follow-up paper. For this reason, we fix the microturbulent velocity $\xi$ to 2~\kms\ and 10~\kms\ for all unevolved ({\sc gssp} and {\sc tlusty}) and blue supergiant ({\sc cmfgen}) models, respectively. Furthermore, we assume solar and about half solar ([M/H]=-0.4~dex) metallicity for all the Galactic and LMC \citep{Choudhury2021} targets, respectively.
The results of the spectroscopic analysis based on this approach are summarised in Table~\ref{tab:results}, where the grid of used atmosphere models for a particular target is indicated with the superscript ``a'', ``b'', and ``c'' for {\sc gssp}, {\sc tlusty}, and {\sc cmfgen}, respectively. The positions of the stars in the spectroscopic 
$\log T_{\text{eff}}$ -- $\log(T_{\text{eff}}^4/g)$
HR diagram are shown in Fig.~\ref{spec_hrd}.  The \vsini\ values inferred from this fitting approach (representing the total line broadening) are also provided in Table~\ref{tab:results_lp}.
As it was suggested in Sect.~\ref{sec:spec_data_reduction} already, the majority of the sample stars are found to be blue supergiants, with just a few objects in their main-sequence stage of evolution. We find four objects that have masses in the range between some 2.5 and 4.0~M$_{\odot}$, with three of them being located within the SPB instability region while the fourth object is found at the border when uncertainties are taken into account. The rest of the stellar sample is consistent with masses in excess of 10~M$_{\odot}$.

\subsection{Inference of the macroturbulent velocity}\label{sec:VmacroInference}

As discussed in Sect.~\ref{sect:VsiniInference}, massive stars are known to show macroturbulent velocity fields that are often comparable to or exceed the spectral line broadening associated with the stellar rotation \citep[e.g.,][]{Aerts2014,Simon-Diaz2014,Simon-Diaz2017}. There is currently no consensus achieved as to the exact physical mechanism(s) responsible for the macroturbulent broadening in the spectral lines of massive stars. Two main hypotheses are (i) the collective effect of gravity waves, either heat-driven \citep{Aerts2009} or stochastically excited at the boundary between the convective core and radiative envelope in these stars \citep{Aerts2015}; and (ii) velocity perturbations due to turbulent pressure in the subsurface convection zone driven by the iron opacity peak at some 150~kK \citep{Cantiello2021}. These two hypotheses are not mutually exclusive \citep{Bowman2020b,Bowman2022e}.

Our stellar sample is well suited to infer the macroturbulent velocity $\Theta$, as well as  to estimate the relative contributions of its radial $\Theta_{\rm R}$ and tangential $\Theta_{\rm T}$ components. Indeed, we were able to acquire high-resolution, high S/N, multi-epoch spectroscopic data for the majority of objects in the sample. Furthermore, the sample is composed of stars spanning a wide range of stellar mass ($M\in$[3, 80]~M$_{\odot}$) and spectroscopic luminosity ($\log{\rm (\mathcal{L}/\mathcal{L}_{\odot})} \in$[1.8, 4.5], cf. Fig.~\ref{spec_hrd}). Therefore, we proceed with the observational characterisation of the macroturbulent velocity field using synthetic spectra to fit \ion{Mg}{ii}~4481~\AA\ line profiles. Macroturbulence is treated with a radial-tangential prescription as defined in \citet{Takeda2017} for an area element, where a locally broadened profile is defined by a convolution of an intrinsic profile and a macroturbulent kernel:
\begin{equation}
     I(v, \theta) = I^{0}(v, \theta) * K(v, \theta). 
 \label{macroeq1}
\end{equation}
Here, $v$ stands for the profile velocity 
from transforming the wavelength with respect to the central wavelength of the line, while $\theta$ defines a direction angle into the observer's line-of-sight. The quantity $K(v,\theta)$ is the convolution kernel due to the macroturbulent broadening and is defined as:
\begin{multline}
  K(v, \theta)  = \frac{a_R}{\pi^{1/2} \Theta_R \cos \theta } \exp \left[ -\frac{v^2}{(\Theta_R \cos \theta)^2} \right] +  \\
   \frac{a_T}{\pi^{1/2} \Theta_T \sin \theta } \exp \left[ -\frac{v^2}{(\Theta_T \sin \theta)^2} \right]. 
 \label{macroeq2}
\end{multline}
The most common approach in the literature to estimate macroturbulence is the one assuming isotropy, that is $a_R = a_T = 0.5$ and $\Theta_R = \Theta_T = \Theta$.

After convolving the local specific intensities $I^0({v,\theta})$ with the kernel $K({v,\theta})$, the final broadened profile is computed by integrating synthetic specific intensities $I(\mu=\cos\theta)$ based on our deduced \teff\ and \logg\ 
over the visible sphere for all individual targets. This way, the limb darkening effect is included directly as opposed to the less detailed yet commonly used approach in the literature relying on parametric approximations. In our method, we make the following assumptions: (i) similar to the case of the parameter inference for \vsini, \teff, and \logg\ (cf. Sects~\ref{sect:VsiniInference} and \ref{sect:TeffLoggInference}), we fix the value of the microturbulent velocity to 2~\kms\ and 10~\kms\ for all main-sequence and supergiant stars in the sample, respectively; (ii) the \vsini\ parameter is varied in the  range determined by its value and uncertainties inferred with the Fourier-based method (cf. Sect.~\ref{sect:VsiniInference}); (iii) in first instance, the macroturbulent parameter $\Theta$ is estimated assuming equal contributions of its radial and tangential components (i.e.\ the isotropic case);
and (iv) 
because specific intensities are not available for the {\sc cmfgen} and {\sc tlusty} models, they are calculated from {\sc gssp} at the grid node closest to the parameters of the star in consideration. We introduce a line depth correction factor and apply it to the {\sc gssp} model when fitting the observed profile to account for the difference between the {\sc gssp} LTE and {\sc cmfgen}/{\sc tlusty} non-LTE models. This approach allows us to properly scale the LTE-based intrinsic profile to the expected depth while any possible difference in the shape of the intrinsic profile between the LTE and non-LTE models is considered negligible compard to the total spectral line broadening due to the combined effect of \vsini, $\xi$, and $\Theta$.

Figure~\ref{macroturbulence} summarises the obtained results in the form of the distribution of stars in the spectroscopic HR diagram colour coded by the inferred value of the macroturbulent velocity (left panel) and a histogram of the estimated values (right panel). In the left panel of Fig.~\ref{macroturbulence}, one can see a tendency of increasing macroturbulent parameter as the effective temperature and spectroscopic luminosity of the star increase. A similar trend has been previously reported in \citet{Simon-Diaz2017} (their figure~5) who also concluded that stars in the upper part of the spectroscopic HR diagram (M$\geq$15~M$_{\odot}$) tend to show line profiles with predominantly macroturbulent broadening as opposed to rotational broadening. In the right panel of Fig.~\ref{macroturbulence}, one can see that the majority of stars in our sample have a macroturbulent broadening parameter $\Theta$ between some 20~\kms\ and 70~\kms, with just a few objects being characterised by $\Theta$ below or above that range of values. Correcting for the sample size, which in our case is about four times smaller, and for larger contribution of more evolved objects, the range of the inferred macroturbulent velocity values is consistent with \citet{Simon-Diaz2017}. This is a particularly interesting observation given that the stellar sample analysed in \citet{Simon-Diaz2017} is fully comprised of galactic objects while ours is dominated by stars in the LMC, hence representing a substantially different metallicity regime.

\begin{figure}
   \centering
   \includegraphics[trim={0.6cm 0.6cm 0cm 0.6cm},clip,width=240px]{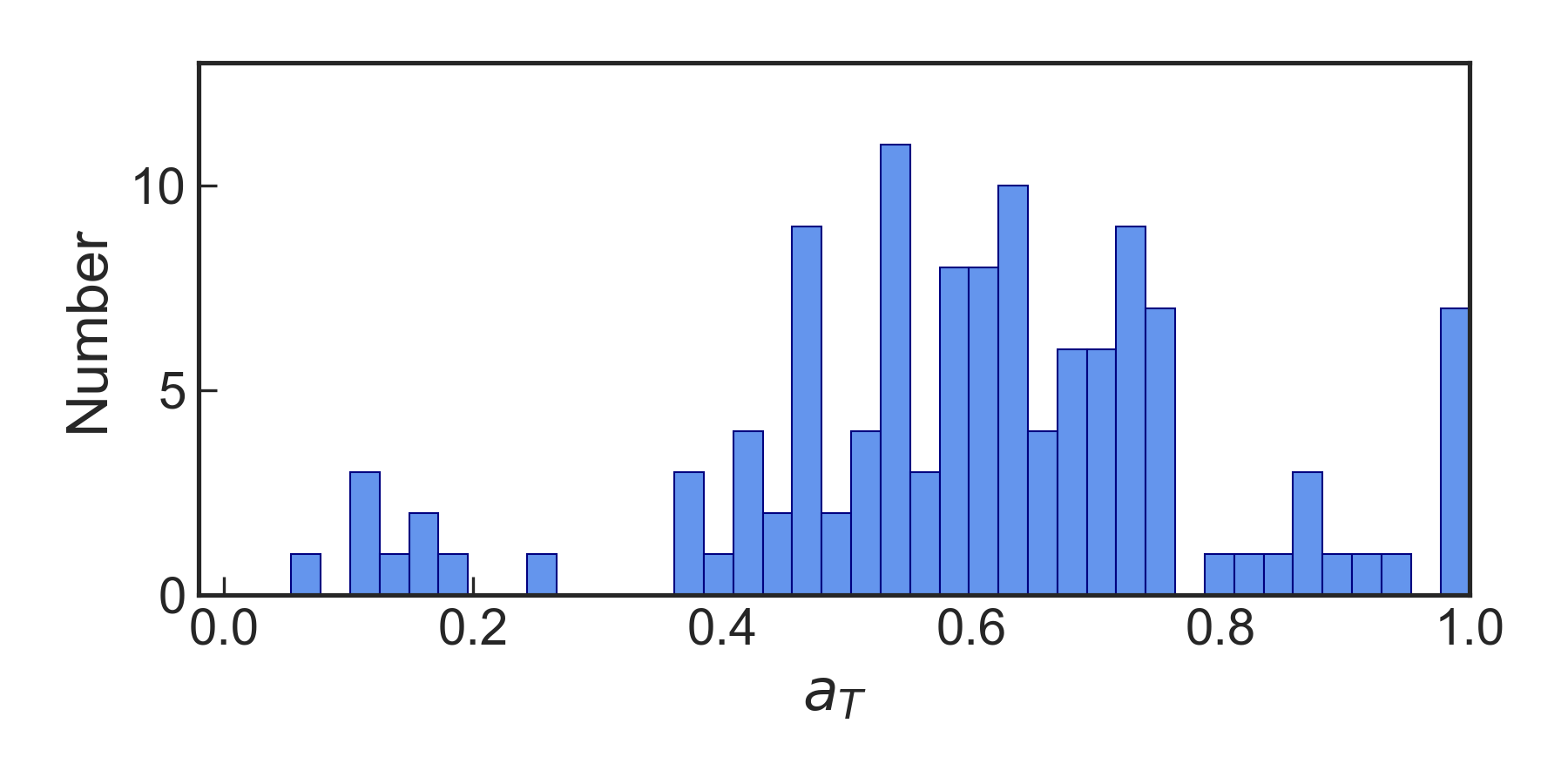}
   \includegraphics[trim={0.6cm 0.6cm 0cm 0.6cm},clip,width=240px]{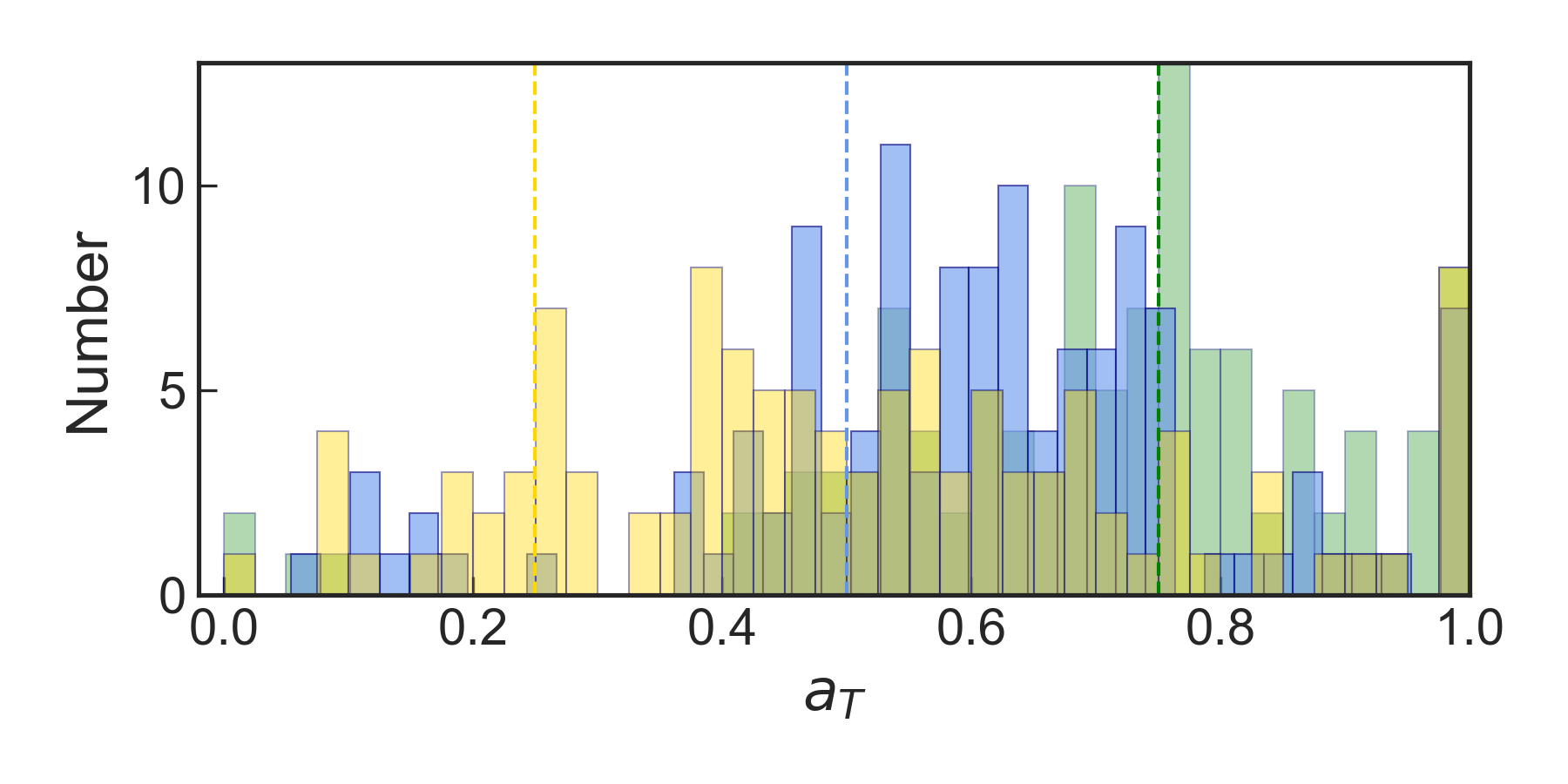}
   \includegraphics[trim={0.35cm 0.6cm 0cm 0.6cm},clip,width=240px]{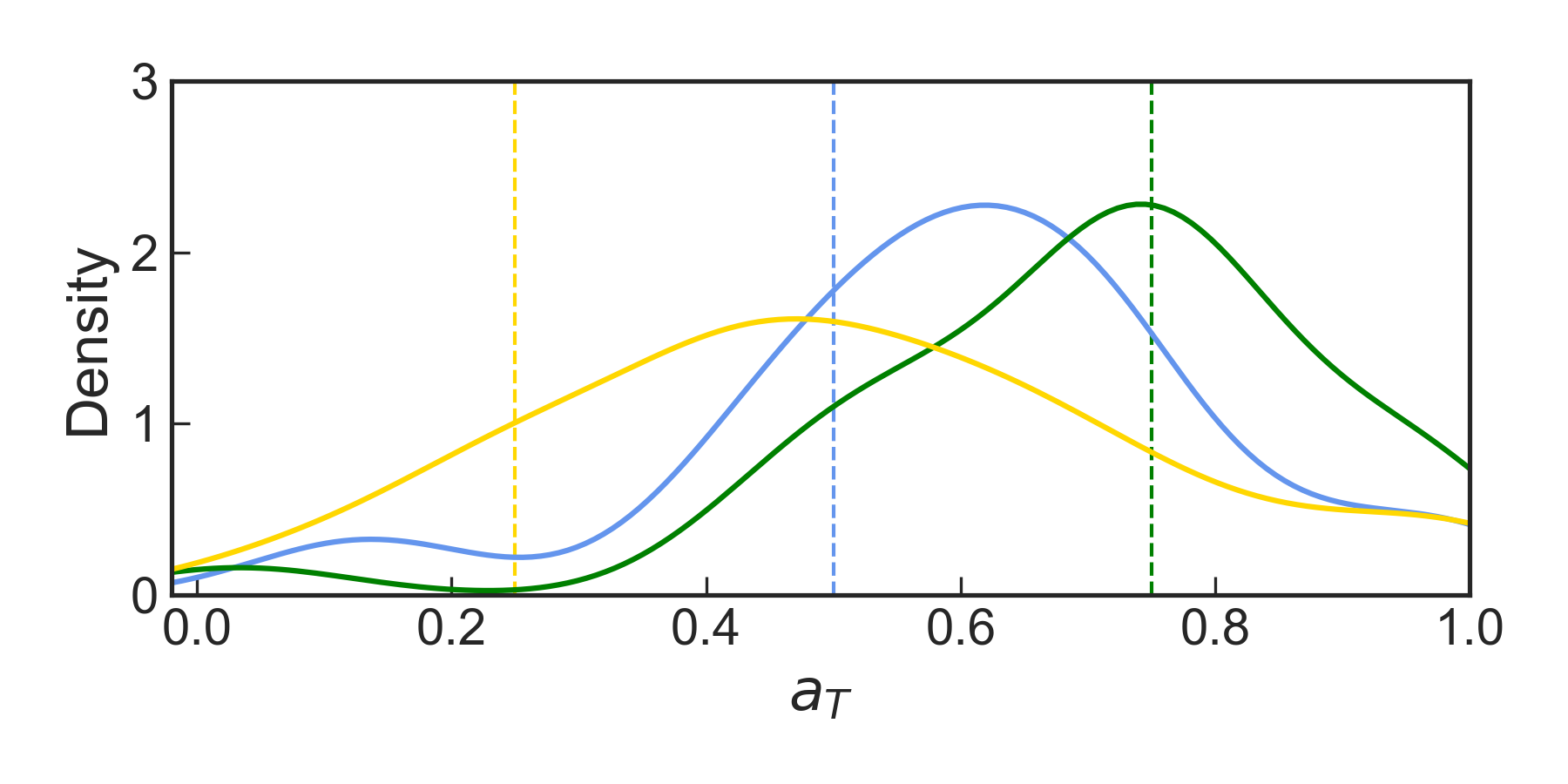}
   \caption{Results of the inference of the tangential component of macroturbulence. Top panel: histogram of $a_{\rm T}$ contributions derived from observed profiles of the entire sample with $a_{\rm T} = 0.5$ as an initial assumption, \vsini~ varied within errors of \vsini$_{\rm FT}$. Middle panel: histograms of $a_{\rm T}$ contributions derived from observed profiles of the entire sample with different initial assumptions: 0.25 (yellow), 0.5 (blue), 0.75 (green), \vsini varied freely. Bottom panel: KDE (kernel density estimate) functions for the same data as in the middle panel. Dashed lines mark initial guesses. The fits demonstrate the results are sensitive to the initial guess. }
              \label{tanar_hist_kde}
\end{figure}

\begin{figure}
   \centering
   \includegraphics[trim={1.2cm 1cm 1cm 2.5cm},clip,width=250px]{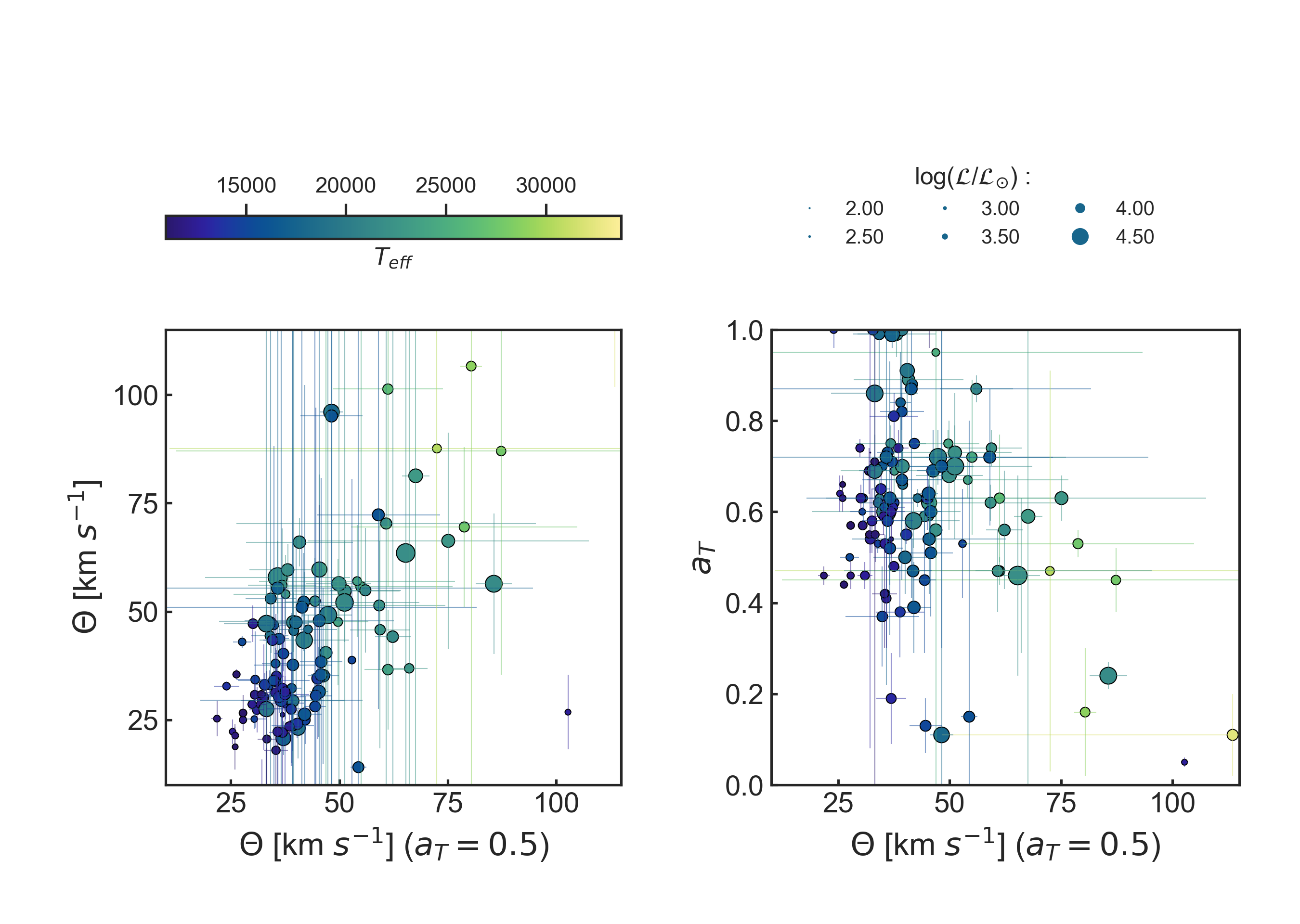}
   \caption{Results of the inference of the tangential component of macroturbulence. Left panel: comparison of $\Theta$ measured with fixed tangential component contribution $a_T = 0.5$ and $\Theta$ measured when $a_T$ varied freely. Right panel: distribution of tangential component contribution $a_T$ over $\Theta$. Colour encodes \teff, while symbol size changes with spectroscopic luminosity. }
              \label{thetas}
\end{figure}

\begin{figure*}

   \begin{center}
   \includegraphics[trim={1cm 0cm 1cm 0cm},clip,width=500px]{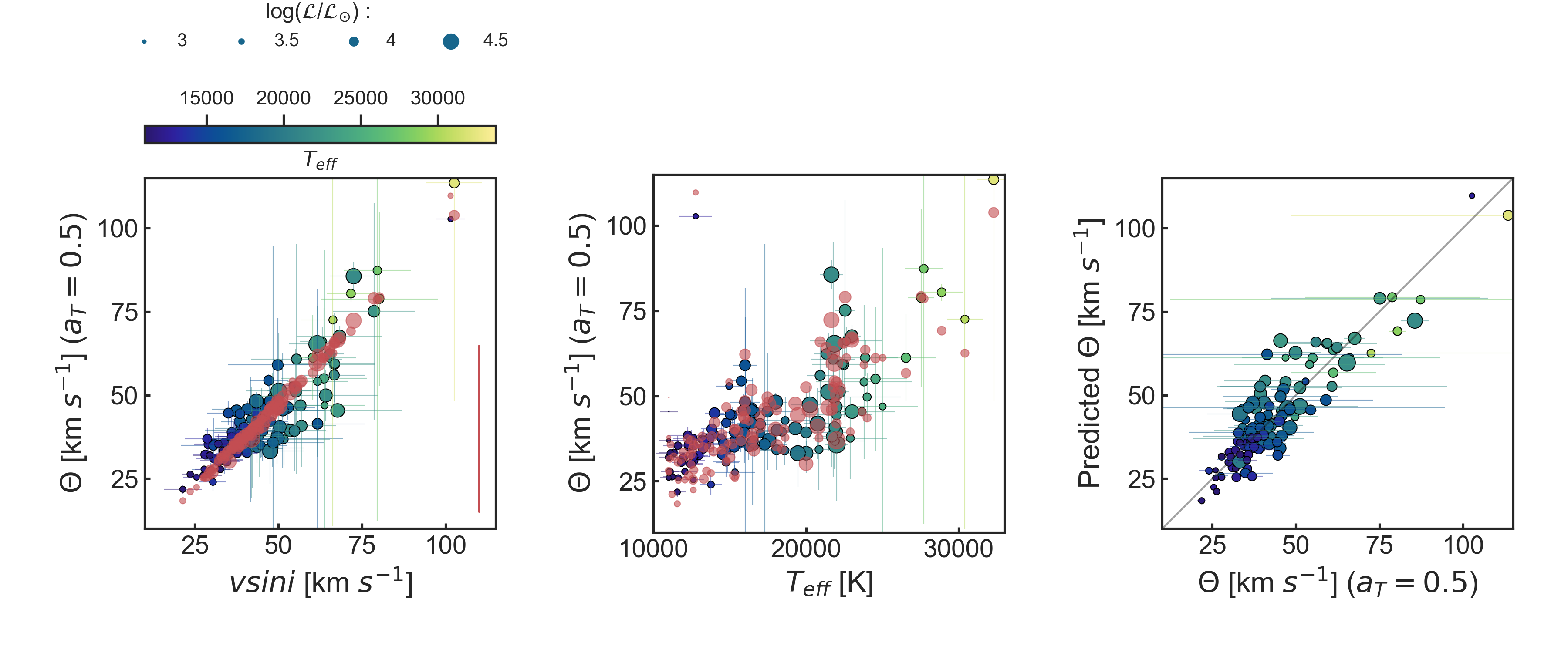}
   \caption{Results of the multivariate linear regression models for $\Theta$. Left panel: comparison between $\Theta$ and \vsini. The colours encode \teff, while the symbol sizes change with spectroscopic luminosity. The red circles show the predicted $\Theta$ value from the best multivariate regression model in Eq.\,(\ref{reg}) for our LMC sample. The red vertical line in the corner indicates a typical error bar for the predicted $\Theta$ values and applies to all three panels. Middle panel: comparison between observed $\Theta$ and predictor \teff, where $\Theta$ predicted from Eq.\,(\ref{reg}) is colour coded in red. Right panel: comparison of $\Theta$ predicted from Eq.\,(\ref{reg}) and the measured values of $\Theta$. }
\label{multivariate}
   \end{center}
\end{figure*}

\begin{figure*}

   \begin{center}
   \includegraphics[trim={1cm 0cm 1cm 0cm},clip,width=500px]{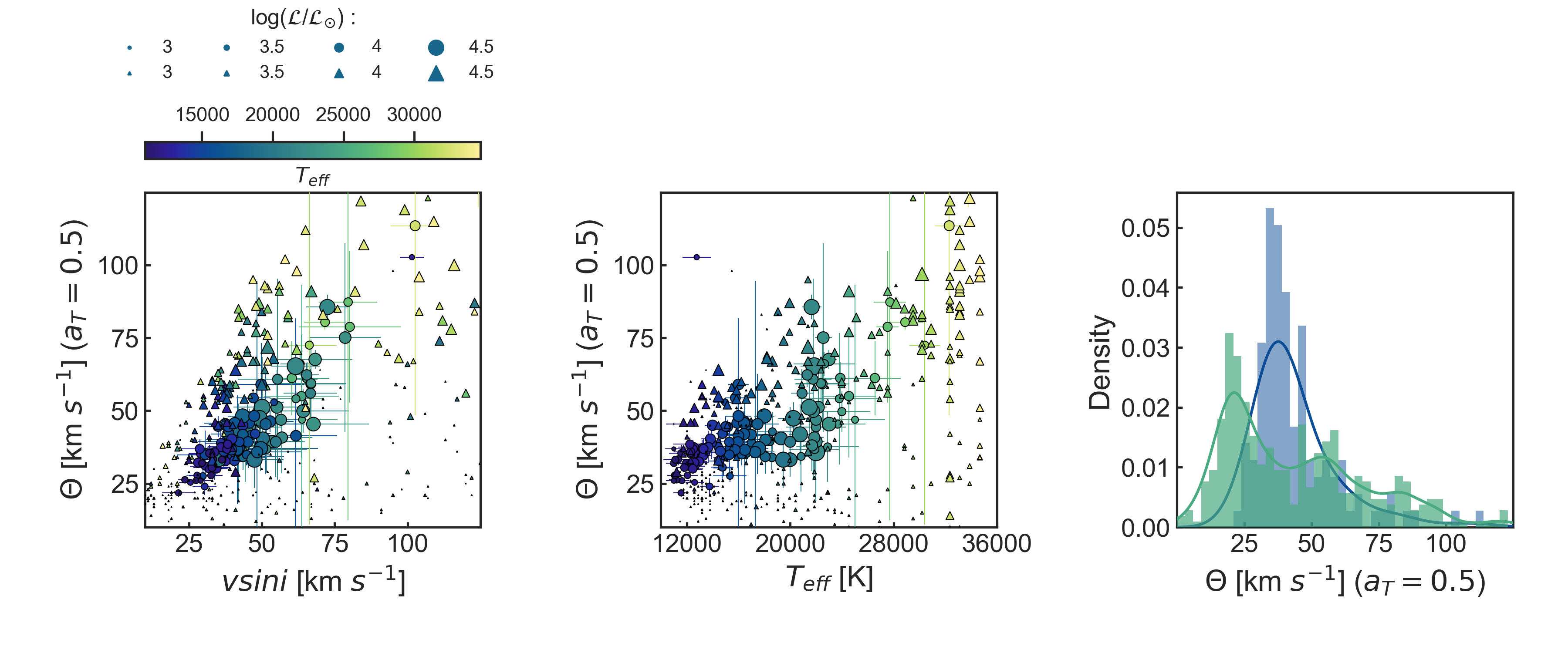}
   \caption{Comparison of $\Theta$ observed in our sample and $\Theta$ for the sample treated by \cite{Simon-Diaz2017}. Circles show the results for our LMC stars while triangles are the results from \cite{Simon-Diaz2017} for their sample of O and B stars in the Galaxy. Left panel: comparison between observed $\Theta$ and \vsini with the same colour and symbol scheme for \teff as in Fig.\,\ref{multivariate}.
Middle panel: comparison between observed $\Theta$ and \teff. Right panel: comparison of histograms and KDE for $\Theta$ in our sample (blue) and the one of \cite{Simon-Diaz2017} (green).  }
\label{sampleSergio}
   \end{center}
\end{figure*}

As a next step, we attempt to estimate from an observational viewpoint the individual contributions of the radial, $\Theta_{\rm R}$, and tangential, $\Theta_{\rm T}$, components  of the macroturbulence. In the radial-tangential prescription of macroturbulence, the contribution of the components may be controlled by parameters $a_{\rm R}$ and $a_{\rm T}$ such that $a_{\rm R} + a_{\rm T} = 1$. The parameter $a$ may be seen as the relative ratio of cells moving tangentially or radially in a unit area. Methodologically, we expand the parameter vector by $a_{\rm T}$, where the contribution of the radial component is assumed to be $a_{\rm R} = 1 - a_{\rm T}$. The top panel in Fig.~\ref{tanar_hist_kde} shows the results obtained from the fits, relying on an initial value of 0.5 for $a_{\rm T}$. The majority of stars tends to show an equal or more dominant tangential component. The middle and bottom panels in Fig.~\ref{tanar_hist_kde} offer a closer look upon the dependence of the obtained results for $a_{\rm T}$ with respect to its assumed initial value. The distributions (in the form of histograms and Kernel Density Estimations) of $a_{\rm T}$ are shown for three values of its initial guess, $a_{\rm T}$ = 0.25 (yellow), 0.5 (blue), and 0.75 (green). The main conclusions that can be drawn from these distributions are: (i) the peak position correlates with the initial guess assumed for $a_{\rm T}$; (ii) the peak height decreases in the KDE plot as the initial guess for the $a_{\rm T}$ parameter decreases; (iii) all distributions are skewed towards higher values of $a_{\rm T}$, where the skewness of the observed distribution is statistically significant as confirmed by a Kolmogorov-Smirnov test 
with respect to a reference Gaussian distribution centered at the peak of the observed distributions; and (iv) visual inspection of observed profiles reveals most of the outlying values ($a_{\rm T}$ < 0.4 and $a_{\rm T}$ > 0.8) to be due to a high level of noise in the spectra and/or strongly asymmetric line profiles. At the same time, spectra of the best quality in the entire sample tend to show $a_{\rm T}/a_{\rm R} \sim 1$, where the estimated uncertainty of $a_{\rm T}$ may reach up to 20\% for the case treated here.

Furthermore, we compare $\Theta$ when $a_{\rm T}$ = 0.5 is fixed and $\Theta$ when $a_{\rm T}$ is included as a free parameter and analyse the trends they demonstrate. The comparison (Fig.~\ref{thetas}, left panel) illustrates that these  $\Theta$ are in agreement to within their uncertainties, while the errors increase significantly when $a_{\rm T}$ is varied. There is no obvious trend in the distribution of $a_{\rm T}$ with respect to the value of $\Theta$ for $a_{\rm T}$ = 0.5 fixed, as shown in the right panel of Fig.~\ref{thetas}.  

Finally, following the statistical approach in \citet{Aerts2023}, 
we investigate observed trends in $\Theta$ quantitatively from multivariate linear regression models relying on five predictors, of the following form:
\begin{equation}
    \Theta = c_0 + c_1 \rm{log(T_{eff})} + c_2 \rm{logg} + c_3 \rm{log(\mathcal{L}/\mathcal{L}_{\odot})} + c_4 \rm{vsini} + c_5 a_{\rm{T}}.    
\end{equation}
After consecutive elimination of the insignificant predictors based on the adjusted $R^2$-statistics taking into account the degrees-of-freedom, we conclude that the best regression model includes only \vsini~ and \teff~ as relevant predictors, with the following coefficients:
\begin{equation}
\label{reg}
    \Theta = 66(24) + 17.7(5.8)\ \rm{log(T_{eff})} + 1.15(0.02) \rm{vsini}.    
\end{equation}
This model explains 96.4\% of the observed variability in $\Theta$. We plot the predicted  $\Theta$ from this regression model together with the observed values for our sample in Fig.~\ref{multivariate}.  There is a clear trend in the \vsini-$\Theta$ plot showing an almost one-to-one correspondence between $\Theta$ and \vsini. 
Our regression results are remarkably similar to those in the recent study by \citet{Aerts2023}, who analysed the correspondence between macroturbulent broadening and \vsini~ for a large sample of $\sim$15\,000 galactic gravity-mode pulsators in core-hydrogen burning from low-resolution Gaia DR3 spectra. Moreover, as 
we mentioned earlier in this section, a similar trend between $\Theta$ and \vsini\ was found from high-resolution spectroscopy by \cite{Simon-Diaz2017}, who analysed a galactic sample of O and B stars. Since our sample consists mostly of LMC objects and thus represents a significantly lower metallicity regime, it is interesting to compare the results for $\Theta$ of our stars with those in the sample of \cite{Simon-Diaz2017}
to check if there is any dependence on metallicity. We plot the macroturbulent values from \cite{Simon-Diaz2017} and from our work in  Fig.~\ref{sampleSergio}. Though the $\Theta$ values in their sample are in general larger for their supergiants than for ours (note that the first green peak in the right panel corresponds to main-sequence stars, which are absent in the LMC sample), the differences remain roughly within the errors of $\Theta$ determined for our sample stars. Since our sample is populated with more luminous stars, and different methods were used to measure $\Theta$, it is difficult to conclude if the shift towards lower $\Theta$ in our LMC sample has a physical or methodological origin, also because there are no errors listed for values of the macroturbulent broadening in the study by  \cite{Simon-Diaz2017}.

\section{Improving the inference of macroturbulence}\label{sec:VmacroSimulations}

In the above parameterised estimation of the macroturbulent velocity and of its individual radial and tangential component contributions to the line broadening of our sample stars, we made two fundamental assumptions: (i) the microturbulent velocity $\xi$ is known a priori (for example, estimated along with \teff\ and \logg\ of the star or fixed to a value typical for a given spectral type and luminosity class), and (ii) the typical S/N$\simeq$100 in the observed spectra is sufficient to distinguish between different line broadening mechanisms. Although these assumptions are motivated by the most commonly used methodologies and/or empirical evidences in the literature (e.g. \citealt{Simon-Diaz2017,Holgado2018}), they may not be valid. To investigate this, we look into the problem in more detail, attempting to find out (i) what are the requirements in terms of S/N, spectral line selection, etc. for a meaningful inference of the macroturbulent parameter $\Theta$ from spectra of O and B stars, (ii) what are the analysis limitations associated with the parameter correlations, in particular between micro- ($\xi$) and macro- ($\Theta$) turbulent velocities and rotational broadening \vsini, and (iii) under what conditions can one empirically estimate individual contributions of the radial, $a_{\rm R}$, and tangential, $a_{\rm T}$, components of the global macroturbulent broadening in the spectra of O and B stars.

\begin{table}[t]
    \centering
    \small
    \caption{Combinations of the $\Theta$, $\xi$, and \vsini\ parameter values used in the analysis of spectral line broadening due to the macroturbulent velocity and its interplay with the two other broadening mechanisms. See text for details.
    }    
    \begin{tabular}{llllll}
    \hline\hline
    S/N & \teff & \logg & $\xi$ & $\Theta$ & \vsini\\
    & (K) & (dex) & \multicolumn{3}{c}{(\kms)}\\
    \hline
    & & & & &\\
    \multirow{8}{*}{100, 150, 250} & \multirow{8}{*}{12\,000, 22\,000} & \multirow{8}{*}{4.0} & \multirow{8}{*}{8} & \multirow{2}{*}{15} & 15\\
    & & & & & 50\vspace{1.5mm}\\
    & & & & \multirow{3}{*}{30} & 15\\
    & & & & & 30\\
    & & & & & 100\vspace{1.5mm}\\
    & & & & \multirow{3}{*}{65} & 15\\
    & & & & & 65\\
    & & & & & 150\\
    & & & & &\\
    \hline
    & & & & &\\
    \multirow{1}{*}{100, 150, 250} & \multirow{1}{*}{12\,000} & \multirow{1}{*}{4.0} & \multirow{1}{*}{2, 8, 16} & \multirow{1}{*}{30} & 15\\
    & & & & &\\
    \hline
    \end{tabular}
    \label{tab:xi_theta_vsini}
\end{table}

The above questions are particularly important to address in light of the high sensitivity of the results obtained in Sect.~\ref{sec:VmacroInference} to the initial guess of $a_{\rm T}$. Moreover, the ability to realistically estimate individual contributions of the radial and tangential components of the macroturbulent broadening is a powerful observational test of various theoretical hypotheses for the origin of macroturbulence as a physical phenomenon. In particular, should wave-driven broadening be at the origin of macroturbulence, one would expect the radial component to dominate over the tangential one in the case of pressure waves, whereas the opposite is true for the case of gravity waves. So far, an appreciable effort has been put in the literature into estimating the macroturbulent velocity parameter $\Theta$ for as many massive stars as possible, and to search for possible correlations with various physical mechanisms at work (e.g \citealt{Simon-Diaz2014,Simon-Diaz2017}). However, the most common approach is to assume equal contributions of the radial and tangential components, while this choice is the least justified in the case of pulsational broadening \citep{Aerts2009,Aerts2014}.

We simulate synthetic spectra of a late and early B-type star with \teff=12\,000~K, \logg = 4.0~dex and \teff=22\,000~K, \logg=4.0~dex, respectively, to analyze them as they are observed ones and quantify deviations from the true parameters. In the case of the late B-type star, we focus on the \ion{Mg}{ii}~4481~\AA\ spectral line, while we use the \ion{Si}{iii}~4552~\AA, 4557~\AA, and 4574~\AA\ triplet in the early B-type star. Spectra are simulated for several combinations of the parameters $\Theta$, $\xi$, and \vsini\ to consider cases where their dominant role is interchanging. Table~\ref{tab:xi_theta_vsini} provides a summary of all combinations used and we repeat the analysis for three values of S/N=100, 150, and 250 to address the influence of the quality of observations on the obtained results. At this stage, we assume equal contributions for the radial $a_{\rm R}$ and tangential $a_{\rm T}$ components of the macroturbulence to isolate possible correlations between the $\Theta$, $\xi$, and \vsini\ parameters.

\begin{figure*}
   \begin{center}
            \includegraphics[clip,scale=0.7]{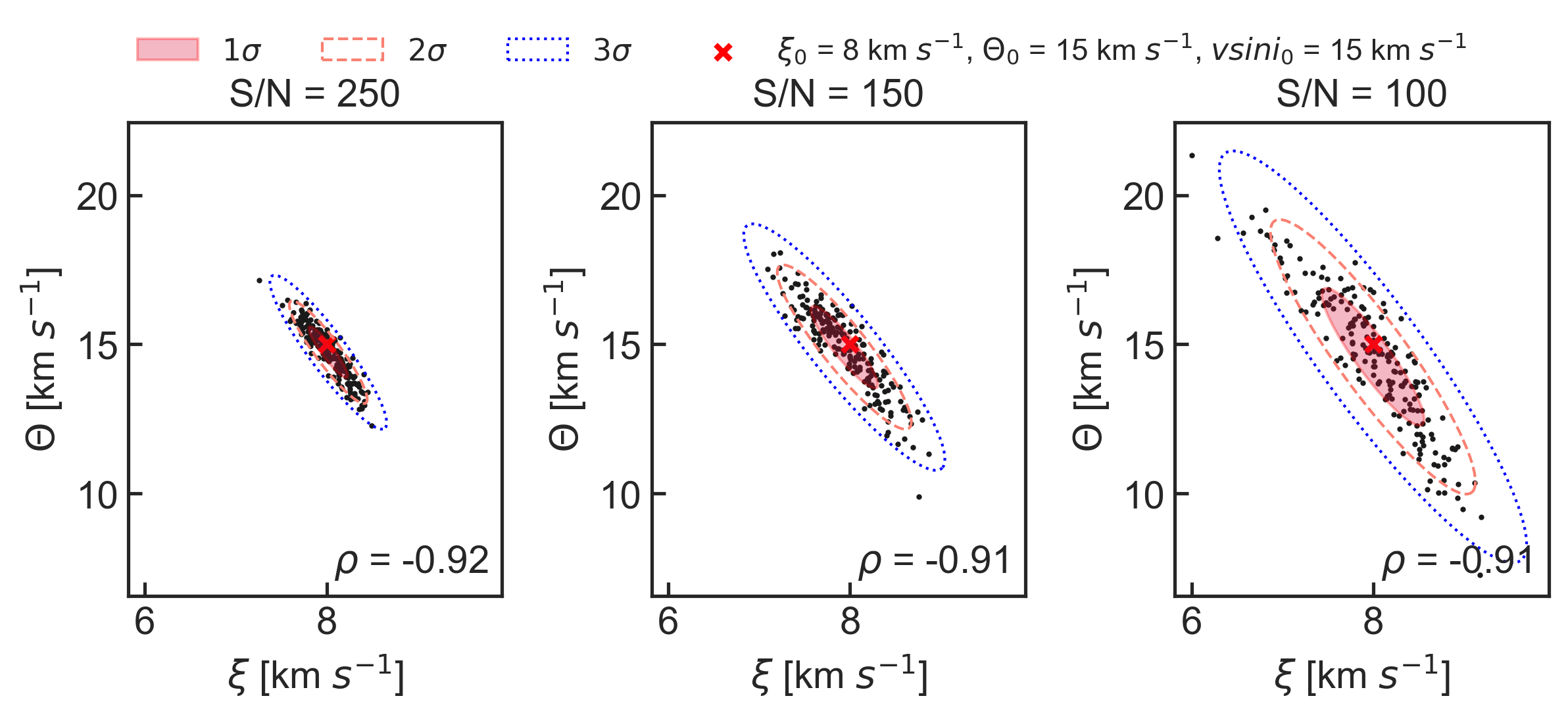}
            \includegraphics[clip,scale=0.7]{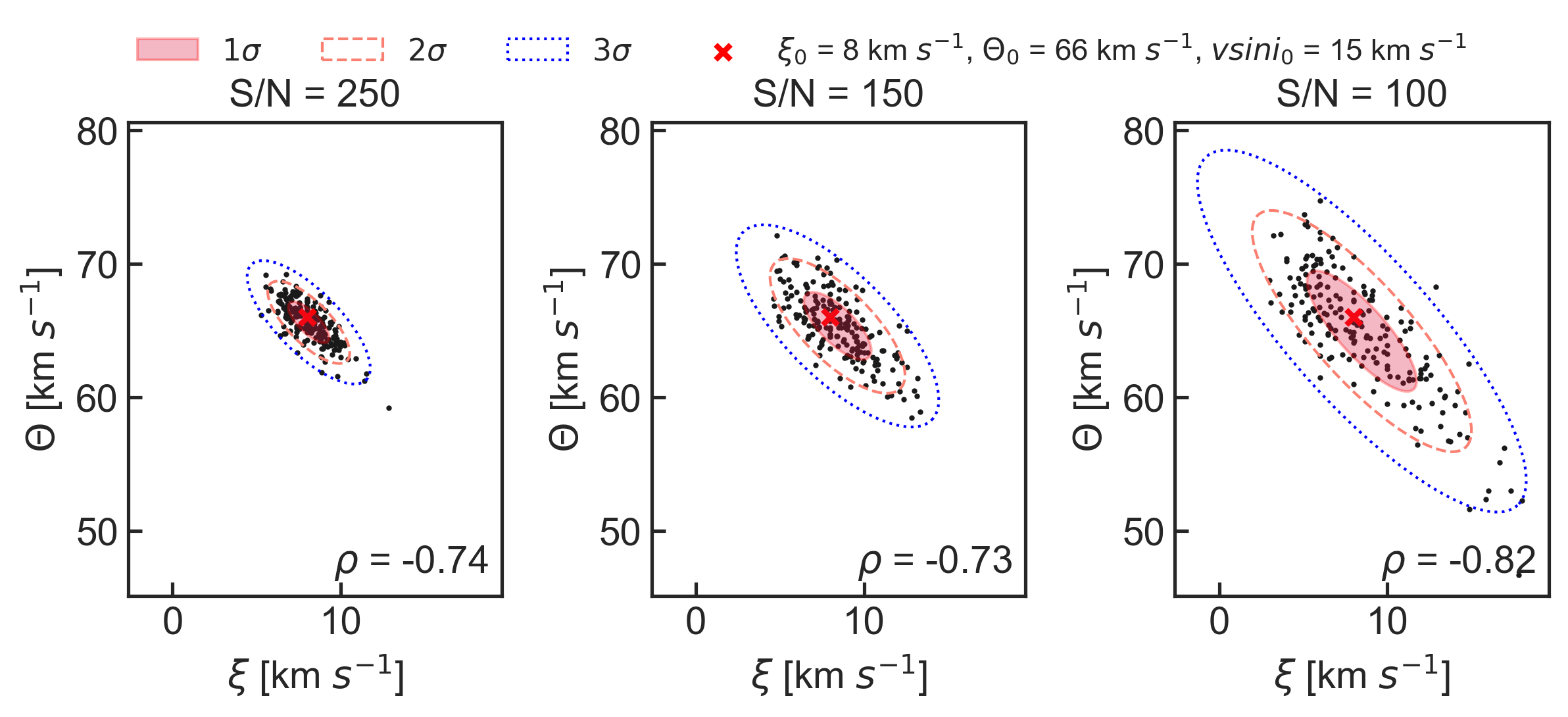}
      \caption{Results of Monte Carlo simulations on synthetic profiles of \ion{Mg}{ii} (top set of panels) and \ion{Si}{iii} (bottom set of panels) with artificial noise (left panels: S/N = 250, middle panels: S/N = 150, and right panels: S/N = 100) with 1-, 2- and 3-$\sigma$ confidence ellipses. Correlation coefficients $\rho$ are indicated in the right corner of each panel. \vsini\ is fixed, while $\xi$ and $\Theta$ are optimised. The red crosses mark true parameters of the synthetic profiles.}
         \label{VmacroVmicroCorrelations}
    \end{center}
 \end{figure*}

The analysis of all 57 profiles (3 values of S/N $\times$ 16 combinations of the $\Theta$ and \vsini\ parameters with one fixed value of $\xi$, plus 3 values of S/N $\times$ 3 combinations with varied $\xi$ for a fixed pair of $\Theta$ and \vsini) is approached with a Monte Carlo formalism. That is, each simulated profile (in other words, each combination of input parameters) is fitted multiple (200) times, where each new fit is done after applying normally-distributed noise to the simulated profile. The fitting is being performed by minimizing a $\chi^2$ merit function between the noisy synthetic profile and the grid of synthetic profiles covering the full parameter space. Such simulations allow for quantitative estimation of the effect of the S/N on the spread of the estimated parameters around their true value, as well as assessment of their correlations. 

 Figure~\ref{VmacroVmicroCorrelations} illustrates a simplified case where \vsini\ is fixed (under the assumption that it can be determined from an independent method, such as the one based on the FT). The minimisation is performed for only two parameters, namely the micro- and macroturbulent velocities ($\xi$ and $\Theta$). 
 Similar figures for other combinations with \teff=12\,000 K (\ion{Mg}{ii}) are presented in Appendix \ref{section:MonteCarloSim} (Figs~\ref{cc2_mg_2_30_15_vmi}-\ref{cc2_mg_8_66_150}). Plots for the case of \teff=22\,000~K  and the \ion{Si}{iii} diagnostic line are left out as they show the same trends. For a given value of the \vsini\ parameter and quality of the observed spectrum characterised by its S/N value, we find a strong correlation between $\xi$ and $\Theta$.
 Moreover, the correlation persists irrespective of whether the rotational broadening dominates over the macroturbulent broadening (top versus bottom row in Fig.~\ref{VmacroVmicroCorrelations}). Furthermore, the 1-, 2-, and 3-$\sigma$ uncertainty intervals for $\xi$ and $\Theta$
are a strong function of the S/N value in the observed spectrum of the star (left, middle, and right columns in Fig.~\ref{VmacroVmicroCorrelations}). In particular, irrespective of the assumed effective temperature of the star and accounting for the parameter correlations, we find that at the 2-$\sigma$ level: (i) $\Theta$ can be inferred with a typical precision ranging from some 40\% to 100\% at S/N$\simeq$150 and 100, respectively, and better than some 30\% at S/N$\simeq$250 when the rotational broadening dominates over macroturbulence; (ii) $\Theta$ can be inferred with precision better than 50\%, 30\%, and 10\% at S/N$\simeq$100, 150, and 250, respectively, when the contributions from rotational broadening and macroturbulence are equal; (iii) $\Theta$ can be inferred with a precision of some 20\%, 10\%, and 5\% at S/N$\simeq$100, 150, and 250, respectively, when macroturbulence dominates over rotational broadening.

Since the described test was performed with the assumption of \vsini\ being known precisely, which is never the case, we now repeat the above test with one of the combinations of $\xi$ and $\Theta$ including optimization of the third free parameter \vsini, allowing it to vary in a range of typical uncertainty of the FT method (10\%). The result is shown in Fig.~\ref{limvsini_34D}. One can see that \vsini\ is not significantly correlated with either $\xi$ or $\Theta$ but introduces a spread of up to 100\% in $\Theta$, even in this case where it is varied in a limited range. Note that \vsini\ dominates over $\Theta$ in this instance of the test, which confirms that determination of $\Theta$ is unreliable unless it dominates over \vsini.

As a further test, we now also include a tangential component of macroturbulence, $a_{\rm T}$, as another free parameter and allow \vsini\ to vary. This implies we use four free parameters when fitting the line profiles:  \vsini, $\xi$, $\Theta$, and $a_{\rm T}$. The result is shown in Fig.~\ref{4D} for the case of S/N=250. The main conclusion is that even at such high S/N one can barely distinguish between a fit with equal radial and tangential contributions ($a_{\rm T} = 0.5$) versus one with complete dominance of the tangential component ($a_{\rm T} = 1.0$) at 1-$\sigma$ confidence level. When allowing for 3-$\sigma$ confidence level, the solution space covers the entire range from 0 to 1 for $a_{\rm T}$. Thus, line profiles of lower S/N do not allow to constrain this parameter observationally in case of absent independent constraints on other parameters.   

\section{Discussion and Conclusions}\label{sec:discussion_conclusions}

In this first paper of a series, we present a study of 124 O and B stars with unique catalog identifiers from two ESO large programmes that employ the high-resolution {\sc feros} and {\sc uves} instruments. The sample is mostly composed of supergiants in the LMC and a few dwarfs in the Galaxy. All stars in our sample are found in the CVZ-S of the TESS space telescope and are thus potential candidates for future asteroseismic modelling, thanks to the 1-year long time base of photometric observations \citep{Garcia2022,Garcia2023}. The same argument applies to 124 galactic dwarf stars analysed in detail in the complementary study by \citet{Gebruers2022}.

We find that the reduced spectra delivered by the ESO pipeline do not meet our stringent requirements in terms of precision of the spectroscopically inferred atmospheric parameters for future asteroseismic modelling. Therefore, we develop a custom normalisation of the obtained spectra starting from the intermediate data product resulting from the ESO pipeline, that is the blaze-corrected non-merged wavelength calibrated spectra. Our tests show that a rather simple recipe of the order-by-order normalisation works well for the blue supergiants in our sample. For dwarfs, however, the same recipe fails to provide a high quality for the spectrum normalisation, most notably in the wavelength regions of the broad hydrogen lines. For those, we develop an even more sophisticated spectrum normalisation method that relies on a grid of synthetic spectra. The latter allows us to locate the local continuum and approximates the normalisation function with a series of Chebyshev polynomials.

We employ the LSD method to compute high S/N average profiles for all epochs and for all stars in our sample. We fit the obtained LSD profiles with asymmetric Gaussian functions to deduce radial velocities and the degree of asymmetry of the line profiles for each epoch. The inferred information is employed to perform a spectroscopic variability classification of the stars, where we consider four variability classes: apparently constant (``const''), spectroscopic binaries (``SB''), stars showing line profiles variations (``LPV''), and stars with line profiles characterised by an appreciable degree of skewness (``asymm''). Considering the {\sc uves} and {\sc feros} samples separately, we find 86 apparently constant stars, 7 stars showing LPV, 28 spectroscopic binary candidates, and 37 stars exhibiting asymmetric line profiles\footnote{The total number of classifications exceeds the sample size of 148 objects because some of the stars are assigned to two variability classes (see Table~\ref{tab:results}).}. Out of 24 stars that have both {\sc uves} and {\sc feros} data, 20 are assigned to the same variability class. The remaining four stars, HD~268653, HD~269145, HD~269766, and HD~268675, are either classified as apparently constant based on one data set and tentatively variable in the other one, or have an extra class assignment based on either data set (see Table~\ref{tab:results} for details). Finally, 24 classifications are tentative and need more spectroscopic epochs for an unambiguous class assignment.

The positions of the sample stars in the spectroscopic HR diagram suggest the majority of them to be blue supergiants, with only a few objects being dwarfs situated within the SPB and $\beta$~Cep instability regions. Detailed analyses of all available sectors of TESS space-based photometry is beyond the scope of this initial spectroscopy paper but will be presented in follow-up publications. Yet, we perform a cross-match of our sample with the O- and B-type stars studied in \citet{Bowman2019a} and \citet{Bowman2020b}. In the observed sample, for 80\% of the blue supergiants situated outside the instability regions in Fig.~\ref{spec_hrd}, these authors report SLF variability. One of those objects was reported to show heat-driven oscillations in TESS photometry. Only 3 of 10 stars situated within the instability regions from our spectroscopic parameters show oscillations. Validation of these initial photometric results from all available TESS sector data will be done in a dedicated follow-up study.

We employ two different methods to infer the projected rotational velocities \vsini\ from the spectra of the sample stars: one based on the FT of line profiles and the other one from fitting synthetic spectra to selected line profiles. Owing to the diversity of mechanisms causing the total line broadening in early-type stars, we find that the method of synthetic spectra delivers \vsini\ values that are either comparable to or exceed those inferred with the FT method. Therefore, we accept the FT-based \vsini\ measurements as the final and most precise ones in our study (see Table~\ref{tab:results_lp}, column ``\vsini$_{\rm FT}$'') and recommend their use in favour of those inferred from the method of synthetic spectra.

The \vsini\ values resulting from the FT method are subsequently used to quantify the amount of spectral line broadening due to macroturbulence for all stars in the sample. In the process, we allow \vsini\ to vary in a narrow range dictated by its uncertainty. We find a tendency for the macroturbulent parameter $\Theta$ to increase as the effective temperature and spectroscopic luminosity of the star increase, and that the majority of stars in the sample show macroturbulent broadening in the range between some 20~\kms\ and 70~\kms. Both these findings are in agreement with the results in \citet{Simon-Diaz2017} based on 430 galactic massive stars of spectral types O and B. Given that the majority of stars in our sample are in the LMC, which is characterised by a significantly lower metallicity than the galactic sample in \citet{Simon-Diaz2017}, we conclude that the observed macroturbulence in spectral lines of massive stars does not depend on their bulk metallicity (taking into account our smaller sample size and dominance by evolved objects). Furthermore, we check for a possible correlation between the inferred macroturbulent velocity parameter and the variability class we have assigned to our targets (cf. Table~\ref{tab:results}). In this exercise, we consider the following groups of stars: (i)``const'' versus all other groups together, (ii) ``const'' versus ``asymm'', (iii) ``const'' versus ``lpv'', and (iv) ``asymm'' versus ``lpv''. We could not identify any correlation between 
the macroturbulent velocity and the class of spectroscopic variability. 

In an attempt to quantify the two individual contributions known as the radial and tangential components of macroturbulence from spectral line fitting, we find the results to be sensitive
to the initial guess for the respective parameters $a_{\rm R}$ and $a_{\rm T}$\footnote{Note that we optimise the $a_{\rm T}$ parameter only while the contribution of the radial component is set according to $a_{\rm R}=1-a_{\rm T}$ (see Sect.~\ref{sec:VmacroInference}).}. Although the obtained distribution is skewed towards $a_{\rm T} \gtrsim 0.6 $, we cannot firmly conclude to observe a dominant tangential component given the limited quality of our spectroscopic data in terms of S/N. When all the assumptions of our approach are met, that is (i) a fixed microturbulent velocity $\xi$ typical for given spectral type applies; (ii) \vsini\ inferred from Fourier transforms of observed profiles varies modestly; (iii) the initial guess for $a_{\rm T}$ is not lower than 0.4, 
one may conclude that the tangential component $a_{\rm T}$ is larger than the radial one in our sample. This may possibly be an indicator of broadening by gravity or gravito-inertial waves \citep{Aerts2023}. This interpretation should be treated with caution since any meaningful inference for the individual radial and tangential components of the macroturbulent broadening requires S/N~$\geq 250$ (see Sect.~\ref{sec:VmacroSimulations}). 

Finally, we add a word of caution to the way uncertainties are estimated for the macroturbulent velocity parameter $\Theta$ in the literature. The commonly accepted approach, and the one employed in this work too, is to decouple the inferences of the macro- and microturbulent broadening parameters, where the latter is usually optimised together with the effective temperature and surface gravity of the star. Our study shows that the $\Theta$ and $\xi$ parameters are strongly correlated for the case of massive stars. Hence, fixing the microturbulent velocity parameter $\xi$ while quantifying contribution of macroturbulence to the line broadening results in underestimation of the uncertainties for the latter. Furthermore, the precision with which the macroturbulent broadening parameter $\Theta$ can be inferred depends strongly on the S/N of the spectra, as well as on the dominant line broadening mechanism among rotation or macroturbulence. The uncertainty on the $\Theta$ parameter increases progressively as the dominance of the projected rotational velocity \vsini\ grows. These findings are in agreement with the results obtained in \citet{Simon-Diaz2017}, where the authors report only the upper limit for the macroturbulence parameter for the majority of stars with \vsini $> \Theta$.

Future studies of all stars in the combined {\sc uves} and {\sc feros} samples will be focused on detailed analyses of their atmospheric chemical compositions and frequency analyses based on all TESS data that will be available by the time. Subsequently, we will also perform detailed asteroseismic modelling for stars with sufficient identified oscillation modes, relying on the spectroscopic  constraints offered in this work and from the future results for the individual elemental abundances.

\begin{acknowledgements}
The research leading to these results has received funding from the KU~Leuven Research Council (grant C16/18/005: PARADISE), from the Research Foundation Flanders (FWO) under grant agreements G089422N (NS, CA, AT), 
K802922N (CA, Sabbatical leave),
11E5620N (SG, PhD Aspirant mandate), 1124321N (LIJ, PhD Aspirant mandate), 12ZB620N (TVR, Junior Postdoctoral Fellowship), and 1286521N (DMB, Senior Postdoctoral Fellowship), as well as from the BELgian federal Science Policy Office (BELSPO) through PRODEX grant PLATO. This research has made use of the SIMBAD database, operated at CDS, Strasbourg, France. The TESS data presented in this paper were obtained from the Mikulski Archive for Space Telescopes (MAST) at the Space Telescope Science  Institute (STScI), which is operated by the Association of Universities for Research in Astronomy, Inc.,  under NASA contract NAS5-26555. Support to MAST for these data is provided by the NASA Office of Space Science via grant NAG5-7584 and by other grants and contracts. Funding for the TESS mission is provided by the NASA Explorer Program. 
\end{acknowledgements}

\bibliographystyle{aa}
\bibliography{bibliography}

\begin{thebibliography}{173}
\expandafter\ifx\csname natexlab\endcsname\relax\def\natexlab#1{#1}\fi

\bibitem[{{Aerts}(2021)}]{Aerts2021}
{Aerts}, C. 2021, Reviews of Modern Physics, 93, 015001

\bibitem[{{Aerts} {et~al.}(2018){Aerts}, {Bowman}, {S{\'\i}mon-D{\'\i}az},
  {Buysschaert}, {Johnston}, {Moravveji}, {Beck}, {De Cat}, {Triana},
  {Aigrain}, {Castro}, {Huber}, \& {White}}]{Aerts2018}
{Aerts}, C., {Bowman}, D.~M., {S{\'\i}mon-D{\'\i}az}, S., {et~al.} 2018,
  \mnras, 476, 1234

\bibitem[{{Aerts} {et~al.}(2011){Aerts}, {Briquet}, {Degroote}, {Thoul}, \&
  {van Hoolst}}]{Aerts2011}
{Aerts}, C., {Briquet}, M., {Degroote}, P., {Thoul}, A., \& {van Hoolst}, T.
  2011, \aap, 534, A98

\bibitem[{{Aerts} {et~al.}(2010{\natexlab{a}}){Aerts}, {Christensen-Dalsgaard},
  \& {Kurtz}}]{asteroseismology}
{Aerts}, C., {Christensen-Dalsgaard}, J., \& {Kurtz}, D.~W. 2010{\natexlab{a}},
  Asteroseismology (Astronomy and Astrophysics Library. ISBN 978-1-4020-5178-4.
  Springer Science+Business Media)

\bibitem[{{Aerts} {et~al.}(1999){Aerts}, {De Cat}, {Peeters}, {Decin}, {De
  Ridder}, {Kolenberg}, {Meeus}, {Van Winckel}, {Cuypers}, \&
  {Waelkens}}]{Aerts1999}
{Aerts}, C., {De Cat}, P., {Peeters}, E., {et~al.} 1999, \aap, 343, 872

\bibitem[{{Aerts} \& {Kolenberg}(2005)}]{Aerts2005}
{Aerts}, C. \& {Kolenberg}, K. 2005, \aap, 431, 615

\bibitem[{{Aerts} {et~al.}(2010{\natexlab{b}}){Aerts}, {Lefever}, {Baglin},
  {Degroote}, {Oreiro}, {Vu{\v{c}}kovi{\'c}}, {Smolders}, {Acke}, {Verhoelst},
  {Desmet}, {Godart}, {Noels}, {Dupret}, {Auvergne}, {Baudin}, {Catala},
  {Michel}, \& {Samadi}}]{Aerts2010}
{Aerts}, C., {Lefever}, K., {Baglin}, A., {et~al.} 2010{\natexlab{b}}, \aap,
  513, L11

\bibitem[{{Aerts} {et~al.}(2019){Aerts}, {Mathis}, \& {Rogers}}]{Aerts2019}
{Aerts}, C., {Mathis}, S., \& {Rogers}, T.~M. 2019, \araa, 57, 35

\bibitem[{{Aerts} {et~al.}(2023){Aerts}, {Molenberghs}, \& {De
  Ridder}}]{Aerts2023}
{Aerts}, C., {Molenberghs}, G., \& {De Ridder}, J. 2023, \aap, 672, A183

\bibitem[{{Aerts} {et~al.}(2009){Aerts}, {Puls}, {Godart}, \&
  {Dupret}}]{Aerts2009}
{Aerts}, C., {Puls}, J., {Godart}, M., \& {Dupret}, M.~A. 2009, \aap, 508, 409

\bibitem[{{Aerts} \& {Rogers}(2015)}]{Aerts2015}
{Aerts}, C. \& {Rogers}, T.~M. 2015, \apjl, 806, L33

\bibitem[{{Aerts} {et~al.}(2017{\natexlab{a}}){Aerts}, {S{\'\i}mon-D{\'\i}az},
  {Bloemen}, {Debosscher}, {P{\'a}pics}, {Bryson}, {Still}, {Moravveji},
  {Williamson}, {Grundahl}, {Fredslund Andersen}, {Antoci}, {Pall{\'e}},
  {Christensen-Dalsgaard}, \& {Rogers}}]{Aerts2017a}
{Aerts}, C., {S{\'\i}mon-D{\'\i}az}, S., {Bloemen}, S., {et~al.}
  2017{\natexlab{a}}, \aap, 602, A32

\bibitem[{{Aerts} {et~al.}(2014){Aerts}, {Sim{\'o}n-D{\'\i}az}, {Groot}, \&
  {Degroote}}]{Aerts2014}
{Aerts}, C., {Sim{\'o}n-D{\'\i}az}, S., {Groot}, P.~J., \& {Degroote}, P. 2014,
  \aap, 569, A118

\bibitem[{{Aerts} {et~al.}(2003){Aerts}, {Thoul}, {Daszy{\'n}ska}, {Scuflaire},
  {Waelkens}, {Dupret}, {Niemczura}, \& {Noels}}]{Aerts2003}
{Aerts}, C., {Thoul}, A., {Daszy{\'n}ska}, J., {et~al.} 2003, Science, 300,
  1926

\bibitem[{{Aerts} {et~al.}(2017{\natexlab{b}}){Aerts}, {Van Reeth}, \&
  {Tkachenko}}]{Aerts2017b}
{Aerts}, C., {Van Reeth}, T., \& {Tkachenko}, A. 2017{\natexlab{b}}, \apjl,
  847, L7

\bibitem[{{Auvergne} {et~al.}(2009){Auvergne}, {Bodin}, {Boisnard}, {Buey},
  {Chaintreuil}, {Epstein}, {Jouret}, {Lam-Trong}, {Levacher}, {Magnan},
  {Perez}, {Plasson}, {Plesseria}, {Peter}, {Steller}, {Tiph{\`e}ne}, {Baglin},
  {Agogu{\'e}}, {Appourchaux}, {Barbet}, {Beaufort}, {Bellenger}, {Berlin},
  {Bernardi}, {Blouin}, {Boumier}, {Bonneau}, {Briet}, {Butler}, {Cautain},
  {Chiavassa}, {Costes}, {Cuvilho}, {Cunha-Parro}, {de Oliveira Fialho},
  {Decaudin}, {Defise}, {Djalal}, {Docclo}, {Drummond}, {Dupuis}, {Exil},
  {Faur{\'e}}, {Gaboriaud}, {Gamet}, {Gavalda}, {Grolleau}, {Gueguen},
  {Guivarc'h}, {Guterman}, {Hasiba}, {Huntzinger}, {Hustaix}, {Imbert},
  {Jeanville}, {Johlander}, {Jorda}, {Journoud}, {Karioty}, {Kerjean},
  {Lafond}, {Lapeyrere}, {Landiech}, {Larqu{\'e}}, {Laudet}, {Le Merrer},
  {Leporati}, {Leruyet}, {Levieuge}, {Llebaria}, {Martin}, {Mazy}, {Mesnager},
  {Michel}, {Moalic}, {Monjoin}, {Naudet}, {Neukirchner}, {Nguyen-Kim},
  {Ollivier}, {Orcesi}, {Ottacher}, {Oulali}, {Parisot}, {Perruchot},
  {Piacentino}, {Pinheiro da Silva}, {Platzer}, {Pontet}, {Pradines},
  {Quentin}, {Rohbeck}, {Rolland}, {Rollenhagen}, {Romagnan}, {Russ}, {Samadi},
  {Schmidt}, {Schwartz}, {Sebbag}, {Smit}, {Sunter}, {Tello}, {Toulouse},
  {Ulmer}, {Vandermarcq}, {Vergnault}, {Wallner}, {Waultier}, \&
  {Zanatta}}]{Auvergne2009}
{Auvergne}, M., {Bodin}, P., {Boisnard}, L., {et~al.} 2009, \aap, 506, 411

\bibitem[{{Balona}(2014)}]{Balona2014}
{Balona}, L.~A. 2014, \mnras, 437, 1476

\bibitem[{{Balona} {et~al.}(2015){Balona}, {Daszy{\'n}ska-Daszkiewicz}, \&
  {Pamyatnykh}}]{Balona2015}
{Balona}, L.~A., {Daszy{\'n}ska-Daszkiewicz}, J., \& {Pamyatnykh}, A.~A. 2015,
  \mnras, 452, 3073

\bibitem[{{Balona} \& {Ozuyar}(2020)}]{Balona2020}
{Balona}, L.~A. \& {Ozuyar}, D. 2020, \mnras, 493, 5871

\bibitem[{{Beck} {et~al.}(2012){Beck}, {Montalban}, {Kallinger}, {De Ridder},
  {Aerts}, {Garc{\'\i}a}, {Hekker}, {Dupret}, {Mosser}, {Eggenberger},
  {Stello}, {Elsworth}, {Frandsen}, {Carrier}, {Hillen}, {Gruberbauer},
  {Christensen-Dalsgaard}, {Miglio}, {Valentini}, {Bedding}, {Kjeldsen},
  {Girouard}, {Hall}, \& {Ibrahim}}]{Beck2012}
{Beck}, P.~G., {Montalban}, J., {Kallinger}, T., {et~al.} 2012, \nat, 481, 55

\bibitem[{Belkacem {et~al.}(2010)Belkacem, Dupret, \& Noels}]{Belkacem2010}
Belkacem, K., Dupret, M.~A., \& Noels, A. 2010, A\&A, 510, A6

\bibitem[{Belkacem {et~al.}(2009)Belkacem, Samadi, Goupil, Lef{\`{e}}vre,
  Baudin, Deheuvels, Dupret, Appourchaux, Scuflaire, Auvergne, Catala, Michel,
  Miglio, Montalban, Thoul, Talon, Baglin, \& Noels}]{Belkacem2009}
Belkacem, K., Samadi, R., Goupil, M.-J., {et~al.} 2009, Science, 324, 1540

\bibitem[{Blomme {et~al.}(2011)Blomme, Mahy, Catala, Cuypers, Gosset, Godart,
  Montalban, Ventura, Rauw, Morel, Degroote, Aerts, Noels, Michel, Baudin,
  Baglin, Auvergne, \& Samadi}]{Blomme2011}
Blomme, R., Mahy, L., Catala, C., {et~al.} 2011, A\&A, 533, A4

\bibitem[{Borucki {et~al.}(2010)Borucki, Koch, Basri, Batalha, Brown, Caldwell,
  Caldwell, Christensen-Dalsgaard, Cochran, DeVore, Dunham, Dupree, Gautier,
  Geary, Gilliland, Gould, Howell, Jenkins, Kondo, Latham, Meibom, Kjeldsen,
  Lissauer, Monet, Morrison, Sasselov, Tarter, Boss, Brownlee, Owen, Buzasi,
  Charbonneau, Doyle, Fortney, Ford, Holman, Seager, Steffen, Welsh, Rowe,
  Anderson, Buchhave, Ciardi, Walkowicz, Sherry, Horch, Isaacson, Everett,
  Fischer, Torres, Johnson, Endl, MacQueen, Bryson, Dotson, Haas,
  Kolodziejczak, Van~Cleve, Chandrasekaran, Twicken, Quintana, Clarke, Allen,
  Li, Wu, Tenenbaum, Verner, Bruhweiler, Barnes, \& Prsa}]{kepler}
Borucki, W.~J., Koch, D., Basri, G., {et~al.} 2010, Science, 327, 977

\bibitem[{{Bouabid} {et~al.}(2013){Bouabid}, {Dupret}, {Salmon},
  {Montalb{\'a}n}, {Miglio}, \& {Noels}}]{Bouabid2013}
{Bouabid}, M.~P., {Dupret}, M.~A., {Salmon}, S., {et~al.} 2013, \mnras, 429,
  2500

\bibitem[{{Bowman}(2020)}]{Bowman2020rev}
{Bowman}, D.~M. 2020, Frontiers in Astronomy and Space Sciences, 7, 70

\bibitem[{Bowman {et~al.}(2019)Bowman, Aerts, Johnston, Pedersen, Rogers,
  Edelmann, Sim{\'{o} }n-D{\'{\i}}az, Reeth, Buysschaert, Tkachenko, \&
  Triana}]{Bowman2019c}
Bowman, D.~M., Aerts, C., Johnston, C., {et~al.} 2019, \aap, 621, A135

\bibitem[{{Bowman} {et~al.}(2019{\natexlab{a}}){Bowman}, {Burssens},
  {Pedersen}, {Johnston}, {Aerts}, {Buysschaert}, {Michielsen}, {Tkachenko},
  {Rogers}, {Edelmann}, {Ratnasingam}, {Sim{\'o}n-D{\'\i}az}, {Castro},
  {Moravveji}, {Pope}, {White}, \& {De Cat}}]{Bowman2019a}
{Bowman}, D.~M., {Burssens}, S., {Pedersen}, M.~G., {et~al.}
  2019{\natexlab{a}}, Nature Astronomy, 3, 760

\bibitem[{{Bowman} {et~al.}(2020){Bowman}, {Burssens}, {Sim{\'o}n-D{\'\i}az},
  {Edelmann}, {Rogers}, {Horst}, {R{\"o}pke}, \& {Aerts}}]{Bowman2020b}
{Bowman}, D.~M., {Burssens}, S., {Sim{\'o}n-D{\'\i}az}, S., {et~al.} 2020,
  \aap, 640, A36

\bibitem[{{Bowman} \& {Dorn-Wallenstein}(2022)}]{Bowman2022e}
{Bowman}, D.~M. \& {Dorn-Wallenstein}, T.~Z. 2022, \aap, 668, A134

\bibitem[{{Bowman} {et~al.}(2019{\natexlab{b}}){Bowman}, {Johnston},
  {Tkachenko}, {Mkrtichian}, {Gunsriwiwat}, \& {Aerts}}]{Bowman2019b}
{Bowman}, D.~M., {Johnston}, C., {Tkachenko}, A., {et~al.} 2019{\natexlab{b}},
  \apjl, 883, L26

\bibitem[{{Bowman} \& {Kurtz}(2018)}]{Bowman2018}
{Bowman}, D.~M. \& {Kurtz}, D.~W. 2018, \mnras, 476, 3169

\bibitem[{{Bowman} \& {Michielsen}(2021)}]{Bowman2021c}
{Bowman}, D.~M. \& {Michielsen}, M. 2021, \aap, 656, A158

\bibitem[{{Brahm} {et~al.}(2017){Brahm}, {Jord{\'a}n}, \&
  {Espinoza}}]{Brahm2017}
{Brahm}, R., {Jord{\'a}n}, A., \& {Espinoza}, N. 2017, \pasp, 129, 034002

\bibitem[{{Briquet} {et~al.}(2011){Briquet}, {Aerts}, {Baglin}, {Nieva},
  {Degroote}, {Przybilla}, {Noels}, {Schiller}, {Vu{\v{c}}kovi{\'c}}, {Oreiro},
  {Smolders}, {Auvergne}, {Baudin}, {Catala}, {Michel}, \&
  {Samadi}}]{Briquet2011}
{Briquet}, M., {Aerts}, C., {Baglin}, A., {et~al.} 2011, \aap, 527, A112

\bibitem[{{Briquet} {et~al.}(2012){Briquet}, {Neiner}, {Aerts}, {Morel},
  {Mathis}, {Reese}, {Lehmann}, {Costero}, {Echevarria}, {Handler}, {Kambe},
  {Hirata}, {Masuda}, {Wright}, {Yang}, {Pintado}, {Mkrtichian}, {Lee}, {Han},
  {Bruch}, {De Cat}, {Uytterhoeven}, {Lefever}, {Vanautgaerden}, {de Batz},
  {Fr{\'e}mat}, {Henrichs}, {Geers}, {Martayan}, {Hubert}, {Thizy}, \&
  {Tijani}}]{Briquet2012}
{Briquet}, M., {Neiner}, C., {Aerts}, C., {et~al.} 2012, \mnras, 427, 483

\bibitem[{{Burssens} {et~al.}(2020){Burssens}, {Sim{\'o}n-D{\'\i}az}, {Bowman},
  {Holgado}, {Michielsen}, {de Burgos}, {Castro}, {Barb{\'a}}, \&
  {Aerts}}]{Burssens2020}
{Burssens}, S., {Sim{\'o}n-D{\'\i}az}, S., {Bowman}, D.~M., {et~al.} 2020,
  \aap, 639, A81

\bibitem[{{Buysschaert} {et~al.}(2015){Buysschaert}, {Aerts}, {Bloemen},
  {Debosscher}, {Neiner}, {Briquet}, {Vos}, {P{\'a}pics}, {Manick}, {Schmid},
  {Van Winckel}, \& {Tkachenko}}]{Buysschaert2015}
{Buysschaert}, B., {Aerts}, C., {Bloemen}, S., {et~al.} 2015, \mnras, 453, 89

\bibitem[{{Buysschaert} {et~al.}(2018){Buysschaert}, {Aerts}, {Bowman},
  {Johnston}, {Van Reeth}, {Pedersen}, {Mathis}, \&
  {Neiner}}]{Buysschaert2018c}
{Buysschaert}, B., {Aerts}, C., {Bowman}, D.~M., {et~al.} 2018, \aap, 616, A148

\bibitem[{{Cantiello} {et~al.}(2009){Cantiello}, {Langer}, {Brott}, {de Koter},
  {Shore}, {Vink}, {Voegler}, {Lennon}, \& {Yoon}}]{Cantiello2009}
{Cantiello}, M., {Langer}, N., {Brott}, I., {et~al.} 2009, \aap, 499, 279

\bibitem[{{Cantiello} {et~al.}(2021){Cantiello}, {Lecoanet}, {Jermyn}, \&
  {Grassitelli}}]{Cantiello2021}
{Cantiello}, M., {Lecoanet}, D., {Jermyn}, A.~S., \& {Grassitelli}, L. 2021,
  \apj, 915, 112

\bibitem[{{Cantiello} {et~al.}(2014){Cantiello}, {Mankovich}, {Bildsten},
  {Christensen-Dalsgaard}, \& {Paxton}}]{Cantiello2014}
{Cantiello}, M., {Mankovich}, C., {Bildsten}, L., {Christensen-Dalsgaard}, J.,
  \& {Paxton}, B. 2014, \apj, 788, 93

\bibitem[{{Carroll}(1933)}]{Carroll1933}
{Carroll}, J.~A. 1933, \mnras, 93, 478

\bibitem[{{Cheng} {et~al.}(2020){Cheng}, {Fuller}, {Guo}, {Lehman}, \&
  {Hambleton}}]{Cheng2020}
{Cheng}, S.~J., {Fuller}, J., {Guo}, Z., {Lehman}, H., \& {Hambleton}, K. 2020,
  \apj, 903, 122

\bibitem[{{Choudhury} {et~al.}(2021){Choudhury}, {de Grijs}, {Bekki}, {Cioni},
  {Ivanov}, {van Loon}, {Miller}, {Niederhofer}, {Oliveira}, {Ripepi}, {Sun},
  \& {Subramanian}}]{Choudhury2021}
{Choudhury}, S., {de Grijs}, R., {Bekki}, K., {et~al.} 2021, \mnras, 507, 4752

\bibitem[{{Claret} \& {Bloemen}(2011)}]{limb_darkening}
{Claret}, A. \& {Bloemen}, S. 2011, A\&A, 529, A75

\bibitem[{{Claret} \& {Torres}(2018)}]{claret}
{Claret}, A. \& {Torres}, G. 2018, The Astrophysical Journal, 859, id.100

\bibitem[{{Claret} \& {Torres}(2019)}]{Claret2019}
{Claret}, A. \& {Torres}, G. 2019, \apj, 876, 134

\bibitem[{Conn {et~al.}(2000)Conn, Gould, \& Toint}]{TRF}
Conn, A.~R., Gould, N. I.~M., \& Toint, P.~L. 2000, Trust Region Methods
  (Society for Industrial and Applied Mathematics)

\bibitem[{{De Cat} \& {Aerts}(2002)}]{DeCatAerts2002}
{De Cat}, P. \& {Aerts}, C. 2002, \aap, 393, 965

\bibitem[{{Deal} {et~al.}(2017){Deal}, {Escobar}, {Vauclair}, {Vauclair},
  {Hui-Bon-Hoa}, \& {Richard}}]{Deal2017}
{Deal}, M., {Escobar}, M.~E., {Vauclair}, S., {et~al.} 2017, \aap, 601, A127

\bibitem[{{Deal} {et~al.}(2016){Deal}, {Richard}, \& {Vauclair}}]{Deal2016}
{Deal}, M., {Richard}, O., \& {Vauclair}, S. 2016, \aap, 589, A140

\bibitem[{{Degroote} {et~al.}(2010{\natexlab{a}}){Degroote}, {Aerts}, {Baglin},
  {Miglio}, {Briquet}, {Noels}, {Niemczura}, {Montalban}, {Bloemen}, {Oreiro},
  {Vu{\v{c}}kovi{\'c}}, {Smolders}, {Auvergne}, {Baudin}, {Catala}, \&
  {Michel}}]{Degroote2010a}
{Degroote}, P., {Aerts}, C., {Baglin}, A., {et~al.} 2010{\natexlab{a}}, \nat,
  464, 259

\bibitem[{{Degroote} {et~al.}(2010{\natexlab{b}}){Degroote}, {Briquet},
  {Auvergne}, {Sim{\'o}n-D{\'\i}az}, {Aerts}, {Noels}, {Rainer}, {Hareter},
  {Poretti}, {Mahy}, {Oreiro}, {Vu{\v{c}}kovi{\'c}}, {Smolders}, {Baglin},
  {Baudin}, {Catala}, {Michel}, \& {Samadi}}]{Degroote2010c}
{Degroote}, P., {Briquet}, M., {Auvergne}, M., {et~al.} 2010{\natexlab{b}},
  \aap, 519, A38

\bibitem[{{Degroote} {et~al.}(2009){Degroote}, {Briquet}, {Catala},
  {Uytterhoeven}, {Lefever}, {Morel}, {Aerts}, {Carrier}, {Auvergne}, {Baglin},
  \& {Michel}}]{Degroote2009}
{Degroote}, P., {Briquet}, M., {Catala}, C., {et~al.} 2009, \aap, 506, 111

\bibitem[{{Deheuvels} {et~al.}(2020){Deheuvels}, {Ballot}, {Eggenberger},
  {Spada}, {Noll}, \& {den Hartogh}}]{Deheuvels2020}
{Deheuvels}, S., {Ballot}, J., {Eggenberger}, P., {et~al.} 2020, \aap, 641,
  A117

\bibitem[{{Deheuvels} {et~al.}(2016){Deheuvels}, {Brand{\~a}o}, {Silva
  Aguirre}, {Ballot}, {Michel}, {Cunha}, {Lebreton}, \&
  {Appourchaux}}]{Deheuvels2016}
{Deheuvels}, S., {Brand{\~a}o}, I., {Silva Aguirre}, V., {et~al.} 2016, \aap,
  589, A93

\bibitem[{{Dekker} {et~al.}(2000){Dekker}, {D'Odorico}, {Kaufer}, {Delabre}, \&
  {Kotzlowski}}]{Dekker2000}
{Dekker}, H., {D'Odorico}, S., {Kaufer}, A., {Delabre}, B., \& {Kotzlowski}, H.
  2000, in Society of Photo-Optical Instrumentation Engineers (SPIE) Conference
  Series, Vol. 4008, Optical and IR Telescope Instrumentation and Detectors,
  ed. M.~{Iye} \& A.~F. {Moorwood}, 534--545

\bibitem[{{Diago} {et~al.}(2009){Diago}, {Guti{\'e}rrez-Soto}, {Auvergne},
  {Fabregat}, {Hubert}, {Floquet}, {Fr{\'e}mat}, {Garrido}, {Andrade}, {de
  Batz}, {Emilio}, {Espinosa Lara}, {Huat}, {Janot-Pacheco}, {Leroy},
  {Martayan}, {Neiner}, {Semaan}, {Suso}, {Catala}, {Poretti}, {Rainer},
  {Uytterhoeven}, {Michel}, \& {Samadi}}]{Diago2009}
{Diago}, P.~D., {Guti{\'e}rrez-Soto}, J., {Auvergne}, M., {et~al.} 2009, \aap,
  506, 125

\bibitem[{{Donati} {et~al.}(1997){Donati}, {Semel}, {Carter}, {Rees}, \&
  {Collier Cameron}}]{Donati1997}
{Donati}, J.~F., {Semel}, M., {Carter}, B.~D., {Rees}, D.~E., \& {Collier
  Cameron}, A. 1997, \mnras, 291, 658

\bibitem[{{Dziembowski} \& {Pamyatnykh}(2008)}]{Dziembowski2008}
{Dziembowski}, W.~A. \& {Pamyatnykh}, A.~A. 2008, \mnras, 385, 2061

\bibitem[{{Edelmann} {et~al.}(2019){Edelmann}, {Ratnasingam}, {Pedersen},
  {Bowman}, {Prat}, \& {Rogers}}]{Edelmann2019}
{Edelmann}, P.~V.~F., {Ratnasingam}, R.~P., {Pedersen}, M.~G., {et~al.} 2019,
  \apj, 876, 4

\bibitem[{{Freudling} {et~al.}(2013){Freudling}, {Romaniello}, {Bramich},
  {Ballester}, {Forchi}, {Garc{\'{\i}}a-Dabl{\'o}}, {Moehler}, \&
  {Neeser}}]{Freudling2013}
{Freudling}, W., {Romaniello}, M., {Bramich}, D.~M., {et~al.} 2013, \aap, 559,
  A96

\bibitem[{{Fuller}(2017)}]{Fuller2017a}
{Fuller}, J. 2017, \mnras, 472, 1538

\bibitem[{{Fuller}(2021)}]{Fuller2021}
{Fuller}, J. 2021, \mnras, 501, 483

\bibitem[{{Fuller} {et~al.}(2017){Fuller}, {Hambleton}, {Shporer}, {Isaacson},
  \& {Thompson}}]{Fuller2017b}
{Fuller}, J., {Hambleton}, K., {Shporer}, A., {Isaacson}, H., \& {Thompson}, S.
  2017, \mnras, 472, L25

\bibitem[{{Fuller} {et~al.}(2020){Fuller}, {Kurtz}, {Handler}, \&
  {Rappaport}}]{Fuller2020}
{Fuller}, J., {Kurtz}, D.~W., {Handler}, G., \& {Rappaport}, S. 2020, \mnras,
  498, 5730

\bibitem[{{Fuller} \& {Lai}(2012)}]{Fuller2012}
{Fuller}, J. \& {Lai}, D. 2012, \mnras, 420, 3126

\bibitem[{{Gaia Collaboration} {et~al.}(2022){Gaia Collaboration}, {De Ridder},
  {Ripepi}, {Aerts}, {Palaversa}, {Eyer}, {Holl}, {Audard}, \&
  {Rimoldini}}]{DeRidder2022}
{Gaia Collaboration}, {De Ridder}, J., {Ripepi}, V., {et~al.} 2022, \aap, in
  press

\bibitem[{{Garcia} {et~al.}(2023){Garcia}, {Van Reeth}, {De Ridder}, \&
  {Aerts}}]{Garcia2023}
{Garcia}, S., {Van Reeth}, T., {De Ridder}, J., \& {Aerts}, C. 2023, \aap, in
  press, arXiv:2210.09526

\bibitem[{{Garcia} {et~al.}(2022){Garcia}, {Van Reeth}, {De Ridder},
  {Tkachenko}, {IJspeert}, \& {Aerts}}]{Garcia2022}
{Garcia}, S., {Van Reeth}, T., {De Ridder}, J., {et~al.} 2022, \aap, 662, A82

\bibitem[{{Gebruers} {et~al.}(2022){Gebruers}, {Tkachenko}, {Bowman}, {Van
  Reeth}, {Burssens}, \& {IJspeert}}]{Gebruers2022}
{Gebruers}, S., {Tkachenko}, A., {Bowman}, D., {et~al.} 2022, submitted to A\&A

\bibitem[{{Gray}(2008)}]{Gray2008}
{Gray}, D.~F. 2008, {The Observation and Analysis of Stellar Photospheres}

\bibitem[{{Gray} \& {Corbally}(2009)}]{Gray2009}
{Gray}, R.~O. \& {Corbally}, Christopher, J. 2009, {Stellar Spectral
  Classification}

\bibitem[{{Grevesse} {et~al.}(2007){Grevesse}, {Asplund}, \&
  {Sauval}}]{Grevesse2007}
{Grevesse}, N., {Asplund}, M., \& {Sauval}, A.~J. 2007, \ssr, 130, 105

\bibitem[{{Guo}(2021)}]{Guo2021}
{Guo}, Z. 2021, Frontiers in Astronomy and Space Sciences, 8, 67

\bibitem[{{Guo} {et~al.}(2019){Guo}, {Fuller}, {Shporer}, {Li}, {Hambleton},
  {Manuel}, {Murphy}, \& {Isaacson}}]{Guo2019}
{Guo}, Z., {Fuller}, J., {Shporer}, A., {et~al.} 2019, \apj, 885, 46

\bibitem[{{Guo} {et~al.}(2020){Guo}, {Shporer}, {Hambleton}, \&
  {Isaacson}}]{Guo2020}
{Guo}, Z., {Shporer}, A., {Hambleton}, K., \& {Isaacson}, H. 2020, \apj, 888,
  95

\bibitem[{{Guti{\'e}rrez-Soto} {et~al.}(2009){Guti{\'e}rrez-Soto}, {Floquet},
  {Samadi}, {Neiner}, {Garrido}, {Fabregat}, {Fr{\'e}mat}, {Diago}, {Huat},
  {Leroy}, {Emilio}, {Hubert}, {Andrade}, {de Batz}, {Janot-Pacheco}, {Espinosa
  Lara}, {Martayan}, {Semaan}, {Suso}, {Auvergne}, {Chaintreuil}, {Michel}, \&
  {Catala}}]{Gutierrez2009}
{Guti{\'e}rrez-Soto}, J., {Floquet}, M., {Samadi}, R., {et~al.} 2009, \aap,
  506, 133

\bibitem[{{Hambleton} {et~al.}(2016){Hambleton}, {Kurtz}, {Pr{\v{s}}a},
  {Quinn}, {Fuller}, {Murphy}, {Thompson}, {Latham}, \&
  {Shporer}}]{Hambleton2016}
{Hambleton}, K., {Kurtz}, D.~W., {Pr{\v{s}}a}, A., {et~al.} 2016, \mnras, 463,
  1199

\bibitem[{{Hambleton} {et~al.}(2013){Hambleton}, {Kurtz}, {Pr{\v{s}}a},
  {Guzik}, {Pavlovski}, {Bloemen}, {Southworth}, {Conroy}, {Littlefair}, \&
  {Fuller}}]{Hambleton2013}
{Hambleton}, K.~M., {Kurtz}, D.~W., {Pr{\v{s}}a}, A., {et~al.} 2013, \mnras,
  434, 925

\bibitem[{{Handler} {et~al.}(2006){Handler}, {Jerzykiewicz}, {Rodr{\'\i}guez},
  {Uytterhoeven}, {Amado}, {Dorokhova}, {Dorokhov}, {Poretti}, {Sareyan},
  {Parrao}, {Lorenz}, {Zsuffa}, {Drummond}, {Daszy{\'n}ska-Daszkiewicz},
  {Verhoelst}, {De Ridder}, {Acke}, {Bourge}, {Movchan}, {Garrido},
  {Papar{\'o}}, {Sahin}, {Antoci}, {Udovichenko}, {Csorba}, {Crowe}, {Berkey},
  {Stewart}, {Terry}, {Mkrtichian}, \& {Aerts}}]{Handler2006}
{Handler}, G., {Jerzykiewicz}, M., {Rodr{\'\i}guez}, E., {et~al.} 2006, \mnras,
  365, 327

\bibitem[{{Handler} {et~al.}(2020){Handler}, {Kurtz}, {Rappaport}, {Saio},
  {Fuller}, {Jones}, {Guo}, {Chowdhury}, {Sowicka}, {Kahraman
  Ali{\c{c}}avu{\c{s}}}, {Streamer}, {Murphy}, {Gagliano}, {Jacobs}, \&
  {Vanderburg}}]{Handler2020}
{Handler}, G., {Kurtz}, D.~W., {Rappaport}, S.~A., {et~al.} 2020, Nature
  Astronomy, 4, 684

\bibitem[{{Handler} {et~al.}(2009){Handler}, {Matthews}, {Eaton},
  {Daszy{\'n}ska-Daszkiewicz}, {Kuschnig}, {Lehmann}, {Rodr{\'\i}guez},
  {Pamyatnykh}, {Zdravkov}, {Lenz}, {Costa}, {D{\'\i}az-Fraile}, {Sota},
  {Kwiatkowski}, {Schwarzenberg-Czerny}, {Borczyk}, {Dimitrov}, {Fagas},
  {Kami{\'n}ski}, {Ro{\.z}ek}, {van Wyk}, {Pollard}, {Kilmartin}, {Weiss},
  {Guenther}, {Moffat}, {Rucinski}, {Sasselov}, \& {Walker}}]{Handler2009}
{Handler}, G., {Matthews}, J.~M., {Eaton}, J.~A., {et~al.} 2009, \apjl, 698,
  L56

\bibitem[{{Hekker} \& {Christensen-Dalsgaard}(2017)}]{Hekker2017}
{Hekker}, S. \& {Christensen-Dalsgaard}, J. 2017, \aapr, 25, 1

\bibitem[{{Hermes} {et~al.}(2017){Hermes}, {G{\"a}nsicke}, {Kawaler}, {Greiss},
  {Tremblay}, {Gentile Fusillo}, {Raddi}, {Fanale}, {Bell}, {Dennihy}, {Fuchs},
  {Dunlap}, {Clemens}, {Montgomery}, {Winget}, {Chote}, {Marsh}, \&
  {Redfield}}]{Hermes2017}
{Hermes}, J.~J., {G{\"a}nsicke}, B.~T., {Kawaler}, S.~D., {et~al.} 2017, \apjs,
  232, 23

\bibitem[{{Hillier} \& {Miller}(1998)}]{Hillier1998}
{Hillier}, D.~J. \& {Miller}, D.~L. 1998, \apj, 496, 407

\bibitem[{{Holgado} {et~al.}(2018){Holgado}, {Sim{\'o}n-D{\'\i}az},
  {Barb{\'a}}, {Puls}, {Herrero}, {Castro}, {Garcia}, {Ma{\'\i}z
  Apell{\'a}niz}, {Negueruela}, \& {Sab{\'\i}n-Sanjuli{\'a}n}}]{Holgado2018}
{Holgado}, G., {Sim{\'o}n-D{\'\i}az}, S., {Barb{\'a}}, R.~H., {et~al.} 2018,
  \aap, 613, A65

\bibitem[{{Horst} {et~al.}(2020){Horst}, {Edelmann}, {Andr{\'a}ssy},
  {R{\"o}pke}, {Bowman}, {Aerts}, \& {Ratnasingam}}]{Horst2020}
{Horst}, L., {Edelmann}, P.~V.~F., {Andr{\'a}ssy}, R., {et~al.} 2020, \aap,
  641, A18

\bibitem[{{Huat} {et~al.}(2009){Huat}, {Hubert}, {Baudin}, {Floquet}, {Neiner},
  {Fr{\'e}mat}, {Guti{\'e}rrez-Soto}, {Andrade}, {de Batz}, {Diago}, {Emilio},
  {Espinosa Lara}, {Fabregat}, {Janot-Pacheco}, {Leroy}, {Martayan}, {Semaan},
  {Suso}, {Auvergne}, {Catala}, {Michel}, \& {Samadi}}]{Huat2009}
{Huat}, A.~L., {Hubert}, A.~M., {Baudin}, F., {et~al.} 2009, \aap, 506, 95

\bibitem[{{Hunter} {et~al.}(2008){Hunter}, {Lennon}, {Dufton}, {Trundle},
  {Sim{\'o}n-D{\'\i}az}, {Smartt}, {Ryans}, \& {Evans}}]{Hunter2008}
{Hunter}, I., {Lennon}, D.~J., {Dufton}, P.~L., {et~al.} 2008, \aap, 479, 541

\bibitem[{{IJspeert} {et~al.}(2021){IJspeert}, {Tkachenko}, {Johnston},
  {Garcia}, {De Ridder}, {Van Reeth}, \& {Aerts}}]{IJspeert2021}
{IJspeert}, L.~W., {Tkachenko}, A., {Johnston}, C., {et~al.} 2021, \aap, 652,
  A120

\bibitem[{{Jayaraman} {et~al.}(2022){Jayaraman}, {Handler}, {Rappaport},
  {Fuller}, {Kurtz}, {Charpinet}, \& {Ricker}}]{Jayaraman2022}
{Jayaraman}, R., {Handler}, G., {Rappaport}, S.~A., {et~al.} 2022, \apjl, 928,
  L14

\bibitem[{{Johnston}(2021)}]{Johnston2021}
{Johnston}, C. 2021, \aap, 655, A29

\bibitem[{{Kaufer} {et~al.}(1999){Kaufer}, {Stahl}, {Tubbesing},
  {N{\o}rregaard}, {Avila}, {Francois}, {Pasquini}, \& {Pizzella}}]{Kaufer1999}
{Kaufer}, A., {Stahl}, O., {Tubbesing}, S., {et~al.} 1999, The Messenger, 95, 8

\bibitem[{{Ko{\l}aczek-Szyma{\'n}ski}
  {et~al.}(2021){Ko{\l}aczek-Szyma{\'n}ski}, {Pigulski}, {Michalska},
  {Mo{\'z}dzierski}, \& {R{\'o}{\.z}a{\'n}ski}}]{Kolaczek2021}
{Ko{\l}aczek-Szyma{\'n}ski}, P.~A., {Pigulski}, A., {Michalska}, G.,
  {Mo{\'z}dzierski}, D., \& {R{\'o}{\.z}a{\'n}ski}, T. 2021, \aap, 647, A12

\bibitem[{{Kurtz}(2022)}]{Kurtz2022}
{Kurtz}, D.~W. 2022, Annual Review of Astronomy and Astrophysics, 60, 31

\bibitem[{{Kurtz} {et~al.}(2020){Kurtz}, {Handler}, {Rappaport}, {Saio},
  {Fuller}, {Jacobs}, {Schmitt}, {Jones}, {Vanderburg}, {LaCourse}, {Nelson},
  {Kahraman Ali{\c{c}}avu{\c{s}}}, \& {Giarrusso}}]{Kurtz2020}
{Kurtz}, D.~W., {Handler}, G., {Rappaport}, S.~A., {et~al.} 2020, \mnras, 494,
  5118

\bibitem[{{Lampens}(2021)}]{Lampens2021}
{Lampens}, P. 2021, Galaxies, 9, 28

\bibitem[{{Lanz} \& {Hubeny}(2003)}]{Lanz2003}
{Lanz}, T. \& {Hubeny}, I. 2003, \apjs, 146, 417

\bibitem[{{Lanz} \& {Hubeny}(2007)}]{Lanz2007}
{Lanz}, T. \& {Hubeny}, I. 2007, \apjs, 169, 83

\bibitem[{{Lecoanet} {et~al.}(2022){Lecoanet}, {Bowman}, \& {Van
  Reeth}}]{Lecoanet2022a}
{Lecoanet}, D., {Bowman}, D.~M., \& {Van Reeth}, T. 2022, \mnras, 512, L16

\bibitem[{{Lecoanet} {et~al.}(2019){Lecoanet}, {Cantiello}, {Quataert},
  {Couston}, {Burns}, {Pope}, {Jermyn}, {Favier}, \& {Le Bars}}]{Lecoanet2019}
{Lecoanet}, D., {Cantiello}, M., {Quataert}, E., {et~al.} 2019, \apjl, 886, L15

\bibitem[{{Lee}(2021)}]{Lee2021}
{Lee}, J.~W. 2021, \pasj, 73, 809

\bibitem[{{Li} {et~al.}(2020){Li}, {Van Reeth}, {Bedding}, {Murphy}, {Antoci},
  {Ouazzani}, \& {Barbara}}]{Li2020}
{Li}, G., {Van Reeth}, T., {Bedding}, T.~R., {et~al.} 2020, \mnras, 491, 3586

\bibitem[{{Maceroni} {et~al.}(2009){Maceroni}, {Montalb{\'a}n}, {Michel},
  {Harmanec}, {Prsa}, {Briquet}, {Niemczura}, {Morel}, {Ladjal}, {Auvergne},
  {Baglin}, {Baudin}, {Catala}, {Samadi}, \& {Aerts}}]{Maceroni2009}
{Maceroni}, C., {Montalb{\'a}n}, J., {Michel}, E., {et~al.} 2009, \aap, 508,
  1375

\bibitem[{{Mahy} {et~al.}(2020{\natexlab{a}}){Mahy}, {Almeida, L. A.}, {Sana,
  H.}, {Clark, J. S.}, {de Koter, A.}, {de Mink, S. E.}, {Evans, C. J.}, {Grin,
  N. J.}, {Langer, N.}, {Moffat, A. F. J.}, {Schneider, F. R. N.}, {Shenar,
  T.}, \& {Tramper, F.}}]{Mahy2020b}
{Mahy}, L., {Almeida, L. A.}, {Sana, H.}, {et~al.} 2020{\natexlab{a}}, A\&A,
  634, A119

\bibitem[{{Mahy} {et~al.}(2011){Mahy}, {Gosset, E.}, {Baudin, F.}, {Rauw, G.},
  {Godart, M.}, {Morel, T.}, {Degroote, P.}, {Aerts, C.}, {Blomme, R.},
  {Cuypers, J.}, {Noels, A.}, {Michel, E.}, {Baglin, A.}, {Auvergne, M.},
  {Catala, C.}, \& {Samadi, R.}}]{Mahy2011}
{Mahy}, L., {Gosset, E.}, {Baudin, F.}, {et~al.} 2011, A\&A, 525, A101

\bibitem[{{Mahy} {et~al.}(2020{\natexlab{b}}){Mahy}, {Sana, H.}, {Abdul-Masih,
  M.}, {Almeida, L. A.}, {Langer, N.}, {Shenar, T.}, {de Koter, A.}, {de Mink,
  S. E.}, {de Wit, S.}, {Grin, N. J.}, {Evans, C. J.}, {Moffat, A. F. J.},
  {Schneider, F. R. N.}, {Barb\'a, R.}, {Clark, J. S.}, {Crowther, P.},
  {Gr\"afener, G.}, {Lennon, D. J.}, {Tramper, F.}, \& {Vink, J.
  S.}}]{Mahy2020a}
{Mahy}, L., {Sana, H.}, {Abdul-Masih, M.}, {et~al.} 2020{\natexlab{b}}, A\&A,
  634, A118

\bibitem[{{Martins} \& {Palacios}(2013)}]{Martins2013}
{Martins}, F. \& {Palacios}, A. 2013, \aap, 560, A16

\bibitem[{{Maxted} {et~al.}(2020){Maxted}, {Gaulme}, {Graczyk}, {He{\l}miniak},
  {Johnston}, {Orosz}, {Pr{\v{s}}a}, {Southworth}, {Torres}, {Davies}, {Ball},
  \& {Chaplin}}]{Maxted2020}
{Maxted}, P.~F.~L., {Gaulme}, P., {Graczyk}, D., {et~al.} 2020, \mnras, 498,
  332

\bibitem[{{Maxted} \& {Hutcheon}(2018)}]{Maxted2018}
{Maxted}, P.~F.~L. \& {Hutcheon}, R.~J. 2018, \aap, 616, A38

\bibitem[{{Michielsen} {et~al.}(2021){Michielsen}, {Aerts}, \&
  {Bowman}}]{Michielsen2021}
{Michielsen}, M., {Aerts}, C., \& {Bowman}, D.~M. 2021, \aap, 650, A175

\bibitem[{{Michielsen} {et~al.}(2019){Michielsen}, {Pedersen}, {Augustson},
  {Mathis}, \& {Aerts}}]{Michielsen2019}
{Michielsen}, M., {Pedersen}, M.~G., {Augustson}, K.~C., {Mathis}, S., \&
  {Aerts}, C. 2019, \aap, 628, A76

\bibitem[{{Miglio} {et~al.}(2008){Miglio}, {Montalb{\'a}n}, {Noels}, \&
  {Eggenberger}}]{Miglio2008}
{Miglio}, A., {Montalb{\'a}n}, J., {Noels}, A., \& {Eggenberger}, P. 2008,
  \mnras, 386, 1487

\bibitem[{{Mombarg} {et~al.}(2022){Mombarg}, {Dotter}, {Rieutord},
  {Michielsen}, {Van Reeth}, \& {Aerts}}]{Mombarg2022}
{Mombarg}, J. S.~G., {Dotter}, A., {Rieutord}, M., {et~al.} 2022, \apj, 925,
  154

\bibitem[{{Mombarg} {et~al.}(2020){Mombarg}, {Dotter}, {Van Reeth},
  {Tkachenko}, {Gebruers}, \& {Aerts}}]{Mombarg2020}
{Mombarg}, J. S.~G., {Dotter}, A., {Van Reeth}, T., {et~al.} 2020, \apj, 895,
  51

\bibitem[{{Mombarg} {et~al.}(2021){Mombarg}, {Van Reeth}, \&
  {Aerts}}]{Mombarg2021}
{Mombarg}, J.~S.~G., {Van Reeth}, T., \& {Aerts}, C. 2021, \aap, 650, A58

\bibitem[{{Mombarg} {et~al.}(2019){Mombarg}, {Van Reeth}, {Pedersen},
  {Molenberghs}, {Bowman}, {Johnston}, {Tkachenko}, \& {Aerts}}]{Mombarg2019}
{Mombarg}, J.~S.~G., {Van Reeth}, T., {Pedersen}, M.~G., {et~al.} 2019, \mnras,
  485, 3248

\bibitem[{{Moravveji} {et~al.}(2015){Moravveji}, {Aerts}, {P{\'a}pics},
  {Triana}, \& {Vandoren}}]{Moravveji2015}
{Moravveji}, E., {Aerts}, C., {P{\'a}pics}, P.~I., {Triana}, S.~A., \&
  {Vandoren}, B. 2015, \aap, 580, A27

\bibitem[{{Moravveji} {et~al.}(2016){Moravveji}, {Townsend}, {Aerts}, \&
  {Mathis}}]{Moravveji2016}
{Moravveji}, E., {Townsend}, R. H.~D., {Aerts}, C., \& {Mathis}, S. 2016, \apj,
  823, 130

\bibitem[{{Mosser} {et~al.}(2012){Mosser}, {Goupil}, {Belkacem}, {Marques},
  {Beck}, {Bloemen}, {De Ridder}, {Barban}, {Deheuvels}, {Elsworth}, {Hekker},
  {Kallinger}, {Ouazzani}, {Pinsonneault}, {Samadi}, {Stello}, {Garc{\'\i}a},
  {Klaus}, {Li}, {Mathur}, \& {Morris}}]{Mosser2012}
{Mosser}, B., {Goupil}, M.~J., {Belkacem}, K., {et~al.} 2012, \aap, 548, A10

\bibitem[{{Mowlavi} {et~al.}(2013){Mowlavi}, {Barblan}, {Saesen}, \&
  {Eyer}}]{Mowlavi2013}
{Mowlavi}, N., {Barblan}, F., {Saesen}, S., \& {Eyer}, L. 2013, \aap, 554, A108

\bibitem[{{Mowlavi} {et~al.}(2016){Mowlavi}, {Saesen}, {Semaan}, {Eggenberger},
  {Barblan}, {Eyer}, {Ekstr{\"o}m}, \& {Georgy}}]{Mowlavi2016}
{Mowlavi}, N., {Saesen}, S., {Semaan}, T., {et~al.} 2016, \aap, 595, L1

\bibitem[{{Mo{\'z}dzierski} {et~al.}(2019){Mo{\'z}dzierski}, {Pigulski},
  {Ko{\l}aczkowski}, {Michalska}, {Kopacki}, {Carrier}, {Walczak}, {Narwid},
  {St{\k{e}}{\'s}licki}, {Fu}, {Jiang}, {Zhang}, {Jackiewicz}, {Telting},
  {Morel}, {Saesen}, {Zahajkiewicz}, {Bru{\'s}}, {{\'S}r{\'o}dka},
  {Vu{\v{c}}kovi{\'c}}, {Verhoelst}, {Van Helshoecht}, {Lefever}, {Gielen},
  {Decin}, {Vanautgaerden}, \& {Aerts}}]{Mozdzierski2019}
{Mo{\'z}dzierski}, D., {Pigulski}, A., {Ko{\l}aczkowski}, Z., {et~al.} 2019,
  \aap, 632, A95

\bibitem[{{Mo{\'z}dzierski} {et~al.}(2014){Mo{\'z}dzierski}, {Pigulski},
  {Kopacki}, {Ko{\l}aczkowski}, \& {St{\k{e}}{\'s}licki}}]{Mozdzierski2014}
{Mo{\'z}dzierski}, D., {Pigulski}, A., {Kopacki}, G., {Ko{\l}aczkowski}, Z., \&
  {St{\k{e}}{\'s}licki}, M. 2014, \actaa, 64, 89

\bibitem[{{Murphy} {et~al.}(2019){Murphy}, {Hey}, {Van Reeth}, \&
  {Bedding}}]{Murphy2019}
{Murphy}, S.~J., {Hey}, D., {Van Reeth}, T., \& {Bedding}, T.~R. 2019, \mnras,
  485, 2380

\bibitem[{{Neiner} {et~al.}(2009){Neiner}, {Guti{\'e}rrez-Soto}, {Baudin}, {de
  Batz}, {Fr{\'e}mat}, {Huat}, {Floquet}, {Hubert}, {Leroy}, {Diago},
  {Poretti}, {Carrier}, {Rainer}, {Catala}, {Thizy}, {Buil}, {Ribeiro},
  {Andrade}, {Emilio}, {Espinosa Lara}, {Fabregat}, {Janot-Pacheco},
  {Martayan}, {Semaan}, {Suso}, {Baglin}, {Michel}, \& {Samadi}}]{Neiner2009}
{Neiner}, C., {Guti{\'e}rrez-Soto}, J., {Baudin}, F., {et~al.} 2009, \aap, 506,
  143

\bibitem[{{Neiner} {et~al.}(2012){Neiner}, {Mathis}, {Saio}, {Lovekin},
  {Eggenberger}, \& {Lee}}]{Neiner2012b}
{Neiner}, C., {Mathis}, S., {Saio}, H., {et~al.} 2012, \aap, 539, A90

\bibitem[{{Ouazzani} {et~al.}(2020){Ouazzani}, {Ligni{\`e}res}, {Dupret},
  {Salmon}, {Ballot}, {Christophe}, \& {Takata}}]{Ouazzani2020}
{Ouazzani}, R.~M., {Ligni{\`e}res}, F., {Dupret}, M.~A., {et~al.} 2020, \aap,
  640, A49

\bibitem[{{Ouazzani} {et~al.}(2019){Ouazzani}, {Marques}, {Goupil},
  {Christophe}, {Antoci}, {Salmon}, \& {Ballot}}]{Ouazzani2019}
{Ouazzani}, R.~M., {Marques}, J.~P., {Goupil}, M.~J., {et~al.} 2019, \aap, 626,
  A121

\bibitem[{{Ouazzani} {et~al.}(2017){Ouazzani}, {Salmon}, {Antoci}, {Bedding},
  {Murphy}, \& {Roxburgh}}]{Ouazzani2017}
{Ouazzani}, R.-M., {Salmon}, S.~J.~A.~J., {Antoci}, V., {et~al.} 2017, \mnras,
  465, 2294

\bibitem[{{P{\'a}pics} {et~al.}(2017){P{\'a}pics}, {Tkachenko}, {Van Reeth},
  {Aerts}, {Moravveji}, {Van de Sande}, {De Smedt}, {Bloemen}, {Southworth},
  {Debosscher}, {Niemczura}, \& {Gameiro}}]{Papics2017}
{P{\'a}pics}, P.~I., {Tkachenko}, A., {Van Reeth}, T., {et~al.} 2017, \aap,
  598, A74

\bibitem[{{Pavlovski} {et~al.}(2022){Pavlovski}, {Hummel}, {Tkachenko},
  {Dervi{\c{s}}o{\u{g}}lu}, {Kayhan}, {Zavala}, {Hutter}, {Tycner},
  {{\c{S}}ahin}, {Audenaert}, {Baeyens}, {Bodensteiner}, {Bowman}, {Gebruers},
  {Jannsen}, \& {Mombarg}}]{Pavlovski2022}
{Pavlovski}, K., {Hummel}, C.~A., {Tkachenko}, A., {et~al.} 2022, \aap, 658,
  A92

\bibitem[{{Pedersen}(2022)}]{Pedersen2022}
{Pedersen}, M.~G. 2022, The Astrophysical Journal, 930, 94

\bibitem[{{Pedersen} {et~al.}(2021){Pedersen}, {Aerts}, {P{\'a}pics},
  {Michielsen}, {Gebruers}, {Rogers}, {Molenberghs}, {Burssens}, {Garcia}, \&
  {Bowman}}]{Pedersen2021}
{Pedersen}, M.~G., {Aerts}, C., {P{\'a}pics}, P.~I., {et~al.} 2021, Nature
  Astronomy, 5, 715

\bibitem[{{Pedersen} {et~al.}(2018){Pedersen}, {Aerts}, {P{\'a}pics}, \&
  {Rogers}}]{Pedersen2018}
{Pedersen}, M.~G., {Aerts}, C., {P{\'a}pics}, P.~I., \& {Rogers}, T.~M. 2018,
  \aap, 614, A128

\bibitem[{{Pedersen} {et~al.}(2019){Pedersen}, {Chowdhury}, {Johnston},
  {Bowman}, {Aerts}, {Handler}, {De Cat}, {Neiner}, {David-Uraz}, {Buzasi},
  {Tkachenko}, {Sim{\'o}n-D{\'\i}az}, {Moravveji}, {Sikora}, {Mirouh},
  {Lovekin}, {Cantiello}, {Daszy{\'n}ska-Daszkiewicz}, {Pigulski},
  {Vanderspek}, \& {Ricker}}]{Pedersen2019}
{Pedersen}, M.~G., {Chowdhury}, S., {Johnston}, C., {et~al.} 2019, \apjl, 872,
  L9

\bibitem[{{Rappaport} {et~al.}(2021){Rappaport}, {Kurtz}, {Handler}, {Jones},
  {Nelson}, {Saio}, {Fuller}, {Holdsworth}, {Vanderburg}, {{\v{Z}}{\'a}k},
  {Skarka}, {Aiken}, {Maxted}, {Stevens}, {Feliz}, \& {Kahraman
  Ali{\c{c}}avu{\c{s}}}}]{Rappaport2021}
{Rappaport}, S.~A., {Kurtz}, D.~W., {Handler}, G., {et~al.} 2021, \mnras, 503,
  254

\bibitem[{{Ratnasingam} {et~al.}(2020){Ratnasingam}, {Edelmann}, \&
  {Rogers}}]{Ratnasingam2020}
{Ratnasingam}, R.~P., {Edelmann}, P.~V.~F., \& {Rogers}, T.~M. 2020, \mnras,
  497, 4231

\bibitem[{Ricker {et~al.}(2015)Ricker, Winn, Vanderspek, Latham, Bakos, Bean,
  Berta-Thompson, Brown, Buchhave, Butler, Butler, Chaplin, Charbonneau,
  Christensen-Dalsgaard, Clampin, Deming, Doty, De~Lee, Dressing, Dunham,
  Fressin, Ge, Henning, Holman, Howard, Ida, Jenkins, Jernigan, Johnson,
  Kaltenegger, Kawai, Kjeldsen, Laughlin, Levine, Lin, Lissauer, MacQueen,
  Marcy, McCullough, Morton, Narita, Paegert, Palle, Pepe, Pepper, Quirrenbach,
  Rinehart, Sasselov, Sato, Seager, Sozzetti, Stassun, Sullivan, Szentgyorgyi,
  Torres, Udry, \& Villasenor}]{tess}
Ricker, G.~R., Winn, J.~N., Vanderspek, R., {et~al.} 2015, Journal of
  Astronomical Telescopes, Instruments, and Systems, 1, id. 014003

\bibitem[{{Rogers} {et~al.}(2013){Rogers}, {Lin}, {McElwaine}, \&
  {Lau}}]{Rogers2013}
{Rogers}, T.~M., {Lin}, D.~N.~C., {McElwaine}, J.~N., \& {Lau}, H.~H.~B. 2013,
  \apj, 772, 21

\bibitem[{{Saesen} {et~al.}(2013){Saesen}, {Briquet}, {Aerts}, {Miglio}, \&
  {Carrier}}]{Saesen2013}
{Saesen}, S., {Briquet}, M., {Aerts}, C., {Miglio}, A., \& {Carrier}, F. 2013,
  \aj, 146, 102

\bibitem[{{Saesen} {et~al.}(2010){Saesen}, {Carrier}, {Pigulski}, {Aerts},
  {Handler}, {Narwid}, {Fu}, {Zhang}, {Jiang}, {Vanautgaerden}, {Kopacki},
  {St{\k{e}}{\'s}licki}, {Acke}, {Poretti}, {Uytterhoeven}, {Gielen},
  {{\O}stensen}, {De Meester}, {Reed}, {Ko{\l}aczkowski}, {Michalska},
  {Schmidt}, {Yakut}, {Leitner}, {Kalomeni}, {Cherix}, {Spano}, {Prins}, {van
  Helshoecht}, {Zima}, {Huygen}, {Vandenbussche}, {Lenz}, {Ladjal}, {Puga
  Antol{\'\i}n}, {Verhoelst}, {De Ridder}, {Niarchos}, {Liakos}, {Lorenz},
  {Dehaes}, {Reyniers}, {Davignon}, {Kim}, {Kim}, {Lee}, {Lee}, {Kwon},
  {Broeders}, {van Winckel}, {Vanhollebeke}, {Waelkens}, {Raskin}, {Blom},
  {Eggen}, {Degroote}, {Beck}, {Puschnig}, {Schmitzberger}, {Gelven},
  {Steininger}, {Blommaert}, {Drummond}, {Briquet}, \&
  {Debosscher}}]{Saesen2010}
{Saesen}, S., {Carrier}, F., {Pigulski}, A., {et~al.} 2010, \aap, 515, A16

\bibitem[{{Salmon} {et~al.}(2014){Salmon}, {Montalb{\'a}n}, {Reese}, {Dupret},
  \& {Eggenberger}}]{Salmon2014}
{Salmon}, S.~J.~A.~J., {Montalb{\'a}n}, J., {Reese}, D.~R., {Dupret}, M.~A., \&
  {Eggenberger}, P. 2014, \aap, 569, A18

\bibitem[{{Serenelli} {et~al.}(2021){Serenelli}, {Weiss}, {Aerts}, {Angelou},
  {Baroch}, {Bastian}, {Beck}, {Bergemann}, {Bestenlehner}, {Czekala},
  {Elias-Rosa}, {Escorza}, {Van Eylen}, {Feuillet}, {Gandolfi}, {Gieles},
  {Girardi}, {Lebreton}, {Lodieu}, {Martig}, {Miller Bertolami}, {Mombarg},
  {Morales}, {Moya}, {Nsamba}, {Pavlovski}, {Pedersen}, {Ribas}, {Schneider},
  {Silva Aguirre}, {Stassun}, {Tolstoy}, {Tremblay}, \&
  {Zwintz}}]{Serenelli2021}
{Serenelli}, A., {Weiss}, A., {Aerts}, C., {et~al.} 2021, \aapr, 29, 4

\bibitem[{{Sharma} {et~al.}(2022){Sharma}, {Bedding}, {Saio}, \&
  {White}}]{Sharma2022}
{Sharma}, A.~N., {Bedding}, T.~R., {Saio}, H., \& {White}, T.~R. 2022, Monthly
  Notices of the Royal Astronomical Society, 515, 828

\bibitem[{{Sim{\'o}n-D{\'\i}az} {et~al.}(2018){Sim{\'o}n-D{\'\i}az}, {Aerts},
  {Urbaneja}, {Camacho}, {Antoci}, {Fredslund Andersen}, {Grundahl}, \&
  {Pall{\'e}}}]{SSD2018}
{Sim{\'o}n-D{\'\i}az}, S., {Aerts}, C., {Urbaneja}, M.~A., {et~al.} 2018, \aap,
  612, A40

\bibitem[{{Sim{\'o}n-D{\'\i}az} {et~al.}(2017){Sim{\'o}n-D{\'\i}az}, {Godart},
  {Castro}, {Herrero}, {Aerts}, {Puls}, {Telting}, \&
  {Grassitelli}}]{Simon-Diaz2017}
{Sim{\'o}n-D{\'\i}az}, S., {Godart}, M., {Castro}, N., {et~al.} 2017, \aap,
  597, A22

\bibitem[{{Sim{\'o}n-D{\'\i}az} \& {Herrero}(2014)}]{Simon-Diaz2014}
{Sim{\'o}n-D{\'\i}az}, S. \& {Herrero}, A. 2014, \aap, 562, A135

\bibitem[{{Southworth}(2021)}]{Southworth2021}
{Southworth}, J. 2021, Universe, 7, 369

\bibitem[{{Straumit} {et~al.}(2022){Straumit}, {Tkachenko}, {Gebruers},
  {Audenaert}, {Xiang}, {Zari}, {Aerts}, {Johnson}, {Kollmeier}, {Rix},
  {Beaton}, {Van Saders}, {Teske}, {Roman-Lopes}, {Ting}, \&
  {Román-Zúñiga}}]{Straumit2022}
{Straumit}, I., {Tkachenko}, A., {Gebruers}, S., {et~al.} 2022, The
  Astronomical Journal, 163, 236

\bibitem[{{Szewczuk} \& {Daszy{\'n}ska-Daszkiewicz}(2018)}]{Szewczuk2018}
{Szewczuk}, W. \& {Daszy{\'n}ska-Daszkiewicz}, J. 2018, \mnras, 478, 2243

\bibitem[{{Szewczuk} {et~al.}(2021){Szewczuk}, {Walczak}, \&
  {Daszy{\'n}ska-Daszkiewicz}}]{Szewczuk2021}
{Szewczuk}, W., {Walczak}, P., \& {Daszy{\'n}ska-Daszkiewicz}, J. 2021, \mnras,
  503, 5894

\bibitem[{{Szewczuk} {et~al.}(2022){Szewczuk}, {Walczak},
  {Daszy{\'n}ska-Daszkiewicz}, \& {Mo{\'z}dzierski}}]{Szewczuk2022}
{Szewczuk}, W., {Walczak}, P., {Daszy{\'n}ska-Daszkiewicz}, J., \&
  {Mo{\'z}dzierski}, D. 2022, \mnras, 511, 1529

\bibitem[{Takeda \& UeNo(2017)}]{Takeda2017}
Takeda, Y. \& UeNo, S. 2017, Publications of the Astronomical Society of Japan,
  69

\bibitem[{{Telting} {et~al.}(2006){Telting}, {Schrijvers}, {Ilyin},
  {Uytterhoeven}, {De Ridder}, {Aerts}, \& {Henrichs}}]{Telting2006}
{Telting}, J.~H., {Schrijvers}, C., {Ilyin}, I.~V., {et~al.} 2006, \aap, 452,
  945

\bibitem[{{Thompson} {et~al.}(2012){Thompson}, {Everett}, {Mullally},
  {Barclay}, {Howell}, {Still}, {Rowe}, {Christiansen}, {Kurtz}, {Hambleton},
  {Twicken}, {Ibrahim}, \& {Clarke}}]{Thompson2012}
{Thompson}, S.~E., {Everett}, M., {Mullally}, F., {et~al.} 2012, \apj, 753, 86

\bibitem[{{Thoul} {et~al.}(2013){Thoul}, {Degroote}, {Catala}, {Aerts},
  {Morel}, {Briquet}, {Hillen}, {Raskin}, {Van Winckel}, {Auvergne}, {Baglin},
  {Baudin}, \& {Michel}}]{Thoul2013}
{Thoul}, A., {Degroote}, P., {Catala}, C., {et~al.} 2013, \aap, 551, A12

\bibitem[{{Tkachenko}(2015)}]{gssp}
{Tkachenko}, A. 2015, A\&A, 581, A129

\bibitem[{{Tkachenko} {et~al.}(2020){Tkachenko}, {Pavlovski}, {Johnston},
  {Pedersen}, {Michielsen}, {Bowman}, {Southworth}, {Tsymbal}, \&
  {Aerts}}]{binTkachenko}
{Tkachenko}, A., {Pavlovski}, K., {Johnston}, C., {et~al.} 2020, A\&A, 637, A60

\bibitem[{{Tkachenko} {et~al.}(2013){Tkachenko}, {Van Reeth}, {Tsymbal},
  {Aerts}, {Kochukhov}, \& {Debosscher}}]{lsd}
{Tkachenko}, A., {Van Reeth}, T., {Tsymbal}, V., {et~al.} 2013, A\&A, 560, A37

\bibitem[{{Triana} {et~al.}(2015){Triana}, {Moravveji}, {P{\'a}pics}, {Aerts},
  {Kawaler}, \& {Christensen-Dalsgaard}}]{Triana2015}
{Triana}, S.~A., {Moravveji}, E., {P{\'a}pics}, P.~I., {et~al.} 2015, \apj,
  810, 16

\bibitem[{{Uytterhoeven} {et~al.}(2011){Uytterhoeven}, {Moya},
  {Grigahc{\`e}ne}, {Guzik}, {Guti{\'e}rrez-Soto}, {Smalley}, {Handler},
  {Balona}, {Niemczura}, {Fox Machado}, {Benatti}, {Chapellier}, {Tkachenko},
  {Szab{\'o}}, {Su{\'a}rez}, {Ripepi}, {Pascual}, {Mathias},
  {Mart{\'\i}n-Ru{\'\i}z}, {Lehmann}, {Jackiewicz}, {Hekker}, {Gruberbauer},
  {Garc{\'\i}a}, {Dumusque}, {D{\'\i}az-Fraile}, {Bradley}, {Antoci}, {Roth},
  {Leroy}, {Murphy}, {De Cat}, {Cuypers}, {Kjeldsen}, {Christensen-Dalsgaard},
  {Breger}, {Pigulski}, {Kiss}, {Still}, {Thompson}, \& {van
  Cleve}}]{Uytterhoeven2011}
{Uytterhoeven}, K., {Moya}, A., {Grigahc{\`e}ne}, A., {et~al.} 2011, \aap, 534,
  A125

\bibitem[{{Van Eylen} {et~al.}(2016){Van Eylen}, {Winn}, \&
  {Albrecht}}]{VanEylen2016}
{Van Eylen}, V., {Winn}, J.~N., \& {Albrecht}, S. 2016, \apj, 824, 15

\bibitem[{{Van Reeth} {et~al.}(2018){Van Reeth}, {Mombarg}, {Mathis},
  {Tkachenko}, {Fuller}, {Bowman}, {Buysschaert}, {Johnston}, {Garc{\'\i}a
  Hern{\'a}ndez}, {Goldstein}, {Townsend}, \& {Aerts}}]{VanReeth2018}
{Van Reeth}, T., {Mombarg}, J.~S.~G., {Mathis}, S., {et~al.} 2018, \aap, 618,
  A24

\bibitem[{{Van Reeth} {et~al.}(2022){Van Reeth}, {Southworth}, {Van Beeck}, \&
  {Bowman}}]{VanReeth2022}
{Van Reeth}, T., {Southworth}, J., {Van Beeck}, J., \& {Bowman}, D.~M. 2022,
  \aap, 659, A177

\bibitem[{{Van Reeth} {et~al.}(2016){Van Reeth}, {Tkachenko}, \&
  {Aerts}}]{VanReeth2016}
{Van Reeth}, T., {Tkachenko}, A., \& {Aerts}, C. 2016, \aap, 593, A120

\bibitem[{{Van Reeth} {et~al.}(2015{\natexlab{a}}){Van Reeth}, {Tkachenko},
  {Aerts}, {P{\'a}pics}, {Degroote}, {Debosscher}, {Zwintz}, {Bloemen}, {De
  Smedt}, {Hrudkova}, {Raskin}, \& {Van Winckel}}]{VanReeth2015a}
{Van Reeth}, T., {Tkachenko}, A., {Aerts}, C., {et~al.} 2015{\natexlab{a}},
  \aap, 574, A17

\bibitem[{{Van Reeth} {et~al.}(2015{\natexlab{b}}){Van Reeth}, {Tkachenko},
  {Aerts}, {P{\'a}pics}, {Triana}, {Zwintz}, {Degroote}, {Debosscher},
  {Bloemen}, {Schmid}, {De Smedt}, {Fremat}, {Fuentes}, {Homan}, {Hrudkova},
  {Karjalainen}, {Lombaert}, {Nemeth}, {{\O}stensen}, {Van De Steene}, {Vos},
  {Raskin}, \& {Van Winckel}}]{VanReeth2015b}
{Van Reeth}, T., {Tkachenko}, A., {Aerts}, C., {et~al.} 2015{\natexlab{b}},
  \apjs, 218, 27

\bibitem[{{Welsh} {et~al.}(2011){Welsh}, {Orosz}, {Aerts}, {Brown},
  {Brugamyer}, {Cochran}, {Gilliland}, {Guzik}, {Kurtz}, {Latham}, {Marcy},
  {Quinn}, {Zima}, {Allen}, {Batalha}, {Bryson}, {Buchhave}, {Caldwell},
  {Gautier}, {Howell}, {Kinemuchi}, {Ibrahim}, {Isaacson}, {Jenkins}, {Prsa},
  {Still}, {Street}, {Wohler}, {Koch}, \& {Borucki}}]{Welsh2011}
{Welsh}, W.~F., {Orosz}, J.~A., {Aerts}, C., {et~al.} 2011, \apjs, 197, 4

\bibitem[{{Wenger} {et~al.}(2000){Wenger}, {Ochsenbein}, {Egret}, {Dubois},
  {Bonnarel}, {Borde}, {Genova}, {Jasniewicz}, {Lalo{\"e}}, {Lesteven}, \&
  {Monier}}]{Wenger2000}
{Wenger}, M., {Ochsenbein}, F., {Egret}, D., {et~al.} 2000, \aaps, 143, 9

\bibitem[{{Zahn}(2013)}]{Zahn2013}
{Zahn}, J.-P. 2013, in Lecture Notes in Physics, Berlin Springer Verlag, ed.
  J.~{Souchay}, S.~{Mathis}, \& T.~{Tokieda}, Vol. 861, 301

\end{thebibliography}

\onecolumn

\appendix

\section{Results of the global stellar parameter estimations}

\small
\addtolength{\tabcolsep}{-1pt}
\begin{longtable}{lllllllll}
    \caption{Results of the spectrum analysis and respective spectroscopic classification of stars in the combined {\sc uves} and {\sc feros} sample. The ``?'' mark indicates ambiguous classifications. For example, the star is classified as ``SB1?'' when the detected RV difference between the two observed epochs exceeds mildly the 3$\sigma$ uncertainty interval and no large asymmetries are detected in the line profiles. Superscirpts ``a'', ``b'', ``c'' refer respectively to the {\sc gssp}, {\sc tlusty}, and {\sc cmfgen} models employed for the spectrum analysis of the sample stars. }
    \label{tab:results} \\
    \hline\hline
    \multirow{2}{*}{Object name} & \multicolumn{1}{c}{Spectral} & \multicolumn{1}{c}{\teff} & \multicolumn{1}{c}{\logg} & \multicolumn{1}{c}{RV$_1$} & \multicolumn{1}{c}{$\Delta$RV} & \multirow{2}{*}{Asymmetry$_1$} & \multirow{2}{*}{Asymmetry$_2$} & \multirow{2}{*}{Variability}\\
    & \multicolumn{1}{c}{class} & \multicolumn{1}{c}{(K)} & \multicolumn{1}{c}{(dex)} & \multicolumn{2}{c}{(km\,s$^{-1}$)} & \\
    \hline
    \endfirsthead
    \caption{continued.} \\
    \hline\hline
    \multirow{2}{*}{Object name} & \multicolumn{1}{c}{Spectral} & \multicolumn{1}{c}{\teff} & \multicolumn{1}{c}{\logg} & \multicolumn{1}{c}{RV$_1$} & \multicolumn{1}{c}{$\Delta$RV} & \multirow{2}{*}{Asymmetry$_1$} & \multirow{2}{*}{Asymmetry$_2$} & \multirow{2}{*}{Variability}\\
    & \multicolumn{1}{c}{class} & \multicolumn{1}{c}{(K)} & \multicolumn{1}{c}{(dex)} & \multicolumn{2}{c}{(km\,s$^{-1}$)} & \\
    \hline
    \endhead
    \hline
    \endfoot
    \multicolumn{9}{c}{{\sc uves}} \\
                 SK -71 51 \textsuperscript {b } & O6V & 47480  $\pm$ 3210 &  3.72  $\pm$ 0.20 & 206.6  $\pm$ 9.6 & 19.7  $\pm$ 18.0 &  0.81  $\pm$ 0.16 &  1.18  $\pm$ 0.24 &  asymm? \\
              HT 83 alf \textsuperscript {b } & O6V & 47440  $\pm$ 3490 &   3.70  $\pm$ 0.20 & 296.2  $\pm$ 8.1 &  3.6  $\pm$ 16.8 &  1.14  $\pm$ 0.18 &  0.97  $\pm$ 0.15 &     asymm? \\
                  BI 42 \textsuperscript {b } & O9V & 33760  $\pm$ 1940 & 3.59  $\pm$ 0.24 & 283.5  $\pm$ 3.5 &   3.7  $\pm$ 6.8 &  1.02  $\pm$ 0.08 &  0.82  $\pm$ 0.06 &      const \\
              HD 269525 \textsuperscript {b } & B0I & 32300  $\pm$ 1100 &  3.24  $\pm$ 0.20 & 309.7  $\pm$ 4.5 &   4.5  $\pm$ 9.4 & -1.01  $\pm$ 0.11 &  -1.0  $\pm$ 0.11 &      const \\
             SK -66 138 \textsuperscript {b } & O9.5III & 30410  $\pm$ 1190 & 3.32  $\pm$ 0.18 & 307.6  $\pm$ 2.5 &   1.5  $\pm$ 4.8 &  1.17  $\pm$ 0.09 &  1.15  $\pm$ 0.09 &      const \\
                 RMC 95 \textsuperscript {b } & B0I & 28890  $\pm$ 1430 & 3.16  $\pm$ 0.16 & 311.6  $\pm$ 1.9 &   0.9  $\pm$ 3.7 &  1.11  $\pm$ 0.06 & -1.17  $\pm$ 0.06 &      const \\
                 RMC 93 \textsuperscript {b } & B0II & 28650  $\pm$ 1620 & 3.13  $\pm$ 0.18 & 297.3  $\pm$ 3.5 &   9.6  $\pm$ 6.7 & -0.63  $\pm$ 0.04 &  1.04  $\pm$ 0.07 &      asymm \\
                 BI 188 \textsuperscript {b } & B0I & 27720  $\pm$ 1240 & 3.11  $\pm$ 0.19 & 250.9  $\pm$ 2.3 &   7.6  $\pm$ 4.3 &  1.37  $\pm$ 0.08 &   1.3  $\pm$ 0.08 &      const \\
              HD 269769 \textsuperscript {b } & B0I &  27540  $\pm$ 860 & 3.04  $\pm$ 0.12 & 282.7  $\pm$ 1.6 &   9.5  $\pm$ 4.0 &  1.22  $\pm$ 0.06 &  1.12  $\pm$ 0.06 &      asymm \\
              SK -70 13 \textsuperscript {b } & O9.5I & 26550  $\pm$ 1990 & 2.96  $\pm$ 0.15 & 258.0  $\pm$ 2.0 &   3.5  $\pm$ 4.1 &  1.03  $\pm$ 0.06 &  1.23  $\pm$ 0.08 &      const \\
                 RMC 97 \textsuperscript {b } & B0.5Ib & 25020  $\pm$ 2310 & 3.11  $\pm$ 0.23 & 325.8  $\pm$ 1.8 &   3.6  $\pm$ 3.6 &  1.36  $\pm$ 0.06 &  1.33  $\pm$ 0.07 & asymm?LPV? \\
              SK -66 15 \textsuperscript {b } & B0.5I & 24550  $\pm$ 2070 &  2.82  $\pm$ 0.20 & 284.0  $\pm$ 1.2 &   3.0  $\pm$ 2.6 &  0.96  $\pm$ 0.04 &  0.92  $\pm$ 0.04 &      const \\
             SK -69 180 \textsuperscript {b } & B1I & 24000  $\pm$ 2210 &   2.90  $\pm$ 0.20 & 272.0  $\pm$ 1.1 &   1.7  $\pm$ 2.4 &  0.94  $\pm$ 0.04 &  0.92  $\pm$ 0.04 &      const \\
              SK -67 54 \textsuperscript {b } & B0.5I & 23880  $\pm$ 2160 & 2.76  $\pm$ 0.22 & 306.6  $\pm$ 1.1 &   0.5  $\pm$ 2.3 &  1.07  $\pm$ 0.04 &  0.94  $\pm$ 0.04 &      const \\
             SK -69 275 \textsuperscript {b } & B1I & 23800  $\pm$ 3160 &  2.90  $\pm$ 0.24 & 252.1  $\pm$ 1.3 &   6.6  $\pm$ 2.7 &  1.26  $\pm$ 0.06 &  1.12  $\pm$ 0.05 &      const \\
\begin{tabular}{r} MCPS 082.95108- \\ 69.23933 \textsuperscript {b } \end{tabular}  & B1I & 23670  $\pm$ 2170 & 2.88  $\pm$ 0.21 & 290.4  $\pm$ 1.2 &   0.7  $\pm$ 2.4 &  1.13  $\pm$ 0.05 &  1.08  $\pm$ 0.05 &      const \\
              HD 269700 \textsuperscript {b } & B1.5 &  23010  $\pm$ 660 &  2.38  $\pm$ 0.20 & 276.8  $\pm$ 1.3 &  15.0  $\pm$ 2.4 &   1.93  $\pm$ 0.1 &   1.95  $\pm$ 0.1 &      asymm \\
             CPD-69 457 \textsuperscript {b } & B0.5I &  23000  $\pm$ 600 &  2.46  $\pm$ 0.20 & 293.1  $\pm$ 1.5 &   6.2  $\pm$ 3.1 &  1.14  $\pm$ 0.06 &  1.33  $\pm$ 0.08 &       SB1? \\
             SK -68 147 \textsuperscript {b } & B1I & 22900  $\pm$ 2420 & 2.84  $\pm$ 0.22 & 291.7  $\pm$ 0.9 &   2.4  $\pm$ 2.0 &  0.97  $\pm$ 0.04 &  0.88  $\pm$ 0.04 &      const \\
         Dachs LMC 2-19 \textsuperscript {b } & B1I & 22640  $\pm$ 2370 & 2.76  $\pm$ 0.21 & 306.0  $\pm$ 1.6 &   3.3  $\pm$ 2.9 &  1.14  $\pm$ 0.06 &  1.16  $\pm$ 0.06 &      const \\
              HD 269314 \textsuperscript {b } & B0I &  22560  $\pm$ 660 &  2.48  $\pm$ 0.20 & 301.6  $\pm$ 1.7 &   3.2  $\pm$ 3.6 &  1.18  $\pm$ 0.06 &  1.05  $\pm$ 0.05 &      const \\
             CPD-69 445 \textsuperscript {b } & B1I & 22480  $\pm$ 2100 & 2.62  $\pm$ 0.18 & 262.0  $\pm$ 1.4 &   1.1  $\pm$ 3.0 &  1.15  $\pm$ 0.05 &  1.05  $\pm$ 0.05 &      const \\
             SK -69 256 \textsuperscript {b } & B1I & 22380  $\pm$ 1960 & 2.66  $\pm$ 0.17 & 241.7  $\pm$ 1.6 &  10.9  $\pm$ 2.9 &   0.9  $\pm$ 0.05 &  1.25  $\pm$ 0.06 &      asymm \\
                RMC 142 \textsuperscript {b } & B0I &  22180  $\pm$ 650 &  2.43  $\pm$ 0.20 & 270.2  $\pm$ 1.3 &   8.2  $\pm$ 2.4 &  1.64  $\pm$ 0.09 &  1.48  $\pm$ 0.08 &      asymm \\
              HD 269859 \textsuperscript {b } & B0I &  22000  $\pm$ 770 &   2.10  $\pm$ 0.20 & 279.6  $\pm$ 0.9 &   1.2  $\pm$ 1.7 &  1.41  $\pm$ 0.06 &  1.57  $\pm$ 0.06 &      asymm \\
              HD 270754 \textsuperscript {b } & B1I &  22000  $\pm$ 700 &  2.48  $\pm$ 0.20 & 321.5  $\pm$ 0.8 &   2.3  $\pm$ 1.8 &   1.4  $\pm$ 0.05 &  1.51  $\pm$ 0.06 &      asymm \\
              HD 270196 \textsuperscript {b } & B1I &  22000  $\pm$ 650 &  2.39  $\pm$ 0.20 & 239.8  $\pm$ 1.0 &   1.3  $\pm$ 2.0 &  1.08  $\pm$ 0.04 &  1.01  $\pm$ 0.04 &      const \\
              SK -67 46 \textsuperscript {b } & B1.5I & 22000  $\pm$ 2300 &  2.62  $\pm$ 0.20 & 310.0  $\pm$ 1.1 &   2.3  $\pm$ 2.3 &  1.03  $\pm$ 0.04 &  1.17  $\pm$ 0.05 &      const \\
             SK -69 167 \textsuperscript {b } & B1I & 21890  $\pm$ 1960 & 2.68  $\pm$ 0.16 & 278.2  $\pm$ 1.4 &   4.9  $\pm$ 2.6 &   1.1  $\pm$ 0.06 &   1.3  $\pm$ 0.06 &      const \\
              HD 269786 \textsuperscript {b } & B1I &  21870  $\pm$ 730 &  2.46  $\pm$ 0.20 & 288.4  $\pm$ 1.0 &   4.3  $\pm$ 1.9 &  1.64  $\pm$ 0.07 &  1.25  $\pm$ 0.05 &      asymm \\
              SK -67 76 \textsuperscript {b } & B0I & 21860  $\pm$ 1040 &  2.11  $\pm$ 0.20 & 322.9  $\pm$ 1.8 &   1.1  $\pm$ 3.6 &  1.16  $\pm$ 0.07 &  1.38  $\pm$ 0.08 &      asymm \\
              HD 269655 \textsuperscript {b } & B0I &  21780  $\pm$ 780 &  2.33  $\pm$ 0.20 & 278.4  $\pm$ 1.0 &  12.2  $\pm$ 2.6 &  1.24  $\pm$ 0.05 &  1.51  $\pm$ 0.07 &      asymm \\
            SK -69- 157 \textsuperscript {b } & B1I & 21770  $\pm$ 1440 &  2.55  $\pm$ 0.20 & 257.8  $\pm$ 1.4 &   4.5  $\pm$ 2.7 &  1.13  $\pm$ 0.05 &   1.1  $\pm$ 0.05 &      const \\
              HD 269660 \textsuperscript {b } & B2I &  21690  $\pm$ 700 &  2.45  $\pm$ 0.20 & 247.4  $\pm$ 1.0 &   0.4  $\pm$ 1.8 &  0.99  $\pm$ 0.04 &  1.15  $\pm$ 0.04 &      const \\
              SK -69 68 \textsuperscript {b } & B1I &  21660  $\pm$ 760 &  2.18  $\pm$ 0.20 & 243.3  $\pm$ 1.2 &  14.5  $\pm$ 2.2 &  1.25  $\pm$ 0.04 &  1.11  $\pm$ 0.03 &      asymm \\
              HD 268809 \textsuperscript {b } & B1I &  21480  $\pm$ 730 &  2.13  $\pm$ 0.20 & 239.5  $\pm$ 1.1 &  15.1  $\pm$ 2.2 &  0.95  $\pm$ 0.04 &  1.21  $\pm$ 0.05 &     asymm? \\
             LH 47-373A \textsuperscript {b } & B1I &  21320  $\pm$ 760 &  2.48  $\pm$ 0.20 & 312.0  $\pm$ 1.2 &   1.6  $\pm$ 2.3 &  0.91  $\pm$ 0.04 &  1.13  $\pm$ 0.04 &      const \\
              HD 268623 \textsuperscript {b } & B1I &  20920  $\pm$ 870 &  2.48  $\pm$ 0.20 & 284.9  $\pm$ 1.0 &   6.9  $\pm$ 2.1 &  1.18  $\pm$ 0.05 &  1.12  $\pm$ 0.05 &       SB1? \\
              SK -68 17 \textsuperscript {b } & B4I & 20850  $\pm$ 1660 & 2.68  $\pm$ 0.18 & 272.4  $\pm$ 1.0 &   2.8  $\pm$ 1.9 &  1.05  $\pm$ 0.04 &   1.1  $\pm$ 0.04 &      const \\
               W61 27-5 \textsuperscript {b } & B1I &  20760  $\pm$ 860 &  2.12  $\pm$ 0.20 & 281.2  $\pm$ 1.0 &   5.1  $\pm$ 2.1 &  0.93  $\pm$ 0.04 &  1.13  $\pm$ 0.05 &      const \\
              HD 268798 \textsuperscript {b } & B2I & 20250  $\pm$ 1210 & 2.04  $\pm$ 0.18 & 278.4  $\pm$ 1.0 &   8.3  $\pm$ 1.8 &  0.96  $\pm$ 0.03 &  1.09  $\pm$ 0.04 & SB1?asymm? \\
             CPD-68 309 \textsuperscript {b } & B2.5I &  19290  $\pm$ 630 &  2.12  $\pm$ 0.20 & 255.4  $\pm$ 0.6 &   4.6  $\pm$ 1.2 &  1.24  $\pm$ 0.04 &  1.18  $\pm$ 0.04 &       SB1? \\
              SK -68 27 \textsuperscript {b } & B3I & 18630  $\pm$ 1880 &  2.50  $\pm$ 0.23 & 285.0  $\pm$ 1.0 &   1.9  $\pm$ 2.1 &  1.14  $\pm$ 0.04 &  0.99  $\pm$ 0.04 &      const \\
              HD 269845 \textsuperscript {b } & B2I & 18590  $\pm$ 1610 &  2.40  $\pm$ 0.16 & 300.0  $\pm$ 0.8 &   0.7  $\pm$ 1.7 &  1.25  $\pm$ 0.04 &  1.01  $\pm$ 0.04 &      const \\
               SK -68 2 \textsuperscript {b } & B9.5Iab & 18420  $\pm$ 1850 &  2.37  $\pm$ 0.20 & 285.6  $\pm$ 1.0 &   1.6  $\pm$ 2.2 &  1.14  $\pm$ 0.04 &  1.26  $\pm$ 0.06 &      const \\
              HD 268653 \textsuperscript {b } & B2I &  18090  $\pm$ 940 & 2.12  $\pm$ 0.15 & 306.1  $\pm$ 0.9 &   0.4  $\pm$ 1.7 &  1.41  $\pm$ 0.05 &  1.31  $\pm$ 0.04 &     asymm? \\
              HD 269992 \textsuperscript {b } & B2I & 18070  $\pm$ 1110 & 2.24  $\pm$ 0.19 & 249.2  $\pm$ 0.7 &   7.1  $\pm$ 1.5 &  1.13  $\pm$ 0.04 &  1.18  $\pm$ 0.04 &     asymm? \\
              HD 269997 \textsuperscript {b } & B2.5I & 17000  $\pm$ 1730 & 2.16  $\pm$ 0.21 & 262.7  $\pm$ 0.8 &   6.7  $\pm$ 1.6 &  1.21  $\pm$ 0.04 &  1.19  $\pm$ 0.04 &       SB1? \\
              HD 269101 \textsuperscript {b } & B3I & 17000  $\pm$ 1530 & 2.26  $\pm$ 0.16 & 303.3  $\pm$ 0.9 &   0.7  $\pm$ 1.8 &   1.2  $\pm$ 0.05 &  1.12  $\pm$ 0.04 &      const \\
             SK -67 133 \textsuperscript {c } & B2.5Iab &  16630  $\pm$ 850 & 2.28  $\pm$ 0.12 & 298.9  $\pm$ 1.0 &   2.2  $\pm$ 2.0 &  1.33  $\pm$ 0.06 &  1.27  $\pm$ 0.05 &      const \\
              HD 268729 \textsuperscript {b } & B5I &  16420  $\pm$ 780 &  1.95  $\pm$ 0.20 & 242.4  $\pm$ 0.8 &  29.3  $\pm$ 1.7 &  0.88  $\pm$ 0.04 &  1.13  $\pm$ 0.04 &        SB1 \\
              HD 271213 \textsuperscript {c } & B2I &  16250  $\pm$ 600 &  2.18  $\pm$ 0.20 & 245.4  $\pm$ 1.0 &   0.4  $\pm$ 2.0 &  1.19  $\pm$ 0.05 &  1.38  $\pm$ 0.06 &      const \\
              HD 270949 \textsuperscript {c } & B5I &  15900  $\pm$ 580 &  1.96  $\pm$ 0.20 & 298.4  $\pm$ 1.0 &   3.8  $\pm$ 1.9 &  1.11  $\pm$ 0.04 &  1.27  $\pm$ 0.05 & SB1?asymm? \\
              HD 269145 \textsuperscript {c } & B5I &  15860  $\pm$ 560 &  1.94  $\pm$ 0.20 & 308.2  $\pm$ 0.9 &  10.2  $\pm$ 1.7 &  -1.0  $\pm$ 0.04 &  1.11  $\pm$ 0.04 &       SB1? \\
              HD 269920 \textsuperscript {c } & B3I &  15560  $\pm$ 560 &  1.97  $\pm$ 0.20 & 280.3  $\pm$ 0.9 &   4.4  $\pm$ 1.8 &  1.13  $\pm$ 0.05 &  1.17  $\pm$ 0.05 &      const \\
              HD 269854 \textsuperscript {b } & B5I &  15340  $\pm$ 690 &  1.98  $\pm$ 0.20 & 294.2  $\pm$ 1.1 &   1.9  $\pm$ 2.3 &  1.17  $\pm$ 0.06 &  1.24  $\pm$ 0.07 &      const \\
              SK -70 53 \textsuperscript {b } & B5I & 15340  $\pm$ 1300 & 2.28  $\pm$ 0.17 & 252.4  $\pm$ 0.9 &   1.1  $\pm$ 1.8 &   1.1  $\pm$ 0.05 &  1.08  $\pm$ 0.05 &      const \\
             SK -67 275 \textsuperscript {c } & B6I &  14960  $\pm$ 890 & 2.23  $\pm$ 0.16 & 294.7  $\pm$ 1.2 &   4.1  $\pm$ 2.4 &  1.45  $\pm$ 0.07 &  1.55  $\pm$ 0.07 &      asymm \\
             SK -67 283 \textsuperscript {c } & B6I &  14880  $\pm$ 800 & 2.26  $\pm$ 0.14 & 295.5  $\pm$ 1.1 &   5.5  $\pm$ 2.3 &  1.27  $\pm$ 0.07 &  1.08  $\pm$ 0.06 &      const \\
               HD 47620 \textsuperscript {a } & B6V &  14850  $\pm$ 740 & 3.97  $\pm$ 0.18 &  38.2  $\pm$ 2.5 &   2.9  $\pm$ 4.7 &   1.3  $\pm$ 0.16 &  0.89  $\pm$ 0.12 &        LPV \\
             SK -66 125 \textsuperscript {c } & B4I &  14830  $\pm$ 760 & 2.34  $\pm$ 0.13 & 309.1  $\pm$ 0.9 &   2.5  $\pm$ 1.8 &  1.13  $\pm$ 0.06 &  1.05  $\pm$ 0.06 &      const \\
              HD 269644 \textsuperscript {c } & B6I &  14770  $\pm$ 500 &  1.94  $\pm$ 0.20 & 306.7  $\pm$ 0.9 &   0.2  $\pm$ 1.8 &  0.98  $\pm$ 0.05 &  0.94  $\pm$ 0.05 &      const \\
             SK -67 127 \textsuperscript {c } & B6I &  14750  $\pm$ 880 & 2.09  $\pm$ 0.14 & 314.4  $\pm$ 1.2 &   1.8  $\pm$ 2.3 &  0.95  $\pm$ 0.06 &  1.07  $\pm$ 0.07 &      const \\
              HD 268907 \textsuperscript {c } & B5I &  14500  $\pm$ 840 & 1.88  $\pm$ 0.12 & 298.1  $\pm$ 1.0 &   6.0  $\pm$ 1.8 &  0.83  $\pm$ 0.04 &  0.98  $\pm$ 0.05 &      const \\
             SK -66 126 \textsuperscript {c } & B5I &  14440  $\pm$ 560 &  1.98  $\pm$ 0.20 & 301.0  $\pm$ 1.1 &   0.1  $\pm$ 2.3 &  1.08  $\pm$ 0.07 &   1.0  $\pm$ 0.06 &      const \\
              HD 268654 \textsuperscript {c } & B9I &  14000  $\pm$ 620 &  1.74  $\pm$ 0.20 & 275.4  $\pm$ 1.3 &  21.2  $\pm$ 2.2 &  2.05  $\pm$ 0.14 &  0.99  $\pm$ 0.06 &      asymm \\
              SK -68 53 \textsuperscript {c } & B5I &  13770  $\pm$ 740 & 2.09  $\pm$ 0.12 & 279.3  $\pm$ 0.9 &   0.8  $\pm$ 1.9 &  0.98  $\pm$ 0.05 &  0.86  $\pm$ 0.05 &      const \\
             SK -66 142 \textsuperscript {c } & B6I &  13640  $\pm$ 740 &  2.00  $\pm$ 0.12 & 304.5  $\pm$ 1.1 &   2.8  $\pm$ 2.1 &  0.99  $\pm$ 0.06 &  0.94  $\pm$ 0.06 &      const \\
              SK -66 92 \textsuperscript {c } & B7I &  13480  $\pm$ 750 & 1.92  $\pm$ 0.11 & 310.7  $\pm$ 1.1 &   6.3  $\pm$ 2.2 &   0.9  $\pm$ 0.06 &  0.92  $\pm$ 0.06 &       SB1? \\
              HD 269801 \textsuperscript {c } & B9I &  13260  $\pm$ 670 & 1.73  $\pm$ 0.14 & 316.7  $\pm$ 0.8 &   0.8  $\pm$ 1.5 &  0.91  $\pm$ 0.04 &  0.98  $\pm$ 0.05 &      const \\
             SK -67 198 \textsuperscript {c } & B7I &  13180  $\pm$ 790 & 1.85  $\pm$ 0.12 & 305.9  $\pm$ 1.1 &   0.7  $\pm$ 2.2 &  0.82  $\pm$ 0.05 &  0.91  $\pm$ 0.06 &      const \\
              HD 270933 \textsuperscript {c } & B8I &  13150  $\pm$ 620 & 1.74  $\pm$ 0.13 & 296.5  $\pm$ 0.6 &   2.0  $\pm$ 1.2 &  0.99  $\pm$ 0.04 &  1.02  $\pm$ 0.04 &      const \\
              HD 269766 \textsuperscript {c } & B8I &  12960  $\pm$ 590 & 1.76  $\pm$ 0.12 & 304.4  $\pm$ 1.0 &   2.7  $\pm$ 2.0 &  0.95  $\pm$ 0.06 &   0.8  $\pm$ 0.05 &      const \\
               HD 41297 \textsuperscript {a } & B8IV &  12950  $\pm$ 730 & 3.88  $\pm$ 0.21 &  17.7  $\pm$ 1.3 &   0.3  $\pm$ 2.6 &    1.2  $\pm$ 0.1 &  1.28  $\pm$ 0.11 &      const \\
              SK -70 81 \textsuperscript {c } & B7Iab & 12770  $\pm$ 1080 & 2.18  $\pm$ 0.21 & 259.9  $\pm$ 2.4 &   6.7  $\pm$ 4.5 &  1.55  $\pm$ 0.09 &  2.09  $\pm$ 0.12 &        LPV \\
             SK -67 171 \textsuperscript {c } & B7I &  12650  $\pm$ 740 &  1.80  $\pm$ 0.12 & 309.8  $\pm$ 0.8 &   1.9  $\pm$ 1.7 &  0.99  $\pm$ 0.06 &  0.88  $\pm$ 0.05 &      const \\
              HD 269619 \textsuperscript {c } & B9I &  12630  $\pm$ 800 &  1.70  $\pm$ 0.14 & 271.1  $\pm$ 0.6 &   2.6  $\pm$ 1.2 &  1.06  $\pm$ 0.04 &  0.93  $\pm$ 0.04 &       LPV? \\
                 GV 375 \textsuperscript {c } & B8I &  12610  $\pm$ 700 & 2.07  $\pm$ 0.14 & 300.7  $\pm$ 0.6 &   0.7  $\pm$ 1.3 &  0.84  $\pm$ 0.05 &  0.79  $\pm$ 0.05 &      const \\
             CPD-69 490 \textsuperscript {c } & B6I &  12600  $\pm$ 760 &  1.70  $\pm$ 0.12 & 255.1  $\pm$ 0.5 &   3.1  $\pm$ 1.0 &  1.01  $\pm$ 0.04 &  0.89  $\pm$ 0.03 &       SB1? \\
              SK -67 72 \textsuperscript {c } & B6I &  12480  $\pm$ 750 & 1.84  $\pm$ 0.14 & 300.9  $\pm$ 0.7 &   1.2  $\pm$ 1.4 &  0.88  $\pm$ 0.04 &  0.88  $\pm$ 0.04 &      const \\
              HD 268675 \textsuperscript {c } & B8I &  12460  $\pm$ 810 &  1.70  $\pm$ 0.15 & 287.4  $\pm$ 0.8 &   2.8  $\pm$ 1.5 & -0.84  $\pm$ 0.04 &   1.1  $\pm$ 0.05 &     asymm? \\
               SK -67 7 \textsuperscript {c } & B9Iab &  12460  $\pm$ 760 & 1.83  $\pm$ 0.14 & 298.7  $\pm$ 0.7 &   2.6  $\pm$ 1.4 &  0.88  $\pm$ 0.04 &  0.88  $\pm$ 0.04 &      const \\
             CPD-69 512 \textsuperscript {c } & B9I &  12300  $\pm$ 780 & 1.72  $\pm$ 0.13 & 241.5  $\pm$ 0.9 &   5.3  $\pm$ 1.8 &  0.81  $\pm$ 0.05 &   0.8  $\pm$ 0.05 &       SB1? \\
              HD 270123 \textsuperscript {c } & B8I &  12150  $\pm$ 770 &  1.70  $\pm$ 0.14 & 292.6  $\pm$ 0.4 &   1.0  $\pm$ 0.9 &  1.01  $\pm$ 0.04 &   1.0  $\pm$ 0.03 &      const \\
              SK -70 31 \textsuperscript {c } & B8Iab &  12060  $\pm$ 790 &  1.90  $\pm$ 0.15 & 250.9  $\pm$ 0.7 &   1.7  $\pm$ 1.3 &  0.82  $\pm$ 0.05 &  0.86  $\pm$ 0.05 &      const \\
               HD 57970 \textsuperscript {a } & B8III &  11970  $\pm$ 720 & 3.73  $\pm$ 0.21 &  87.9  $\pm$ 1.5 &   4.7  $\pm$ 3.0 &  1.31  $\pm$ 0.07 &  1.08  $\pm$ 0.06 &      const \\
             SK -69 250 \textsuperscript {c } & B7I &  11960  $\pm$ 730 &  1.70  $\pm$ 0.15 & 245.3  $\pm$ 0.6 &   8.0  $\pm$ 1.2 &  1.03  $\pm$ 0.05 &  1.12  $\pm$ 0.05 &        SB1 \\
              SK -69 31 \textsuperscript {c } & B5I &  11830  $\pm$ 600 &  1.70  $\pm$ 0.12 & 250.1  $\pm$ 0.5 &   3.6  $\pm$ 1.0 &  1.01  $\pm$ 0.04 &   0.9  $\pm$ 0.04 &      const \\
              SK -70 26 \textsuperscript {c } & B8I &  11630  $\pm$ 570 &  1.90  $\pm$ 0.13 & 235.6  $\pm$ 0.5 &   0.4  $\pm$ 1.0 &  0.89  $\pm$ 0.04 &  0.94  $\pm$ 0.05 &      const \\
              SK -67 42 \textsuperscript {c } & B8I &  11560  $\pm$ 580 & 1.88  $\pm$ 0.13 & 307.7  $\pm$ 0.4 &   1.0  $\pm$ 0.7 &  -0.9  $\pm$ 0.03 &  0.95  $\pm$ 0.04 &      const \\
              SK -69 47 \textsuperscript {c } & B8Ib &  11540  $\pm$ 600 & 1.78  $\pm$ 0.14 & 251.1  $\pm$ 0.6 &   0.3  $\pm$ 1.1 &  0.87  $\pm$ 0.04 &  0.87  $\pm$ 0.04 &      const \\
             SK -67 279 \textsuperscript {c } & B7I &  11330  $\pm$ 600 &  1.90  $\pm$ 0.15 & 280.4  $\pm$ 0.4 &   2.6  $\pm$ 0.8 &  0.91  $\pm$ 0.04 &  0.94  $\pm$ 0.04 &      const \\
              SK -67 88 \textsuperscript {c } & B8I &  11220  $\pm$ 600 & 1.78  $\pm$ 0.13 & 289.8  $\pm$ 0.3 &   1.0  $\pm$ 0.7 &  0.91  $\pm$ 0.03 &   0.8  $\pm$ 0.02 &      const \\
              HD 269721 \textsuperscript {c } & B9I &  11080  $\pm$ 600 &  1.70  $\pm$ 0.13 & 300.6  $\pm$ 0.3 &   2.1  $\pm$ 0.5 &  0.98  $\pm$ 0.02 &  0.95  $\pm$ 0.02 &      const \\
               HD 45527 \textsuperscript {a } & B9IV &  11010  $\pm$ 600 & 3.88  $\pm$ 0.22 &   1.8  $\pm$ 1.0 &   2.1  $\pm$ 1.9 &  0.91  $\pm$ 0.05 &   0.9  $\pm$ 0.05 &       LPV? \\
              HD 269639 \textsuperscript {c } & B6I &  11000  $\pm$ 670 &  1.70  $\pm$ 0.15 & 304.2  $\pm$ 0.4 &   5.2  $\pm$ 0.7 & -0.93  $\pm$ 0.03 &  0.85  $\pm$ 0.02 &       SB1? \\
                SOI 404 \textsuperscript {c } & B9I &  11000  $\pm$ 660 & 2.14  $\pm$ 0.17 & 283.2  $\pm$ 2.4 &   5.5  $\pm$ 0.9 &  1.19  $\pm$ 0.04 &  0.86  $\pm$ 0.03 &        LPV \\
              HD 269510 \textsuperscript {c } & B8I &  11000  $\pm$ 620 &  1.70  $\pm$ 0.14 & 257.4  $\pm$ 0.2 &   4.0  $\pm$ 0.4 &  0.94  $\pm$ 0.02 &  0.92  $\pm$ 0.02 &       SB1? \\
             SK -69 174 \textsuperscript {c } & B9I &  11000  $\pm$ 590 & 1.93  $\pm$ 0.15 & 282.8  $\pm$ 0.2 &   1.2  $\pm$ 0.4 &  0.83  $\pm$ 0.02 &  0.98  $\pm$ 0.02 &      const \\
                SOI 247 \textsuperscript {c } & B9.5III &  11000  $\pm$ 540 &  2.50  $\pm$ 0.15 & 305.2  $\pm$ 0.1 &   1.3  $\pm$ 0.2 &  0.94  $\pm$ 0.03 &  0.77  $\pm$ 0.02 &      const \\
              SK -70 77 \textsuperscript {c } & B9Ib &  11000  $\pm$ 530 & 2.02  $\pm$ 0.15 & 235.7  $\pm$ 0.2 &   1.2  $\pm$ 0.4 &  0.94  $\pm$ 0.02 &  1.05  $\pm$ 0.02 &      const \\
             SK -68 152 \textsuperscript {c } & B8I &  11000  $\pm$ 530 & 2.01  $\pm$ 0.13 & 271.0  $\pm$ 0.2 &   1.0  $\pm$ 0.4 &  0.78  $\pm$ 0.02 &  0.82  $\pm$ 0.02 &      const \\
             SK -67 151 \textsuperscript {c } & B9Ib &  11000  $\pm$ 510 & 2.24  $\pm$ 0.15 & 303.3  $\pm$ 0.2 &   0.2  $\pm$ 0.4 &  0.82  $\pm$ 0.02 & -0.94  $\pm$ 0.02 &      const \\
                SOI 574 \textsuperscript {c } & B9II &  11000  $\pm$ 500 &  2.20  $\pm$ 0.13 & 245.1  $\pm$ 0.2 &   2.1  $\pm$ 0.3 &  1.04  $\pm$ 0.02 &  0.87  $\pm$ 0.02 &        SB1 \\
                                   SOI 166  & B9Ib &            - &           - &   9.6  $\pm$ 3.4 &   1.9  $\pm$ 6.7 &  0.89  $\pm$ 0.08 &  0.92  $\pm$ 0.08 &      const \\
                                   
\multicolumn{9}{c}{{\sc feros}} \\           

HD 269676 \textsuperscript{b} & O7 & 47480 $\pm$ 3790 &  3.56 $\pm$ 0.20 & 267.6 $\pm$ 10.7 &   3.7 $\pm$ 20.9 &  1.26 $\pm$ 0.43 &  1.53 $\pm$ 0.5 &      const \\
HD 269321 \textsuperscript{b} & B5I &  20000 $\pm$ 800 &  2.13 $\pm$ 0.20 &  281.5 $\pm$ 0.9 &   11.0 $\pm$ 2.0 &  1.14 $\pm$ 0.07 &  1.0 $\pm$ 0.06 & SB1?asymm? \\
HD 269440 \textsuperscript{b} & B1I & 20000 $\pm$ 1780 & 2.33 $\pm$ 0.23 &  314.6 $\pm$ 1.3 &    1.2 $\pm$ 2.4 &  1.03 $\pm$ 0.05 & 1.45 $\pm$ 0.06 &     asymm? \\
HD 269660 \textsuperscript{b} & B2I &  19460 $\pm$ 700 &   2.00 $\pm$ 0.20 &  246.1 $\pm$ 0.9 &    4.5 $\pm$ 1.8 &  1.12 $\pm$ 0.05 &  1.2 $\pm$ 0.05 &      const \\
HD 269700 \textsuperscript{c} & B1.5 &  18000 $\pm$ 780 &  1.92 $\pm$ 0.20 &  281.7 $\pm$ 2.4 &   19.7 $\pm$ 4.8 &  6.06 $\pm$ 0.87 & 4.68 $\pm$ 0.63 &      asymm \\
HD 269832 \textsuperscript{b} & B6I &  17580 $\pm$ 730 &  1.92 $\pm$ 0.20 &  261.5 $\pm$ 0.9 &    3.7 $\pm$ 1.6 &  1.14 $\pm$ 0.06 & 1.16 $\pm$ 0.05 &      const \\
HD 268798 \textsuperscript{b} & B2I & 17470 $\pm$ 1660 & 2.08 $\pm$ 0.21 &  286.7 $\pm$ 1.3 &    2.4 $\pm$ 2.7 &  1.68 $\pm$ 0.09 & 1.38 $\pm$ 0.07 & SB1?asymm? \\
HD 268726 \textsuperscript{b} & B3I & 17290 $\pm$ 1490 & 2.05 $\pm$ 0.21 &  310.7 $\pm$ 1.4 &    4.5 $\pm$ 2.9 &  1.47 $\pm$ 0.08 &  1.57 $\pm$ 0.1 &      const \\
HD 269992 \textsuperscript{b} & B2I & 16980 $\pm$ 1980 & 2.07 $\pm$ 0.23 &  252.2 $\pm$ 1.2 &    5.1 $\pm$ 2.6 &  1.56 $\pm$ 0.08 & 1.44 $\pm$ 0.08 &     asymm? \\
HD 268654 \textsuperscript{b} & B9I &  16580 $\pm$ 880 &  1.94 $\pm$ 0.20 &  259.6 $\pm$ 1.2 &    7.4 $\pm$ 2.1 &  0.78 $\pm$ 0.05 & 0.96 $\pm$ 0.05 &     asymm? \\
HD 269997 \textsuperscript{b} & B2.5Iab &  16500 $\pm$ 910 &  1.98 $\pm$ 0.20 &  262.2 $\pm$ 1.4 &    7.3 $\pm$ 2.8 &  0.91 $\pm$ 0.06 & 0.95 $\pm$ 0.05 &       SB1? \\
HD 268907 \textsuperscript{b} & B5I &  16430 $\pm$ 950 &  1.96 $\pm$ 0.20 &  304.5 $\pm$ 0.9 &    4.0 $\pm$ 1.8 &  0.95 $\pm$ 0.05 & 0.97 $\pm$ 0.05 &      const \\
HD 268729 \textsuperscript{b} & B5I &  16430 $\pm$ 930 &  1.98 $\pm$ 0.20 &  248.1 $\pm$ 1.1 &    9.5 $\pm$ 2.2 &   0.8 $\pm$ 0.05 & 0.73 $\pm$ 0.04 &        SB1 \\
HD 269786 \textsuperscript{c} & B1I &  16000 $\pm$ 780 &  1.95 $\pm$ 0.20 &  292.1 $\pm$ 1.4 &    5.9 $\pm$ 3.8 &   1.9 $\pm$ 0.12 &  1.78 $\pm$ 0.2 &      asymm \\
HD 268653 \textsuperscript{c} & B2I & 16000 $\pm$ 1000 &  1.96 $\pm$ 0.20 &  308.1 $\pm$ 1.8 &    9.1 $\pm$ 3.4 &  1.08 $\pm$ 0.08 & 1.27 $\pm$ 0.08 & SB1?asymm? \\
HD 269859 \textsuperscript{c} & B0I & 16000 $\pm$ 1000 &  1.94 $\pm$ 0.20 &  257.9 $\pm$ 1.7 &   23.9 $\pm$ 3.6 &  2.31 $\pm$ 0.18 & 2.75 $\pm$ 0.22 &      asymm \\
HD 269145 \textsuperscript{c} & B5I &  15960 $\pm$ 770 &  1.98 $\pm$ 0.20 &  318.0 $\pm$ 1.2 &    0.1 $\pm$ 2.4 &  1.17 $\pm$ 0.07 & 1.16 $\pm$ 0.07 &      const \\
HD 271163 \textsuperscript{c} & B3I &  15760 $\pm$ 770 &   2.00 $\pm$ 0.20 &  304.6 $\pm$ 2.5 &    0.1 $\pm$ 4.0 &  1.13 $\pm$ 0.13 & 0.99 $\pm$ 0.07 &      const \\
HD 270949 \textsuperscript{b} & B5I & 15730 $\pm$ 1000 &  1.96 $\pm$ 0.20 &  295.7 $\pm$ 1.3 &   11.1 $\pm$ 2.7 &  0.99 $\pm$ 0.06 & 0.99 $\pm$ 0.07 & SB1?asymm? \\
HD 269854 \textsuperscript{c} & B5I &  15300 $\pm$ 790 &  1.98 $\pm$ 0.20 &  291.2 $\pm$ 1.3 &    0.6 $\pm$ 2.6 &   1.0 $\pm$ 0.06 & 0.91 $\pm$ 0.06 &      const \\
HD 269606 \textsuperscript{c} & B5I &  15290 $\pm$ 760 &  1.99 $\pm$ 0.20 &  292.7 $\pm$ 1.2 &   10.9 $\pm$ 2.4 &  0.83 $\pm$ 0.05 & 0.88 $\pm$ 0.05 & SB1?asymm? \\
HD 269644 \textsuperscript{c} & B6I &  15170 $\pm$ 770 &  1.96 $\pm$ 0.20 &  312.9 $\pm$ 1.0 &    5.8 $\pm$ 2.0 &  1.06 $\pm$ 0.06 & 0.94 $\pm$ 0.05 &      const \\
HD 269593 \textsuperscript{c} & B7I &  15020 $\pm$ 770 &  1.92 $\pm$ 0.20 &  316.4 $\pm$ 1.2 &    5.6 $\pm$ 2.4 &  0.92 $\pm$ 0.05 & 0.88 $\pm$ 0.05 &      const \\
HD 269238 \textsuperscript{c} & B6I &  14600 $\pm$ 750 &  1.86 $\pm$ 0.20 &  248.8 $\pm$ 0.7 &    6.5 $\pm$ 1.5 &  0.98 $\pm$ 0.05 & 0.96 $\pm$ 0.05 &       SB1? \\
CD-65 288 \textsuperscript{c} & B6I &  14510 $\pm$ 790 &   1.90 $\pm$ 0.20 &  288.3 $\pm$ 1.1 &    4.0 $\pm$ 2.3 &  0.91 $\pm$ 0.05 & 0.86 $\pm$ 0.06 &      const \\
HD 269766 \textsuperscript{c} & B8I &  13940 $\pm$ 750 &  1.86 $\pm$ 0.20 &  303.1 $\pm$ 1.0 &    5.2 $\pm$ 2.1 &   0.8 $\pm$ 0.04 & 1.01 $\pm$ 0.07 &     asymm? \\
HD 269801 \textsuperscript{c} & B9I &  13830 $\pm$ 720 &  1.76 $\pm$ 0.20 &  317.6 $\pm$ 0.8 &    1.4 $\pm$ 1.5 & -1.13 $\pm$ 0.07 & -1.0 $\pm$ 0.05 &      const \\
HD 269619 \textsuperscript{c} & B9I &  13660 $\pm$ 720 &  1.78 $\pm$ 0.20 &  269.1 $\pm$ 0.9 &    1.9 $\pm$ 1.8 &  0.86 $\pm$ 0.05 & 0.79 $\pm$ 0.04 &       LPV? \\
HD  53048 \textsuperscript{c} & B5III & 13590 $\pm$ 2350 & 2.48 $\pm$ 0.36 &  277.2 $\pm$ 8.3 &  10.9 $\pm$ 16.8 &   1.11 $\pm$ 0.2 &   1.1 $\pm$ 0.2 &      const \\
HD 268675 \textsuperscript{c} & B8I &  13430 $\pm$ 700 &  1.78 $\pm$ 0.20 &  285.1 $\pm$ 0.7 &   10.2 $\pm$ 1.5 &  0.81 $\pm$ 0.04 & 1.06 $\pm$ 0.06 &  SB1asymm? \\
HD 270933 \textsuperscript{c} & B8I &  13300 $\pm$ 700 &  1.79 $\pm$ 0.20 &  291.7 $\pm$ 0.9 &    1.6 $\pm$ 1.7 &  0.86 $\pm$ 0.05 & 1.01 $\pm$ 0.05 &      const \\
HD  61950 \textsuperscript{c} & B8III & 12830 $\pm$ 2470 &  2.50 $\pm$ 0.45 &  314.8 $\pm$ 4.7 & 189.6 $\pm$ 10.5 &   1.4 $\pm$ 0.31 & 1.66 $\pm$ 0.46 &     asymm? \\
HD 270296 \textsuperscript{c} & B6I & 12570 $\pm$ 1280 &   1.70 $\pm$ 0.20 &  287.1 $\pm$ 1.0 &    2.4 $\pm$ 2.3 &  0.94 $\pm$ 0.06 & 0.98 $\pm$ 0.08 &      const \\
HD 269684 \textsuperscript{c} & B9I & 12480 $\pm$ 1240 &   1.70 $\pm$ 0.20 &  327.0 $\pm$ 0.8 &    3.4 $\pm$ 1.7 & -0.98 $\pm$ 0.06 & -1.0 $\pm$ 0.06 &      const \\
HD 269195 \textsuperscript{c} & B9I & 12300 $\pm$ 1250 &  1.70 $\pm$ 0.22 &  293.8 $\pm$ 1.0 &    5.6 $\pm$ 1.8 &  0.95 $\pm$ 0.07 & 0.95 $\pm$ 0.06 &       SB1? \\
HD 270123 \textsuperscript{c} & B8I & 12240 $\pm$ 1280 &  1.70 $\pm$ 0.22 &  289.0 $\pm$ 0.8 &    1.2 $\pm$ 1.6 &  0.79 $\pm$ 0.05 & 0.77 $\pm$ 0.05 &      const \\
HD 269639 \textsuperscript{c} & B7I &  11600 $\pm$ 960 &   1.70 $\pm$ 0.20 &  305.5 $\pm$ 0.6 &    5.6 $\pm$ 1.2 &  1.18 $\pm$ 0.06 & 0.93 $\pm$ 0.05 &       SB1? \\
HD 269510 \textsuperscript{c} & B8I &  11580 $\pm$ 930 &  1.70 $\pm$ 0.19 &  264.4 $\pm$ 0.5 &    7.5 $\pm$ 1.0 &  1.12 $\pm$ 0.04 & 0.89 $\pm$ 0.04 &       SB1? \\
HD 269721 \textsuperscript{c} & B9I &  11330 $\pm$ 970 &   1.70 $\pm$ 0.20 &  300.0 $\pm$ 0.5 &    3.5 $\pm$ 1.1 &   0.9 $\pm$ 0.04 &  0.9 $\pm$ 0.05 &      const \\
HD 269664 \textsuperscript{c} & B9I &  11120 $\pm$ 930 &  1.70 $\pm$ 0.19 &  248.3 $\pm$ 0.4 &    3.7 $\pm$ 0.8 &  0.95 $\pm$ 0.04 & 0.98 $\pm$ 0.04 & SB1?asymm? \\
                   HD 268835 & -- &            - &           - &            - &            - &            - &           - &      const \\
                   HD 269582 & -- &            - &           - &            - &            - &            - &           - &      const \\
                   HD 268718 & -- &            - &           - &            - &            - &            - &           - &      const \\
\end{longtable}

\clearpage
\section{Results of spectral line profiles characterisation}

\begin{longtable}{llllllll}
    \caption{Characteristics of the spectral line profile broadening in the combined {\sc uves} and {\sc feros} sample. }
    \label{tab:results_lp} \\
    \hline\hline
    \multirow{2}{*}{Object name}  & \multicolumn{1}{c}{\vsini$_{\rm SP}$} & \multicolumn{1}{c}{\vsini$_{\rm FT}$} & \multicolumn{1}{c}{$\Theta$ ($a_T = 0.5$)} & \multicolumn{1}{c}{$\Theta$ ($a_T$  varied)} & \multirow{2}{*}{$a_{\rm T}$}\\
    &  \multicolumn{4}{c}{(km\,s$^{-1}$)} & \\
    \hline
    \endfirsthead
    \caption{continued.} \\
    \hline\hline
    \multirow{2}{*}{Object name}  & \multicolumn{1}{c}{\vsini$_{\rm SP}$} & \multicolumn{1}{c}{\vsini$_{\rm FT}$} & \multicolumn{1}{c}{$\Theta$ ($a_T = 0.5$)} & \multicolumn{1}{c}{$\Theta$ ($a_T$  varied)} & \multirow{2}{*}{$a_{\rm T}$}\\
    &  \multicolumn{4}{c}{(km\,s$^{-1}$)} & \\
    \hline
    \endhead
    \hline
    \endfoot
    \multicolumn{6}{c}{{\sc uves}} \\
SK -71 51 & 170.5$\pm$ 23.4 &  131.3$\pm$ 18.2 &   183.8$\pm$ 189.0 &              - &           - \\
HT 83 alf & 269.0$\pm$ 24.3 &   202.6$\pm$ 7.8 &    28.0$\pm$ 117.5 &   38.2$\pm$ 370.6 & 0.02$\pm$ 0.11 \\
BI 42 & 141.0$\pm$ 18.3 &  124.3$\pm$ 24.2 &    67.8$\pm$ 232.7 &   53.3$\pm$ 136.2 & 0.54$\pm$ 0.07 \\
HD 269525 & 174.0$\pm$ 19.8 &   102.6$\pm$ 8.4 &    145.2$\pm$ 70.5 &   140.1$\pm$ 38.3 & 0.11$\pm$ 0.09 \\
SK -66 138 &  72.5$\pm$ 11.7 &    66.3$\pm$ 9.4 &     84.4$\pm$ 87.3 &   87.6$\pm$ 802.0 & 0.47$\pm$ 0.44 \\
RMC 95 &  83.5$\pm$ 12.2 &    71.7$\pm$ 7.2 &    107.6$\pm$ 12.1 &  106.6$\pm$ 134.9 & 0.16$\pm$ 0.14 \\
RMC 93 & 139.5$\pm$ 16.2 &  144.1$\pm$ 20.1 &     50.5$\pm$ 63.6 &    57.6$\pm$ 56.1 & 0.43$\pm$ 0.08 \\
BI 188 &  99.0$\pm$ 15.2 &   79.6$\pm$ 10.0 &    85.3$\pm$ 156.0 &    87.0$\pm$ 51.6 & 0.45$\pm$ 0.07 \\
HD 269769 & 107.0$\pm$ 13.8 &   80.2$\pm$ 17.5 &     73.7$\pm$ 56.1 &    69.5$\pm$ 18.4 & 0.53$\pm$ 0.03 \\
SK -70 13 &  93.5$\pm$ 14.0 &    60.3$\pm$ 4.5 &     98.3$\pm$ 25.8 &   101.3$\pm$ 45.6 & 0.63$\pm$ 0.14 \\
RMC 97 &  84.0$\pm$ 12.3 &   63.8$\pm$ 12.2 &    124.5$\pm$ 20.9 & 127.0$\pm$ - & 0.95$\pm$ 1.24 \\
SK -66 15 &  62.5$\pm$ 10.0 &   63.6$\pm$ 12.8 &     58.0$\pm$ 43.4 &   55.7$\pm$ 233.4 & 0.72$\pm$ 0.17 \\
SK -69 180 &   56.5$\pm$ 8.4 &    48.2$\pm$ 4.6 &     49.5$\pm$ 16.9 &    47.6$\pm$ 14.6 & 0.75$\pm$ 0.03 \\
SK -67 54 &   60.5$\pm$ 9.5 &    65.5$\pm$ 7.6 &     43.1$\pm$ 29.1 &    36.6$\pm$ 13.8 & 0.47$\pm$ 0.03 \\
SK -69 275 &   63.5$\pm$ 9.5 &    61.7$\pm$ 9.0 &     59.9$\pm$ 41.1 &    57.0$\pm$ 12.9 & 0.67$\pm$ 0.04 \\
MCPS 082.95108-69.23933 &   53.5$\pm$ 8.0 &    49.6$\pm$ 9.2 &      47.8$\pm$ 4.2 &    48.0$\pm$ 10.4 & 0.59$\pm$ 0.05 \\
HD 269700 & 101.5$\pm$ 13.9 &   67.7$\pm$ 19.1 &      59.8$\pm$ 4.5 &    59.7$\pm$ 21.9 & 0.62$\pm$ 0.07 \\
CPD-69 457 &  68.5$\pm$ 10.8 &   68.3$\pm$ 12.7 &     88.9$\pm$ 23.8 &  81.3$\pm$ - & 0.59$\pm$ 0.68 \\
SK -68 147 &   51.0$\pm$ 8.2 &    44.3$\pm$ 9.4 &     51.2$\pm$ 15.9 &     54.0$\pm$ 9.1 & 0.69$\pm$ 0.06 \\
Dachs LMC 2-19 &   62.0$\pm$ 9.5 &    63.1$\pm$ 6.7 &     37.4$\pm$ 12.7 &   36.9$\pm$ 181.8 & 0.46$\pm$ 0.17 \\
HD 269314 &  95.5$\pm$ 14.3 &   78.6$\pm$ 12.2 &     70.9$\pm$ 67.6 &    66.3$\pm$ 25.0 & 0.63$\pm$ 0.05 \\
CPD-69 445 &  68.0$\pm$ 10.4 &   66.8$\pm$ 11.9 &     56.0$\pm$ 38.2 &    51.4$\pm$ 14.9 & 0.62$\pm$ 0.04 \\
SK -69 256 &   62.5$\pm$ 9.9 &   66.9$\pm$ 12.1 &     52.5$\pm$ 28.0 &    45.8$\pm$ 27.4 & 0.74$\pm$ 0.04 \\
RMC 142 &  65.0$\pm$ 10.4 &    55.2$\pm$ 8.7 &     52.2$\pm$ 27.4 &    54.7$\pm$ 11.7 & 0.73$\pm$ 0.06 \\
HD 269859 &   52.5$\pm$ 8.5 &    41.9$\pm$ 5.6 &     61.2$\pm$ 14.2 &     57.9$\pm$ 8.4 &  0.6$\pm$ 0.07 \\
HD 270754 &   55.5$\pm$ 9.0 &    56.6$\pm$ 8.2 &      44.4$\pm$ 5.2 &     40.5$\pm$ 8.5 & 0.56$\pm$ 0.04 \\
HD 270196 &   54.5$\pm$ 8.5 &   53.6$\pm$ 10.6 &      52.8$\pm$ 2.2 &     47.5$\pm$ 6.9 &  0.7$\pm$ 0.03 \\
SK -67 46 &   52.5$\pm$ 8.1 &   46.2$\pm$ 10.5 &      51.3$\pm$ 5.2 &    52.4$\pm$ 10.0 & 0.59$\pm$ 0.04 \\
SK -69 167 &   58.5$\pm$ 9.2 &   51.6$\pm$ 13.9 &      61.5$\pm$ 8.0 &    56.1$\pm$ 13.1 & 0.75$\pm$ 0.04 \\
HD 269786 &   58.0$\pm$ 9.2 &   56.9$\pm$ 10.4 &     64.2$\pm$ 11.4 &     66.0$\pm$ 5.6 & 0.89$\pm$ 0.03 \\
SK -67 76 &  82.5$\pm$ 14.2 &    61.6$\pm$ 8.4 &      61.8$\pm$ 9.0 &   63.5$\pm$ 363.1 & 0.46$\pm$ 0.22 \\
HD 269655 &  73.0$\pm$ 11.0 &   64.2$\pm$ 15.7 &     60.0$\pm$ 14.4 &   56.4$\pm$ 193.5 & 0.68$\pm$ 0.12 \\
SK -69- 157 &  72.5$\pm$ 12.0 &   55.4$\pm$ 13.7 &     70.4$\pm$ 58.8 &   70.3$\pm$ 134.0 & 0.47$\pm$ 0.09 \\
HD 269660 &   47.5$\pm$ 7.4 &    49.6$\pm$ 6.5 &     54.0$\pm$ 10.1 &     59.6$\pm$ 6.0 & 0.99$\pm$ 0.05 \\
SK -69 68 &  70.5$\pm$ 10.4 &    72.5$\pm$ 7.2 &     59.8$\pm$ 29.8 &    56.4$\pm$ 16.2 & 0.24$\pm$ 0.03 \\
HD 268809 &   58.5$\pm$ 8.7 &    50.1$\pm$ 9.8 &     55.5$\pm$ 34.0 &   52.1$\pm$ 203.6 &  0.7$\pm$ 0.16 \\
LH 47-373A &   61.5$\pm$ 9.2 &    65.4$\pm$ 4.2 &      47.4$\pm$ 9.3 &   44.2$\pm$ 156.4 & 0.56$\pm$ 0.13 \\
HD 268623 &   59.0$\pm$ 9.4 &    66.7$\pm$ 9.2 &     53.2$\pm$ 20.2 &    54.9$\pm$ 14.4 & 0.87$\pm$ 0.03 \\
SK -68 17 &   44.0$\pm$ 7.2 &    42.1$\pm$ 6.8 &     51.9$\pm$ 11.7 &     47.7$\pm$ 7.3 & 0.63$\pm$ 0.04 \\
W61 27-5 &   56.0$\pm$ 8.4 &    45.0$\pm$ 3.8 &     54.6$\pm$ 14.3 &    43.4$\pm$ 15.8 & 0.58$\pm$ 0.06 \\
HD 268798 &   56.0$\pm$ 8.8 &    49.2$\pm$ 7.8 &      55.2$\pm$ 9.4 &    49.2$\pm$ 82.1 & 0.72$\pm$ 0.06 \\
CPD-68 309 &   46.5$\pm$ 7.8 &    50.4$\pm$ 4.9 &      21.7$\pm$ 4.1 &     23.1$\pm$ 7.0 &  0.91$\pm$ 0.1 \\
SK -68 27 &   55.0$\pm$ 9.9 &    41.5$\pm$ 5.2 &     47.2$\pm$ 11.7 &     45.9$\pm$ 3.3 & 0.63$\pm$ 0.02 \\
HD 269845 &   48.5$\pm$ 8.3 &    43.6$\pm$ 7.7 &      54.2$\pm$ 5.4 &     44.4$\pm$ 2.2 & 0.62$\pm$ 0.01 \\
SK -68 2 &  55.5$\pm$ 10.6 &   38.4$\pm$ 10.4 &      46.9$\pm$ 5.4 &    45.6$\pm$ 75.5 & 0.66$\pm$ 0.08 \\
HD 268653 &   54.0$\pm$ 9.0 &    39.3$\pm$ 8.4 &      46.0$\pm$ 2.5 &    47.5$\pm$ 19.5 &  0.5$\pm$ 0.08 \\
HD 269992 &   45.5$\pm$ 8.2 &    48.3$\pm$ 9.3 &      48.0$\pm$ 6.4 &    53.0$\pm$ 64.8 & 0.99$\pm$ 0.13 \\
HD 269997 &   44.5$\pm$ 9.3 &    46.3$\pm$ 6.9 &      45.0$\pm$ 2.2 &     51.5$\pm$ 2.6 & 0.88$\pm$ 0.02 \\
HD 269101 &   46.0$\pm$ 8.8 &    44.4$\pm$ 8.9 &      36.4$\pm$ 5.2 &    32.3$\pm$ 17.3 & 0.84$\pm$ 0.11 \\
SK -67 133 &  53.5$\pm$ 10.4 &    40.3$\pm$ 7.7 &      40.3$\pm$ 4.9 &     38.0$\pm$ 3.7 & 0.61$\pm$ 0.01 \\
HD 268729 &  53.5$\pm$ 10.4 &    52.8$\pm$ 8.1 &      39.2$\pm$ 3.6 &    35.2$\pm$ 20.3 & 0.69$\pm$ 0.09 \\
HD 271213 &   54.0$\pm$ 9.6 &    44.8$\pm$ 7.9 &      46.0$\pm$ 4.6 &     46.9$\pm$ 4.5 &  0.7$\pm$ 0.01 \\
HD 270949 &  67.0$\pm$ 11.5 &    45.7$\pm$ 6.3 &      38.6$\pm$ 4.8 &    38.4$\pm$ 42.4 & 0.51$\pm$ 0.14 \\
HD 269145 &  58.5$\pm$ 11.2 &    46.5$\pm$ 4.6 &      39.8$\pm$ 1.5 &    36.6$\pm$ 20.0 & 0.73$\pm$ 0.08 \\
HD 269920 &  55.5$\pm$ 10.7 &    41.7$\pm$ 6.1 &      45.3$\pm$ 3.0 &     43.7$\pm$ 3.0 & 0.73$\pm$ 0.01 \\
HD 269854 &  54.5$\pm$ 10.4 &    42.4$\pm$ 4.8 &      27.2$\pm$ 2.0 &     40.3$\pm$ 4.1 & 0.71$\pm$ 0.01 \\
SK -70 53 &   40.5$\pm$ 9.0 &    29.8$\pm$ 5.6 &      43.2$\pm$ 5.1 &     43.0$\pm$ 1.3 &  0.5$\pm$ 0.01 \\
SK -67 275 &  63.5$\pm$ 12.1 &    54.2$\pm$ 3.3 &      39.9$\pm$ 2.8 &    38.8$\pm$ 41.8 & 0.53$\pm$ 0.12 \\
SK -67 283 &  46.5$\pm$ 11.9 &    38.6$\pm$ 5.3 &      34.6$\pm$ 1.0 &     32.8$\pm$ 3.6 & 0.53$\pm$ 0.02 \\
HD 47620 &  23.5$\pm$ 10.0 &      34.3$\pm$ - &               - &              - &           - \\
SK -66 125 &  36.5$\pm$ 11.5 &    34.2$\pm$ 3.9 &      25.0$\pm$ 2.5 &     25.2$\pm$ 2.3 &  0.6$\pm$ 0.01 \\
HD 269644 &  57.5$\pm$ 12.0 &    47.7$\pm$ 5.2 &      28.9$\pm$ 2.2 &     25.0$\pm$ 5.6 & 0.75$\pm$ 0.01 \\
SK -67 127 &  44.5$\pm$ 11.9 &    33.8$\pm$ 5.5 &      36.2$\pm$ 5.9 &    34.3$\pm$ 11.1 & 0.63$\pm$ 0.06 \\
HD 268907 &  72.0$\pm$ 14.0 &    40.6$\pm$ 3.7 &      44.4$\pm$ 2.5 &     33.1$\pm$ 2.8 &  1.0$\pm$ 0.01 \\
SK -66 126 &  46.5$\pm$ 12.8 &    36.5$\pm$ 4.9 &      30.8$\pm$ 4.1 &     28.9$\pm$ 3.8 & 0.62$\pm$ 0.01 \\
HD 268654 & 112.5$\pm$ 21.0 &    43.4$\pm$ 4.6 &      24.9$\pm$ 3.4 &    34.6$\pm$ 14.0 & 0.63$\pm$ 0.07 \\
SK -68 53 &  47.5$\pm$ 11.5 &    30.4$\pm$ 4.1 &      45.9$\pm$ 1.9 &     32.8$\pm$ 1.1 &  1.0$\pm$ 0.04 \\
SK -66 142 &  49.0$\pm$ 11.4 &      33.2$\pm$ - &               - &              - &           - \\
SK -66 92 &  51.5$\pm$ 11.0 &    30.8$\pm$ 3.2 &      31.8$\pm$ 1.1 &     33.6$\pm$ 3.9 & 0.42$\pm$ 0.01 \\
HD 269801 &  71.5$\pm$ 14.2 &    36.7$\pm$ 3.4 &      38.1$\pm$ 3.3 &     43.4$\pm$ 1.1 & 0.65$\pm$ 0.03 \\
SK -67 198 &  56.0$\pm$ 13.2 &    39.5$\pm$ 4.1 &      34.7$\pm$ 2.4 &     29.5$\pm$ 2.5 & 0.61$\pm$ 0.01 \\
HD 270933 &  57.0$\pm$ 11.8 &      35.1$\pm$ - &               - &              - &           - \\
HD 269766 &  44.5$\pm$ 11.9 &    37.6$\pm$ 6.1 &      30.9$\pm$ 1.5 &    30.2$\pm$ 12.2 & 0.58$\pm$ 0.07 \\
HD 41297 &  29.5$\pm$ 11.6 &    32.2$\pm$ 5.3 &       8.4$\pm$ 2.5 &     5.5$\pm$ 10.4 & 0.73$\pm$ 0.65 \\
SK -70 81 & 105.0$\pm$ 24.0 &   101.5$\pm$ 4.2 &      29.8$\pm$ 4.1 &     26.8$\pm$ 8.6 & 0.05$\pm$ 0.01 \\
SK -67 171 &  47.5$\pm$ 11.5 &    30.6$\pm$ 5.2 &      27.1$\pm$ 4.0 &     27.3$\pm$ 5.2 & 0.46$\pm$ 0.03 \\
HD 269619 &  52.5$\pm$ 13.0 &    37.0$\pm$ 2.9 &      35.4$\pm$ 2.9 &     35.1$\pm$ 6.0 & 0.53$\pm$ 0.02 \\
GV 375 &   43.5$\pm$ 9.4 &    25.5$\pm$ 3.9 &      23.7$\pm$ 0.3 &     22.3$\pm$ 2.8 & 0.64$\pm$ 0.04 \\
CPD-69 490 &  45.0$\pm$ 10.9 &    28.1$\pm$ 3.6 &      35.1$\pm$ 2.3 &     30.0$\pm$ 0.5 & 0.54$\pm$ 0.03 \\
SK -67 72 &  46.5$\pm$ 11.8 &    31.9$\pm$ 2.2 &      31.9$\pm$ 2.1 &     28.6$\pm$ 0.5 & 0.74$\pm$ 0.02 \\
HD 268675 &  53.0$\pm$ 11.6 &    34.5$\pm$ 3.8 &      48.0$\pm$ 0.2 &     47.2$\pm$ 4.2 & 0.63$\pm$ 0.03 \\
SK -67 7 &  46.5$\pm$ 11.8 &      19.2$\pm$ - &      35.4$\pm$ 2.6 &     26.0$\pm$ 4.6 & 0.74$\pm$ 0.02 \\
CPD-69 512 &  48.5$\pm$ 12.0 &    29.0$\pm$ 4.2 &      34.3$\pm$ 4.8 &     31.3$\pm$ 0.9 & 0.59$\pm$ 0.04 \\
HD 270123 &  45.0$\pm$ 10.6 &    30.0$\pm$ 5.0 &      30.3$\pm$ 1.7 &     - &           - \\
SK -70 31 &  33.0$\pm$ 10.6 &    27.8$\pm$ 2.8 &      27.2$\pm$ 0.2 &     25.0$\pm$ 0.5 & 0.46$\pm$ 0.03 \\
HD 57970 &  61.5$\pm$ 22.0 &      67.8$\pm$ - &               - &              - &           - \\
SK -69 250 &  34.5$\pm$ 11.8 &    34.8$\pm$ 5.6 &      27.9$\pm$ 3.0 &     28.6$\pm$ 0.7 & 0.69$\pm$ 0.05 \\
SK -69 31 &   44.0$\pm$ 9.7 &    32.9$\pm$ 6.4 &      32.2$\pm$ 1.5 &     30.8$\pm$ 4.4 & 0.57$\pm$ 0.01 \\
SK -70 26 &   39.5$\pm$ 9.8 &    27.4$\pm$ 3.9 &      21.1$\pm$ 2.0 &     21.4$\pm$ 2.0 & 0.63$\pm$ 0.03 \\
SK -67 42 &   41.5$\pm$ 9.0 &    21.4$\pm$ 5.6 &      24.4$\pm$ 0.2 &     25.3$\pm$ 4.1 & 0.46$\pm$ 0.02 \\
SK -69 47 &   42.5$\pm$ 9.5 &      31.9$\pm$ - &               - &              - &           - \\
SK -67 279 &  23.0$\pm$ 10.2 &    23.8$\pm$ 3.5 &      21.2$\pm$ 0.2 &     20.5$\pm$ 2.2 & 0.56$\pm$ 0.03 \\
SK -67 88 &  44.0$\pm$ 10.0 &    23.6$\pm$ 2.3 &      32.0$\pm$ 1.2 &     35.5$\pm$ 1.1 & 0.44$\pm$ 0.01 \\
HD 269721 &   40.0$\pm$ 8.9 &    32.6$\pm$ 4.0 &      27.6$\pm$ 0.1 &     26.6$\pm$ 4.2 & 0.57$\pm$ 0.01 \\
HD 45527 &  44.5$\pm$ 13.0 &    48.2$\pm$ 2.9 &      44.5$\pm$ 2.1 &    52.4$\pm$ 22.2 &  1.0$\pm$ 0.04 \\
HD 269639 &  26.5$\pm$ 15.0 &    34.8$\pm$ 5.5 &      26.0$\pm$ 1.9 &     - &           - \\
SOI 404 &  44.0$\pm$ 11.4 &    38.0$\pm$ 3.1 &      27.6$\pm$ 0.4 &     26.2$\pm$ 0.8 & 0.54$\pm$ 0.06 \\
HD 269510 &  39.0$\pm$ 10.6 &    29.5$\pm$ 5.1 &      30.7$\pm$ 0.1 &     30.9$\pm$ 0.2 & 0.55$\pm$ 0.01 \\
SK -69 174 &  18.0$\pm$ 12.0 &    29.0$\pm$ 4.2 &      21.9$\pm$ 1.3 &     18.8$\pm$ 5.2 & 0.66$\pm$ 0.02 \\
SOI 247 &   36.0$\pm$ 8.9 &      18.2$\pm$ - &               - &              - &           - \\
SK -70 77 &   34.5$\pm$ 8.4 &      22.8$\pm$ - &               - &              - &           - \\
SK -68 152 &   34.0$\pm$ 8.0 &      14.4$\pm$ - &               - &              - &           - \\
SK -67 151 &   8.0$\pm$ 19.8 &      21.8$\pm$ - &               - &              - &           - \\
SOI 574 &   32.0$\pm$ 7.2 &      22.2$\pm$ - &               - &              - &           - \\
SOI 166 &            - &  138.8$\pm$ 28.0 & 200.0$\pm$ 18142.1 & 199.9$\pm$ 6207.7 & 0.11$\pm$ 1.53 \\

\multicolumn{6}{c}{{\sc feros}} \\         

HD 269676 & 181.5 $\pm$ 33.3 &             - &             - &             - &           - \\
HD 269321 &  78.0 $\pm$ 16.4 &    35.3 $\pm$ 7.6 &    30.7 $\pm$ 7.7 &   27.5 $\pm$ 29.3 & 0.69 $\pm$ 0.08 \\
HD 269440 &  51.5 $\pm$ 11.0 &   54.8 $\pm$ 10.0 &   24.9 $\pm$ 59.8 &  29.4 $\pm$ 173.4 &  1.0 $\pm$ 0.46 \\
HD 269660 &  50.0 $\pm$ 11.5 &   47.5 $\pm$ 10.0 &   52.2 $\pm$ 12.5 &  47.2 $\pm$ 255.3 & 0.86 $\pm$ 0.25 \\
HD 269700 &  80.5 $\pm$ 19.5 &    43.4 $\pm$ 5.1 &  100.0 $\pm$ 35.0 &  96.0 $\pm$ 251.6 & 0.11 $\pm$ 1.73 \\
HD 269832 &  96.0 $\pm$ 21.3 &   49.7 $\pm$ 19.5 &    32.6 $\pm$ 4.7 &    20.8 $\pm$ 3.9 & 0.99 $\pm$ 0.03 \\
HD 268798 &  57.5 $\pm$ 13.6 &   51.3 $\pm$ 12.7 &   45.6 $\pm$ 48.3 &  47.9 $\pm$ 120.7 & 0.54 $\pm$ 0.12 \\
HD 268726 &  60.0 $\pm$ 14.2 &   48.4 $\pm$ 15.4 &   56.1 $\pm$ 53.9 &  55.4 $\pm$ 475.4 &  0.72 $\pm$ 0.5 \\
HD 269992 &  51.0 $\pm$ 12.5 &    43.5 $\pm$ 3.3 &   52.6 $\pm$ 11.6 &   52.2 $\pm$ 11.3 & 0.47 $\pm$ 0.07 \\
HD 268654 &  80.0 $\pm$ 18.7 &    37.5 $\pm$ 5.7 &    32.4 $\pm$ 8.4 &   31.5 $\pm$ 10.4 & 0.64 $\pm$ 0.03 \\
HD 269997 &  50.5 $\pm$ 12.0 &   41.6 $\pm$ 11.5 &   42.8 $\pm$ 36.0 &  29.5 $\pm$ 136.3 &  0.63 $\pm$ 0.3 \\
HD 268907 &  64.5 $\pm$ 15.3 &    38.5 $\pm$ 5.8 &    24.0 $\pm$ 7.7 &   26.3 $\pm$ 75.9 &  0.39 $\pm$ 0.1 \\
HD 268729 &  55.5 $\pm$ 13.2 &   40.7 $\pm$ 10.5 &   23.4 $\pm$ 16.9 &    35.4 $\pm$ 6.7 &  0.6 $\pm$ 0.02 \\
HD 269786 &  64.0 $\pm$ 17.8 &   49.8 $\pm$ 14.8 &  63.3 $\pm$ 132.8 &   72.3 $\pm$ 44.7 & 0.72 $\pm$ 0.15 \\
HD 268653 &  65.0 $\pm$ 17.6 &   61.7 $\pm$ 14.1 &   47.3 $\pm$ 66.0 &  51.0 $\pm$ 591.1 & 0.87 $\pm$ 0.46 \\
HD 269859 &  67.5 $\pm$ 17.4 &    47.4 $\pm$ 5.6 &   85.4 $\pm$ 18.3 &  95.1 $\pm$ 185.3 &   0.7 $\pm$ 0.4 \\
HD 269145 &  65.0 $\pm$ 17.8 &    45.3 $\pm$ 3.6 &   36.2 $\pm$ 21.3 &  37.7 $\pm$ 170.2 & 0.67 $\pm$ 0.22 \\
HD 271163 &  57.5 $\pm$ 17.5 &    47.1 $\pm$ 5.8 &   11.9 $\pm$ 61.8 &  14.1 $\pm$ 787.2 & 0.15 $\pm$ 0.65 \\
HD 270949 &  55.5 $\pm$ 13.2 &    39.6 $\pm$ 6.9 &   32.6 $\pm$ 11.3 &    30.6 $\pm$ 6.9 & 0.52 $\pm$ 0.08 \\
HD 269854 &  72.0 $\pm$ 19.6 &    30.5 $\pm$ 9.7 &   34.4 $\pm$ 18.3 &   34.1 $\pm$ 54.0 & 0.37 $\pm$ 0.11 \\
HD 269606 &  56.0 $\pm$ 16.8 &    42.3 $\pm$ 6.6 &    30.9 $\pm$ 6.6 &    23.7 $\pm$ 5.8 & 0.82 $\pm$ 0.04 \\
HD 269644 &  62.0 $\pm$ 17.6 &    38.5 $\pm$ 3.9 &   26.0 $\pm$ 13.8 &  28.1 $\pm$ 102.2 & 0.45 $\pm$ 0.16 \\
HD 269593 &  55.0 $\pm$ 17.1 &    35.0 $\pm$ 6.4 &   36.2 $\pm$ 13.3 &   30.6 $\pm$ 66.4 & 0.13 $\pm$ 0.06 \\
HD 269238 &  64.5 $\pm$ 20.0 &    37.0 $\pm$ 5.4 &    31.5 $\pm$ 4.9 &    30.3 $\pm$ 3.6 & 0.58 $\pm$ 0.01 \\
CD-65 288 &  59.5 $\pm$ 19.0 &    33.4 $\pm$ 6.8 &    30.1 $\pm$ 3.0 &   32.9 $\pm$ 28.7 & 0.37 $\pm$ 0.05 \\
HD 269766 &  58.0 $\pm$ 18.0 &    36.0 $\pm$ 5.3 &    26.9 $\pm$ 7.3 &   27.5 $\pm$ 13.2 &  0.38 $\pm$ 0.1 \\
HD 269801 &  68.5 $\pm$ 20.3 &    39.7 $\pm$ 2.3 &    18.6 $\pm$ 3.0 &    24.1 $\pm$ 3.2 & 0.55 $\pm$ 0.02 \\
HD 269619 &  62.0 $\pm$ 19.6 &    36.9 $\pm$ 4.4 &    33.9 $\pm$ 5.8 &   30.9 $\pm$ 24.8 & 0.81 $\pm$ 0.05 \\
HD  53048 & 344.0 $\pm$ 73.6 &  231.1 $\pm$ 68.8 & 82.5 $\pm$ 1272.7 & 69.5 $\pm$ - & 0.48 $\pm$ 0.32 \\
HD 268675 &  52.5 $\pm$ 17.5 &      36.9 $\pm$ - &             - &             - &           - \\
HD 270933 &  52.5 $\pm$ 17.5 &    28.7 $\pm$ 6.8 &    26.5 $\pm$ 4.4 &   32.3 $\pm$ 10.8 &  0.19 $\pm$ 0.1 \\
HD  61950 & 344.0 $\pm$ 74.0 &  298.9 $\pm$ 63.5 &  40.6 $\pm$ 146.8 &  51.9 $\pm$ 753.3 & 0.75 $\pm$ 1.44 \\
HD 270296 &  53.5 $\pm$ 19.5 &    36.0 $\pm$ 4.9 &    30.2 $\pm$ 2.5 &    31.5 $\pm$ 3.6 & 0.48 $\pm$ 0.01 \\
HD 269684 &  49.0 $\pm$ 17.8 &    38.1 $\pm$ 2.5 &    24.1 $\pm$ 4.6 &   22.1 $\pm$ 31.4 &   0.6 $\pm$ 0.1 \\
HD 269195 &  48.0 $\pm$ 18.1 &    33.8 $\pm$ 3.1 &    27.7 $\pm$ 3.5 &   22.3 $\pm$ 13.3 & 0.41 $\pm$ 0.11 \\
HD 270123 &  52.0 $\pm$ 17.8 &    38.3 $\pm$ 3.3 &    21.3 $\pm$ 5.6 &    23.5 $\pm$ 3.2 & 0.74 $\pm$ 0.04 \\
HD 269639 &  43.5 $\pm$ 14.5 &    32.0 $\pm$ 7.0 &    18.7 $\pm$ 5.6 &   18.0 $\pm$ 31.4 & 0.42 $\pm$ 0.12 \\
HD 269510 &  38.5 $\pm$ 15.4 &    36.5 $\pm$ 3.9 &    15.3 $\pm$ 3.4 &    12.9 $\pm$ 3.8 & 0.75 $\pm$ 0.14 \\
HD 269721 &  43.0 $\pm$ 16.0 &    33.3 $\pm$ 5.9 &    19.6 $\pm$ 3.5 &   20.6 $\pm$ 15.7 & 0.55 $\pm$ 0.06 \\
HD 269664 &  35.5 $\pm$ 15.2 &    36.3 $\pm$ 2.3 &    11.9 $\pm$ 8.1 &    5.9 $\pm$ 12.6 & 0.71 $\pm$ 0.72 \\
HD 268835 &            - &             - &             - &             - &           - \\
HD 269582 &            - &             - &             - &             - &           - \\
HD 268718 &            - &   36.7 $\pm$ 12.4 &             - &             - &           - \\

\end{longtable}

\onecolumn
\section{Results of Monte Carlo simulations of broadened line profiles}
\label{section:MonteCarloSim}
   
 \begin{figure}[H]
   \begin{center}
            {\includegraphics[clip,scale=0.67]{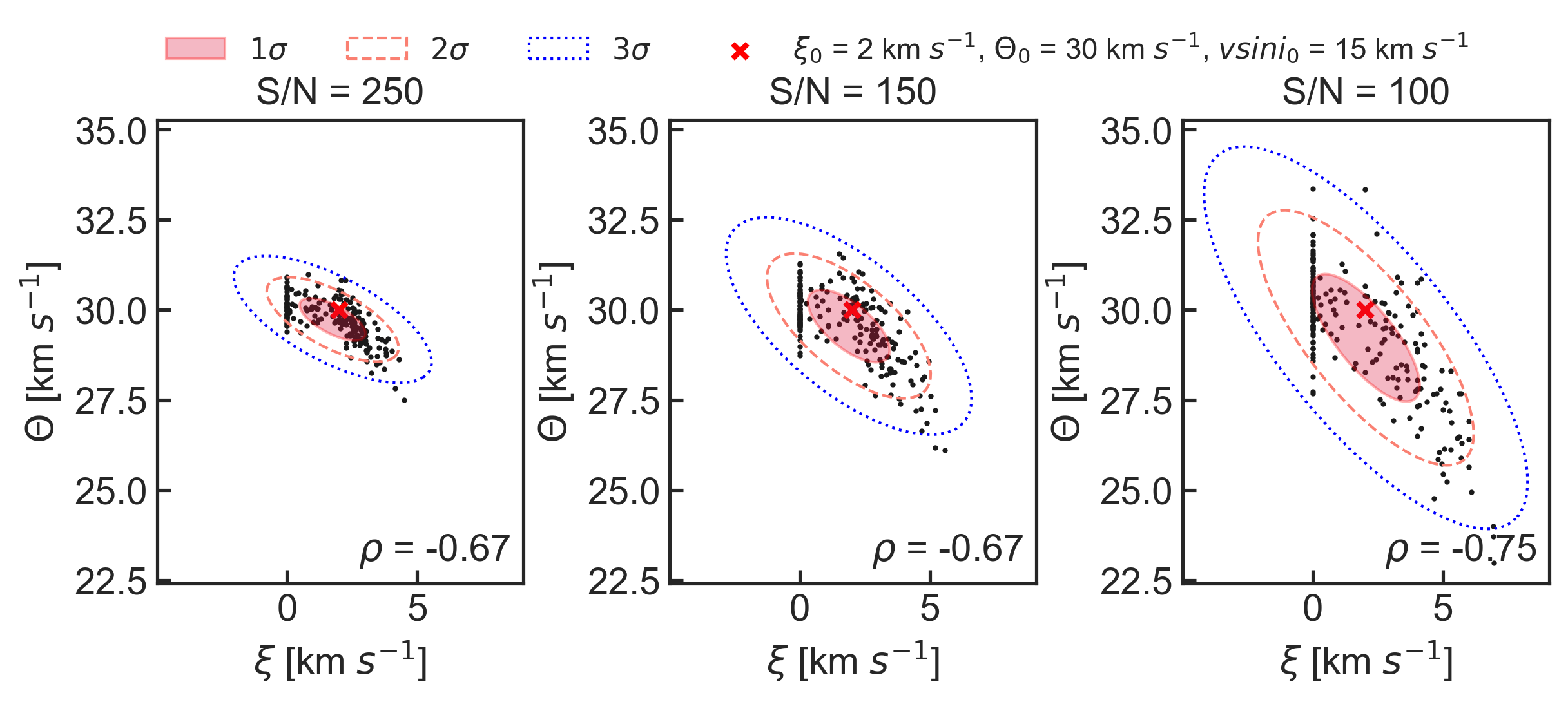}}
      \caption{ Results of Monte Carlo simulations on synthetic profiles of \ion{Mg}{ii}~4481~\AA\ line with artificial noise (left panel: S/N = 250, middle panel: S/N = 150, and right panel: S/N = 100) with 1-, 2- and 3-$\sigma$ confidence ellipses. Correlation coefficients $\rho$ are indicated in right corner of each panel. \vsini\ is fixed, while $\Theta$ and $\xi$ are optimised. The red crosses mark true parameters of the synthetic profiles. }
         \label{cc2_mg_2_30_15_vmi}
    \end{center}
 \end{figure}
      
 \begin{figure}[H]
   \begin{center}
            {\includegraphics[clip,scale=0.67]{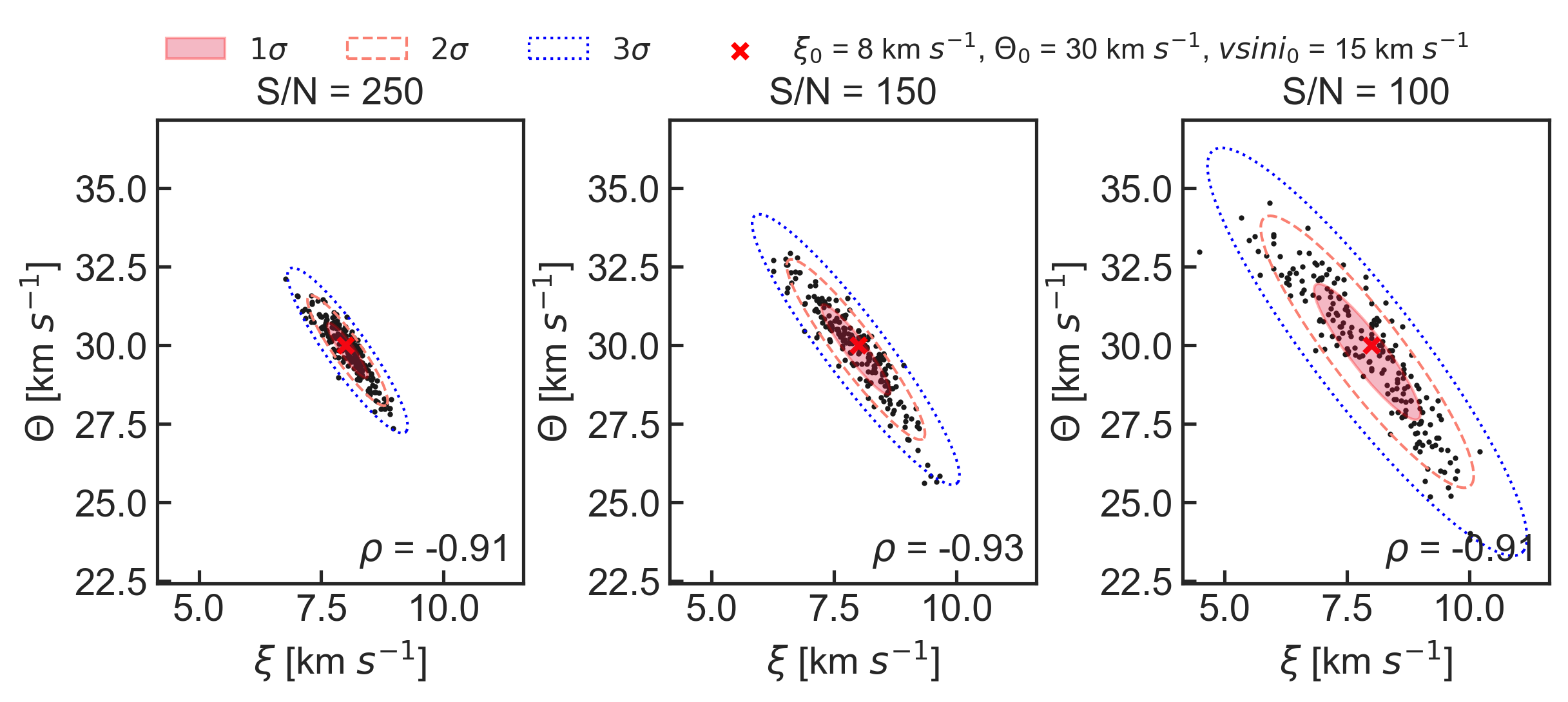}}
      \caption{ Same as Fig. \ref{cc2_mg_2_30_15_vmi} for another set of parameters \vsini, $\xi$, and $\Theta$ in the synthetic profile. }
         \label{cc2_mg_8_30_15_vmi}
    \end{center}
 \end{figure}   
 
 \begin{figure}[H]
   \begin{center}
            {\includegraphics[clip,scale=0.67]{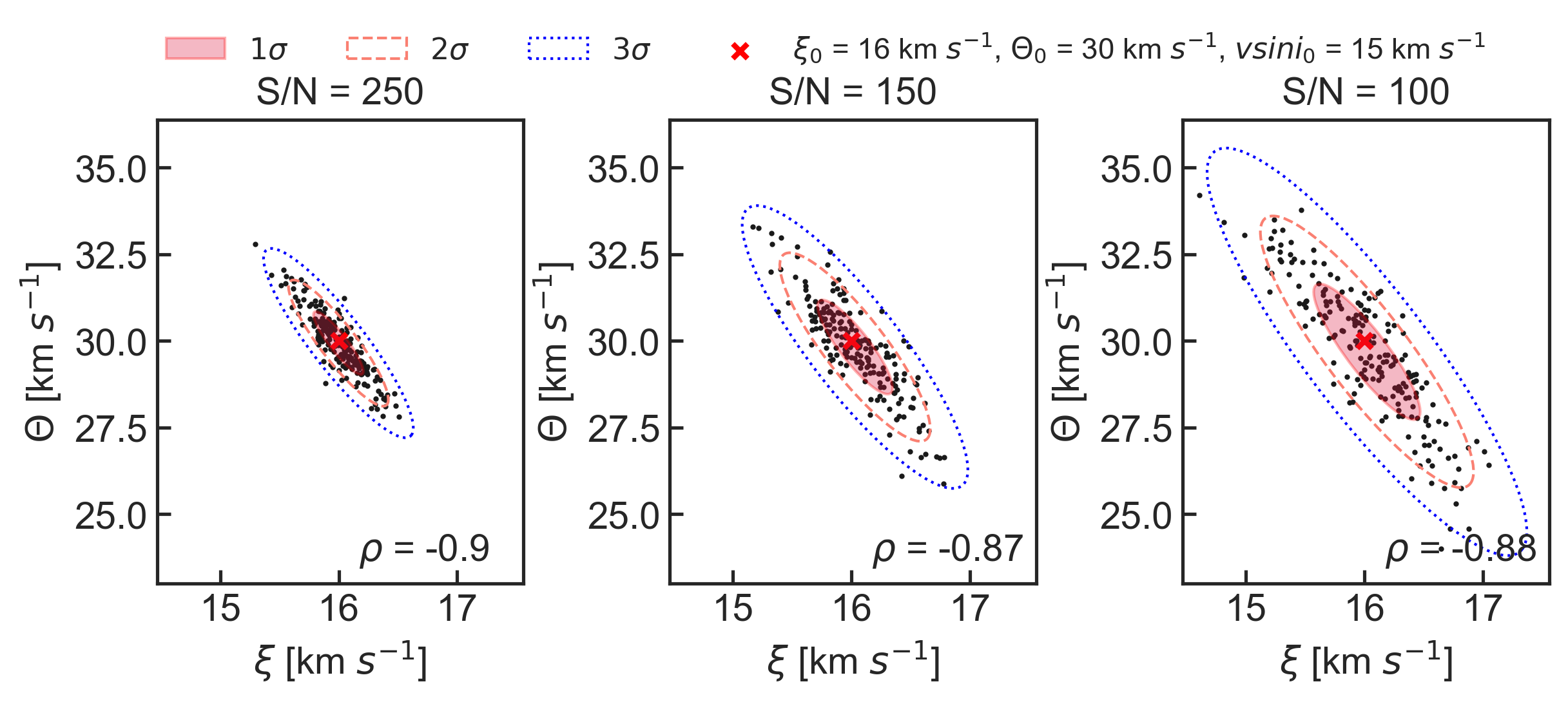}}
      \caption{ Same as Fig. \ref{cc2_mg_2_30_15_vmi} for another set of parameters \vsini, $\xi$, and $\Theta$ in the synthetic profile. }
         \label{cc2_mg_16_30_15_vmi}
    \end{center}
 \end{figure}

   
 \begin{figure}
   \begin{center}
            {\includegraphics[clip,scale=0.7]{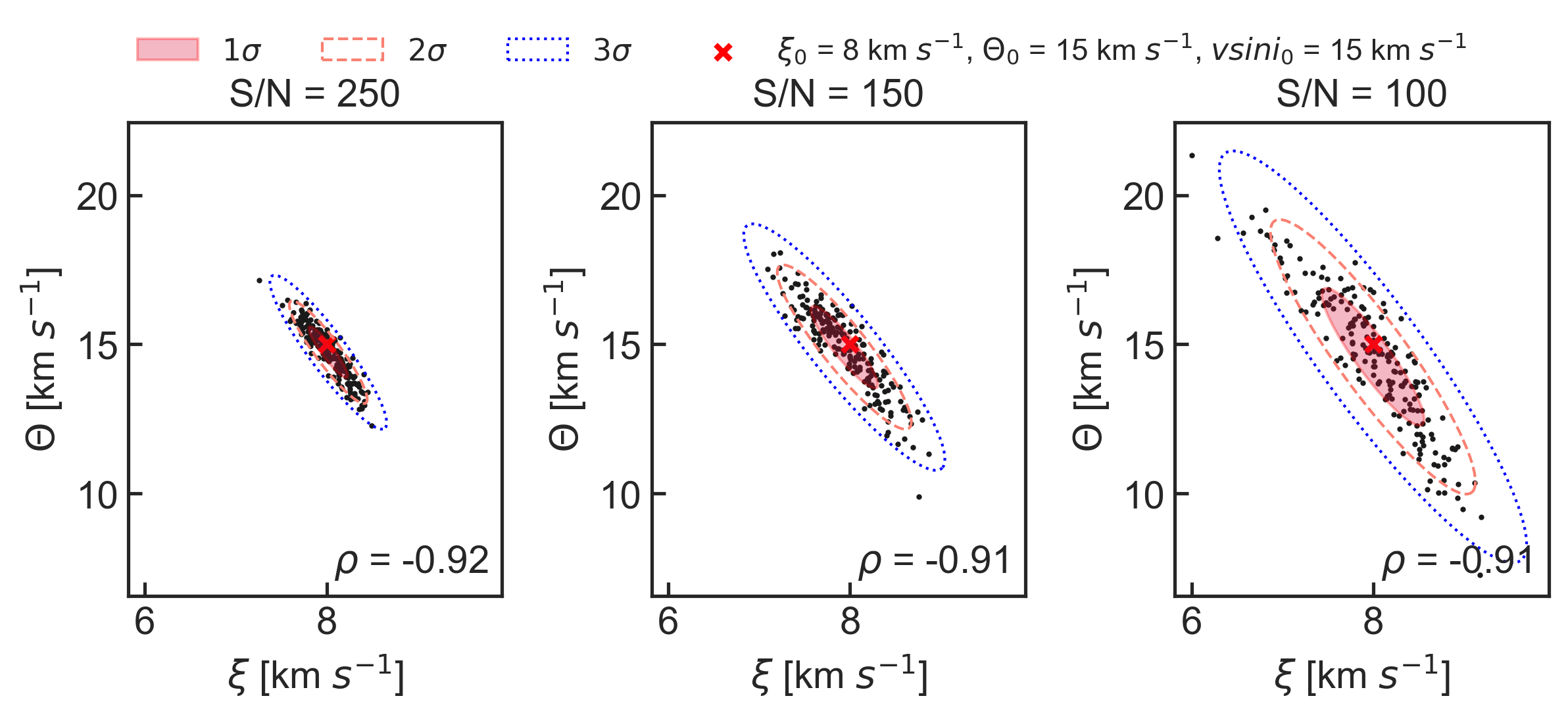}}
      \caption{ Same as Fig. \ref{cc2_mg_2_30_15_vmi} for another set of parameters \vsini, $\xi$, and $\Theta$ in the synthetic profile. }
         \label{cc2_mg_8_15_15}
    \end{center}
 \end{figure}

 \begin{figure*}
   \begin{center}
            {\includegraphics[clip,scale=0.7]{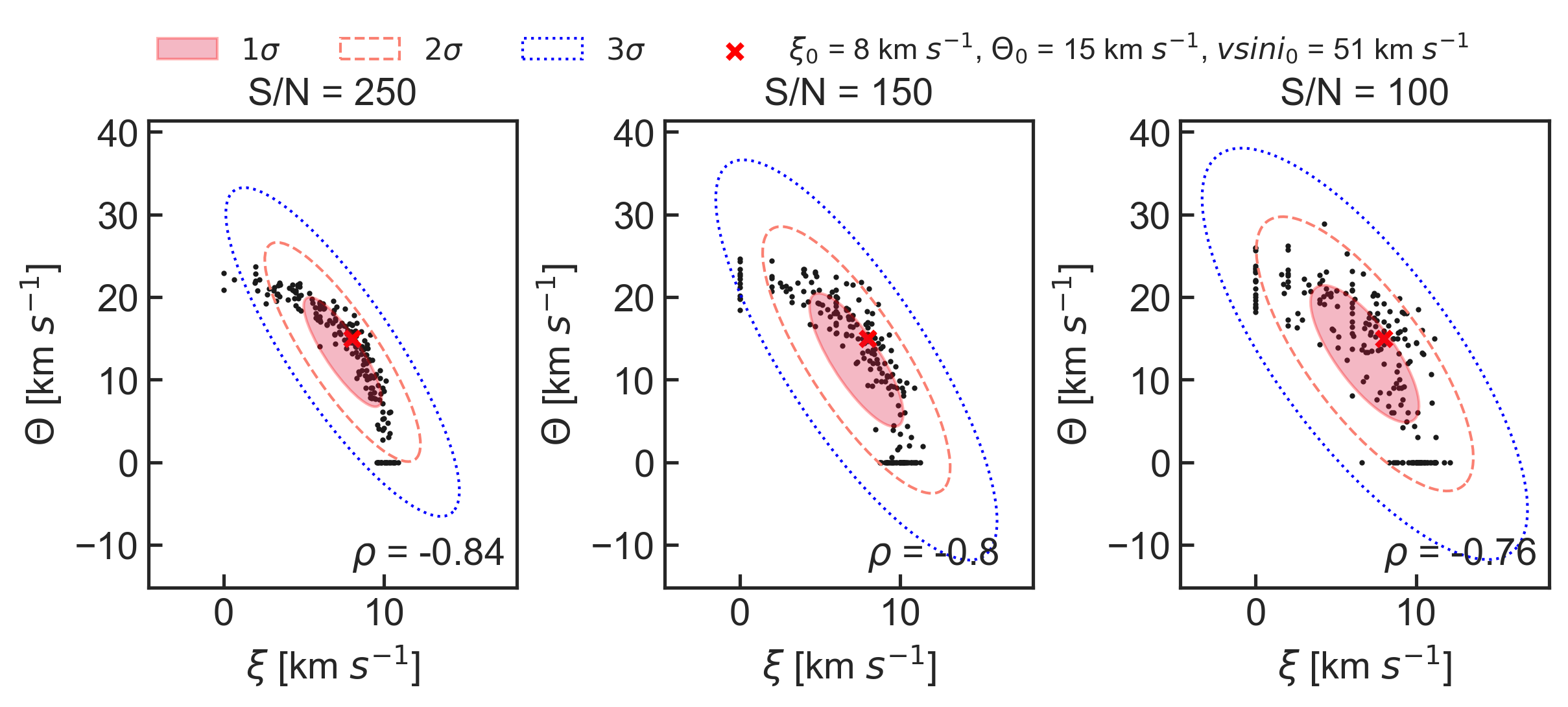}}
      \caption{ Same as Fig. \ref{cc2_mg_2_30_15_vmi} for another set of parameters \vsini, $\xi$, and $\Theta$ in the synthetic profile. }
         \label{cc2_mg_8_15_51}
    \end{center}
 \end{figure*}

    
 \begin{figure*}
   \begin{center}
            {\includegraphics[clip,scale=0.7]{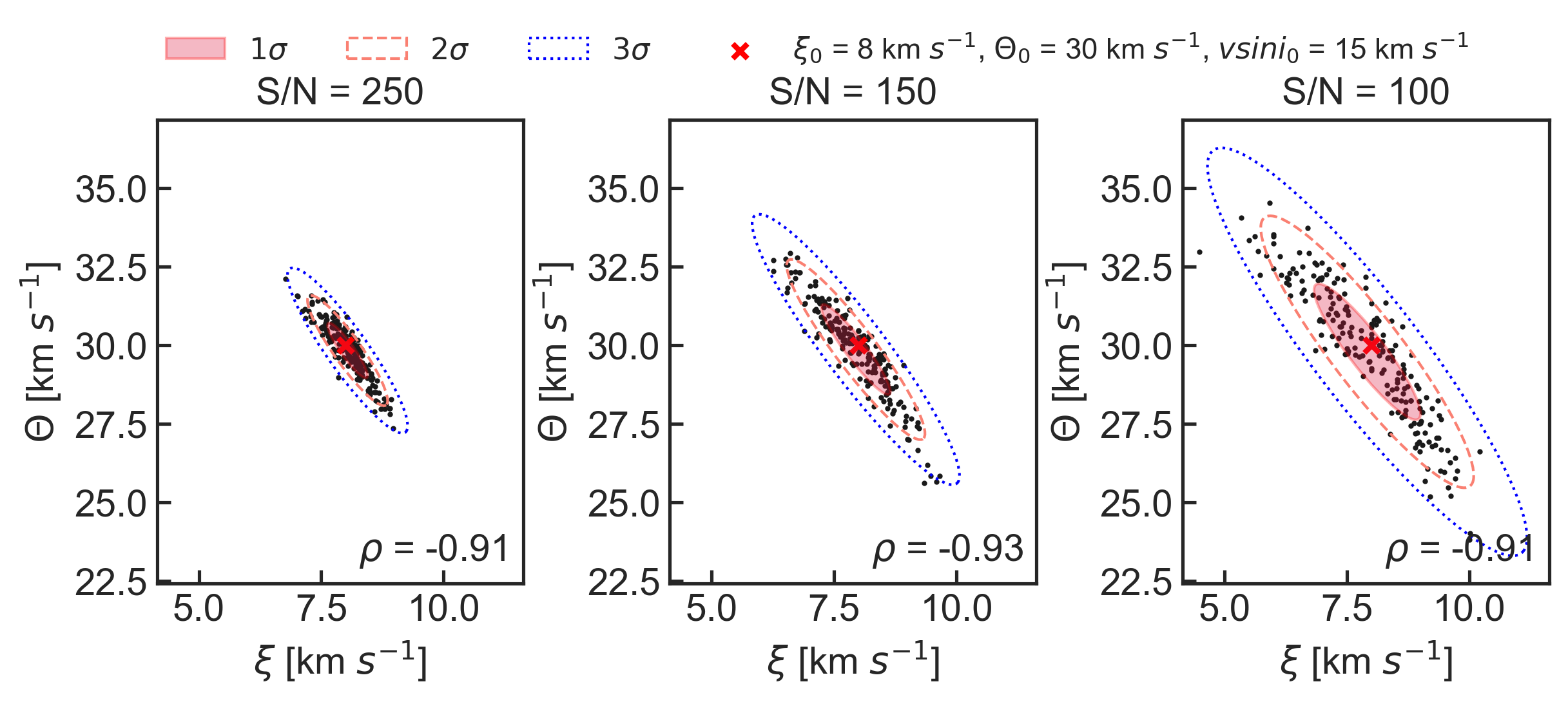}}
      \caption{ Same as Fig. \ref{cc2_mg_2_30_15_vmi} for another set of parameters \vsini, $\xi$, and $\Theta$ in the synthetic profile. }
         \label{cc2_mg_8_30_15}
    \end{center}
 \end{figure*}
    
 \begin{figure*}
   \begin{center}
            {\includegraphics[clip,scale=0.7]{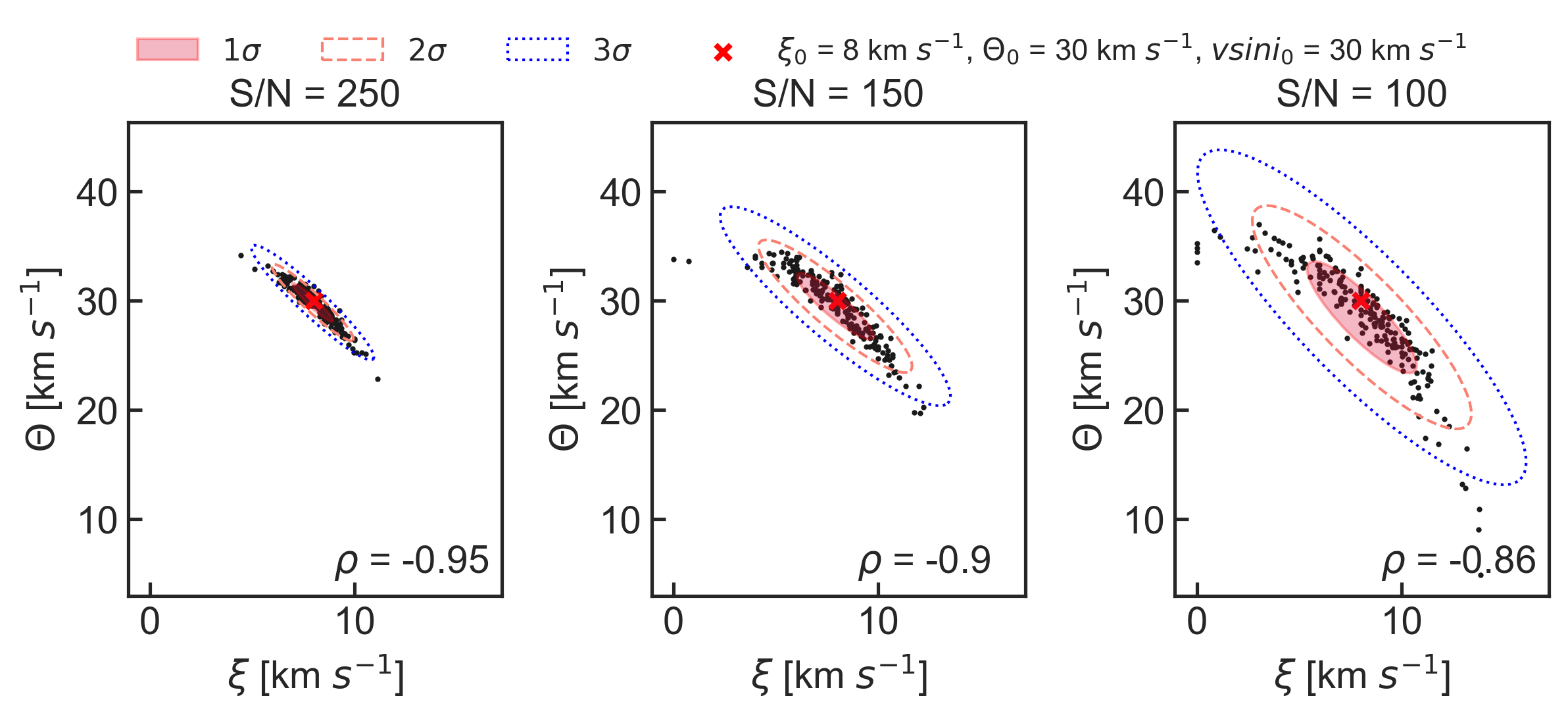}}
      \caption{ Same as Fig. \ref{cc2_mg_2_30_15_vmi} for another set of parameters \vsini, $\xi$, and $\Theta$ in the synthetic profile. }
         \label{cc2_mg_8_30_30}
    \end{center}
 \end{figure*}
    
 \begin{figure*}
   \begin{center}
            {\includegraphics[clip,scale=0.7]{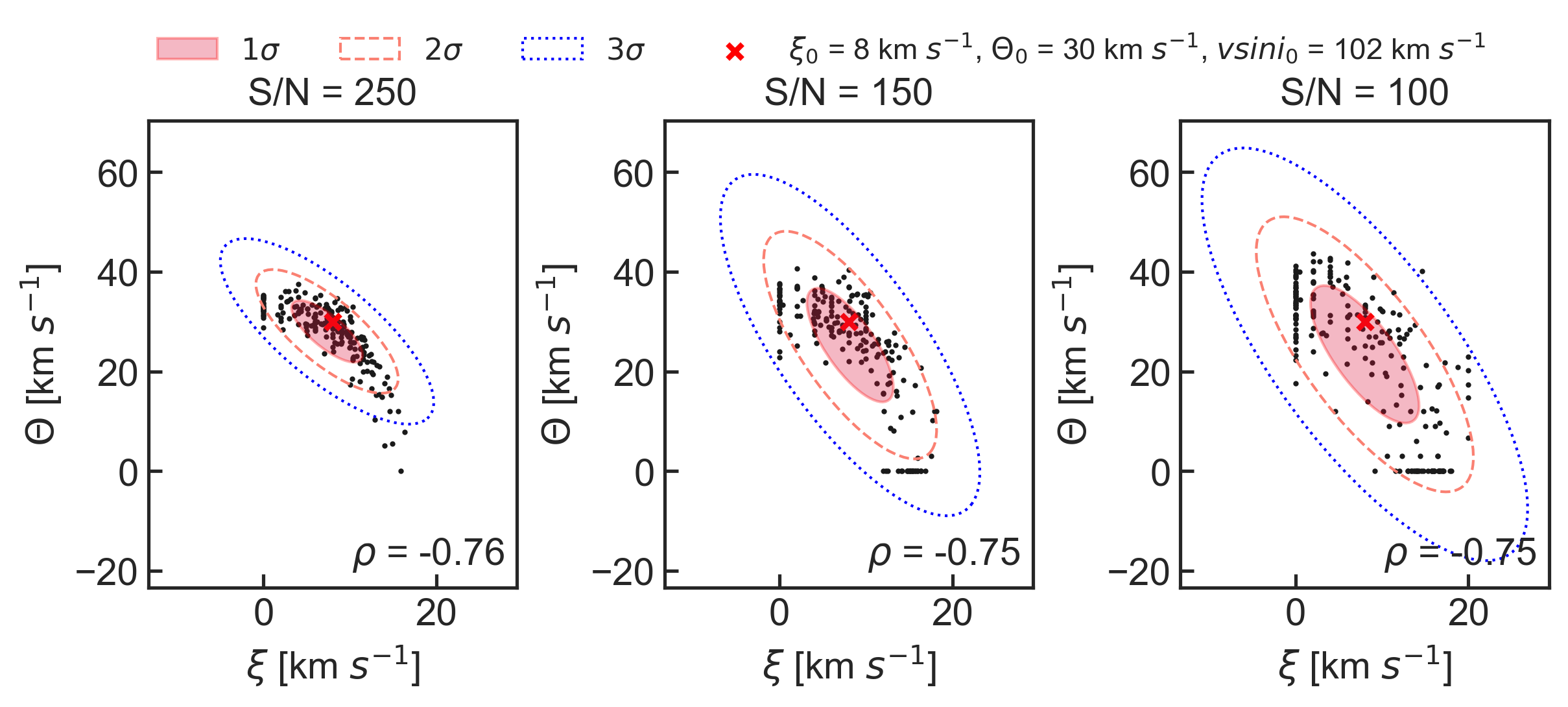}}
      \caption{ Same as Fig. \ref{cc2_mg_2_30_15_vmi} for another set of parameters \vsini, $\xi$, and $\Theta$ in the synthetic profile. }
         \label{cc2_mg_8_30_102}
    \end{center}
 \end{figure*}
 
    
 \begin{figure*}
   \begin{center}
            {\includegraphics[clip,scale=0.7]{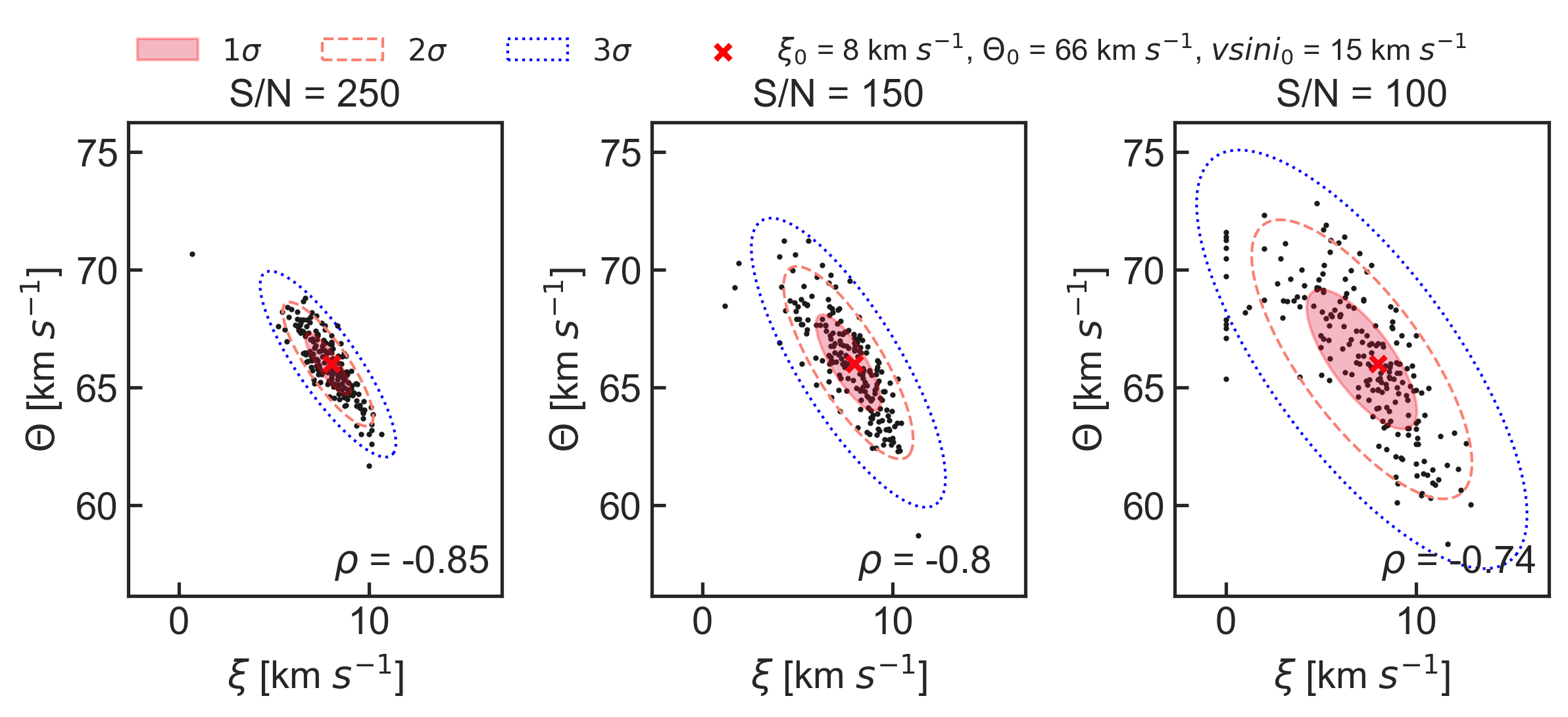}}
      \caption{ Same as Fig. \ref{cc2_mg_2_30_15_vmi} for another set of parameters \vsini, $\xi$, and $\Theta$ in the synthetic profile. }
         \label{cc2_mg_8_66_15}
    \end{center}
 \end{figure*}
    
 \begin{figure*}
   \begin{center}
            {\includegraphics[clip,scale=0.7]{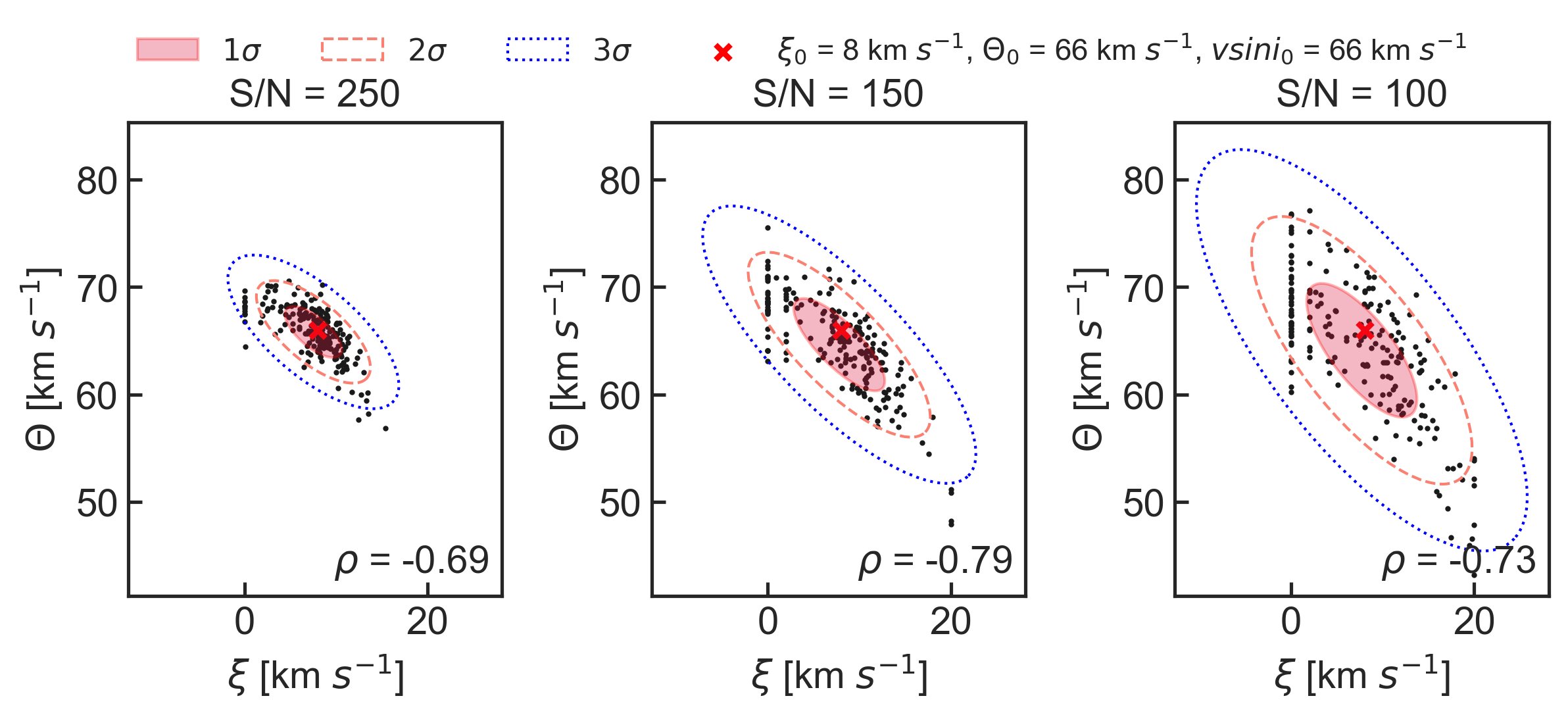}}
      \caption{ Same as Fig. \ref{cc2_mg_2_30_15_vmi} for another set of parameters \vsini, $\xi$, and $\Theta$ in the synthetic profile. }
         \label{cc2_mg_8_66_66}
    \end{center}
 \end{figure*}
    
 \begin{figure*}
   \begin{center}
            {\includegraphics[clip,scale=0.7]{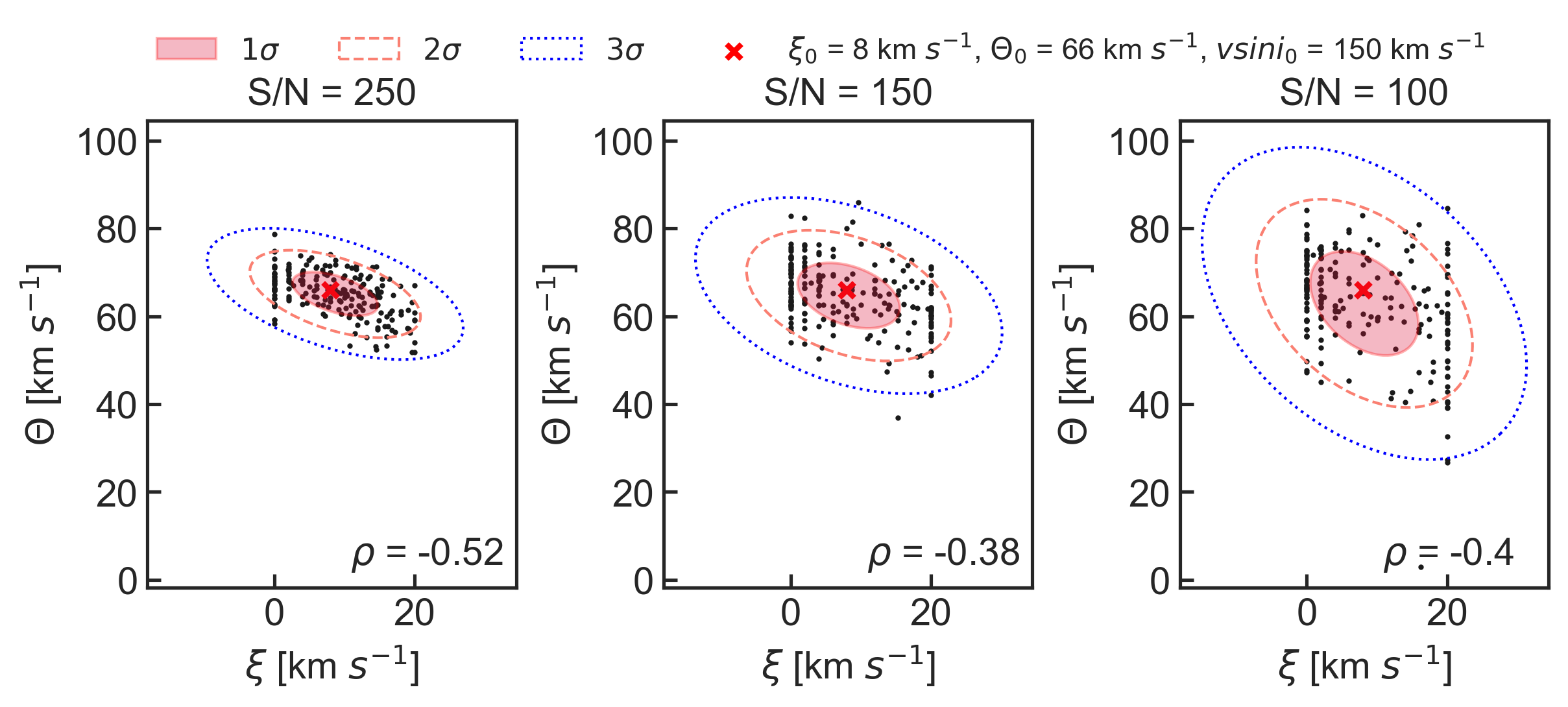}}
      \caption{ Same as Fig. \ref{cc2_mg_2_30_15_vmi} for another set of parameters \vsini, $\xi$, and $\Theta$ in the synthetic profile. }
         \label{cc2_mg_8_66_150}
    \end{center}
 \end{figure*}

 \begin{figure*}
   \begin{center}
            {\includegraphics[clip,scale=0.7]{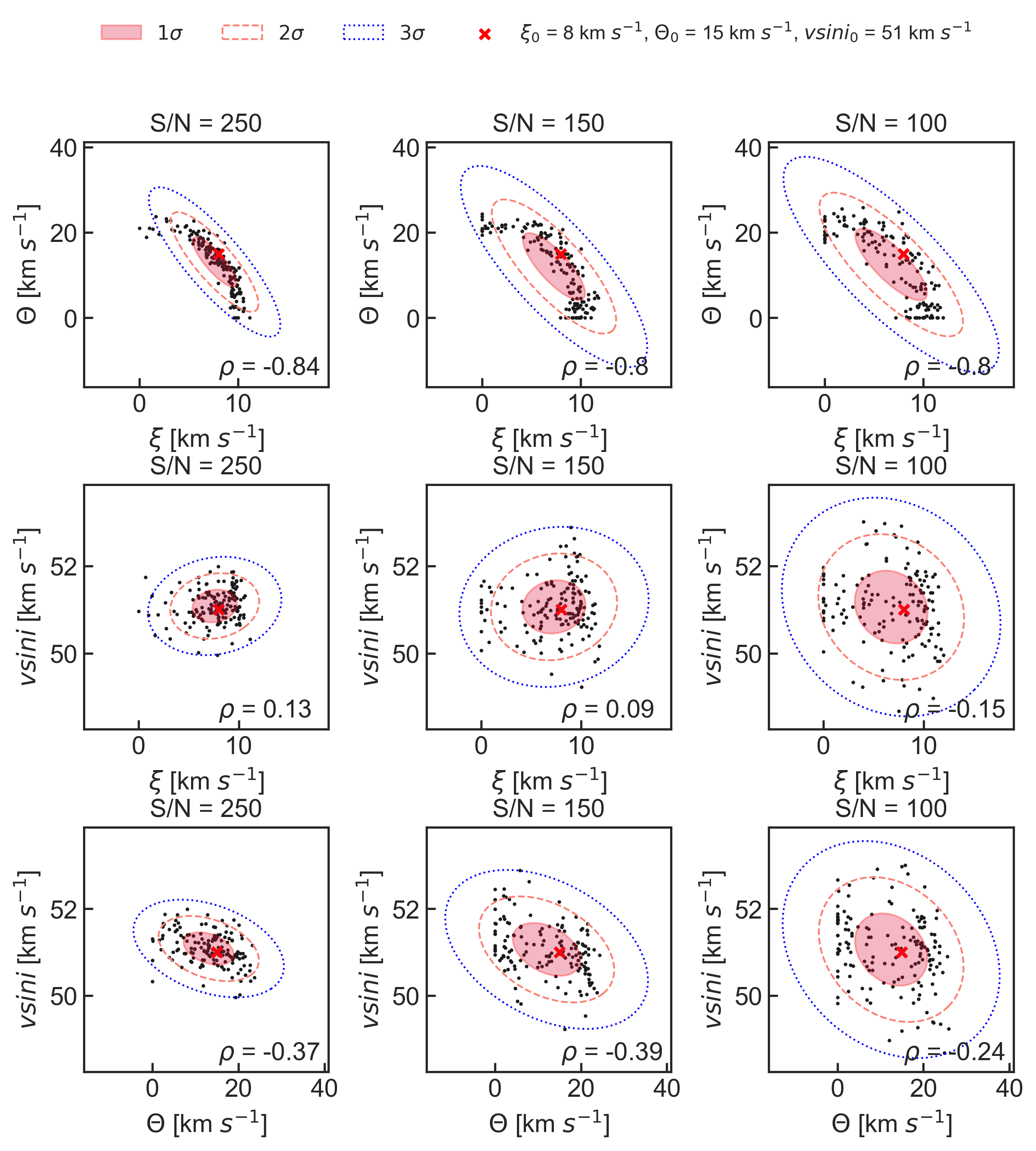}}
      \caption{ Same as Fig. \ref{cc2_mg_2_30_15_vmi} but \vsini\ is optimised as well as $\xi$ and $\Theta$, resulting in three free parameters. }
         \label{limvsini_34D}
    \end{center}
 \end{figure*}   
 
 \begin{figure*}
   \begin{center}
            {\includegraphics[clip,scale=0.75]{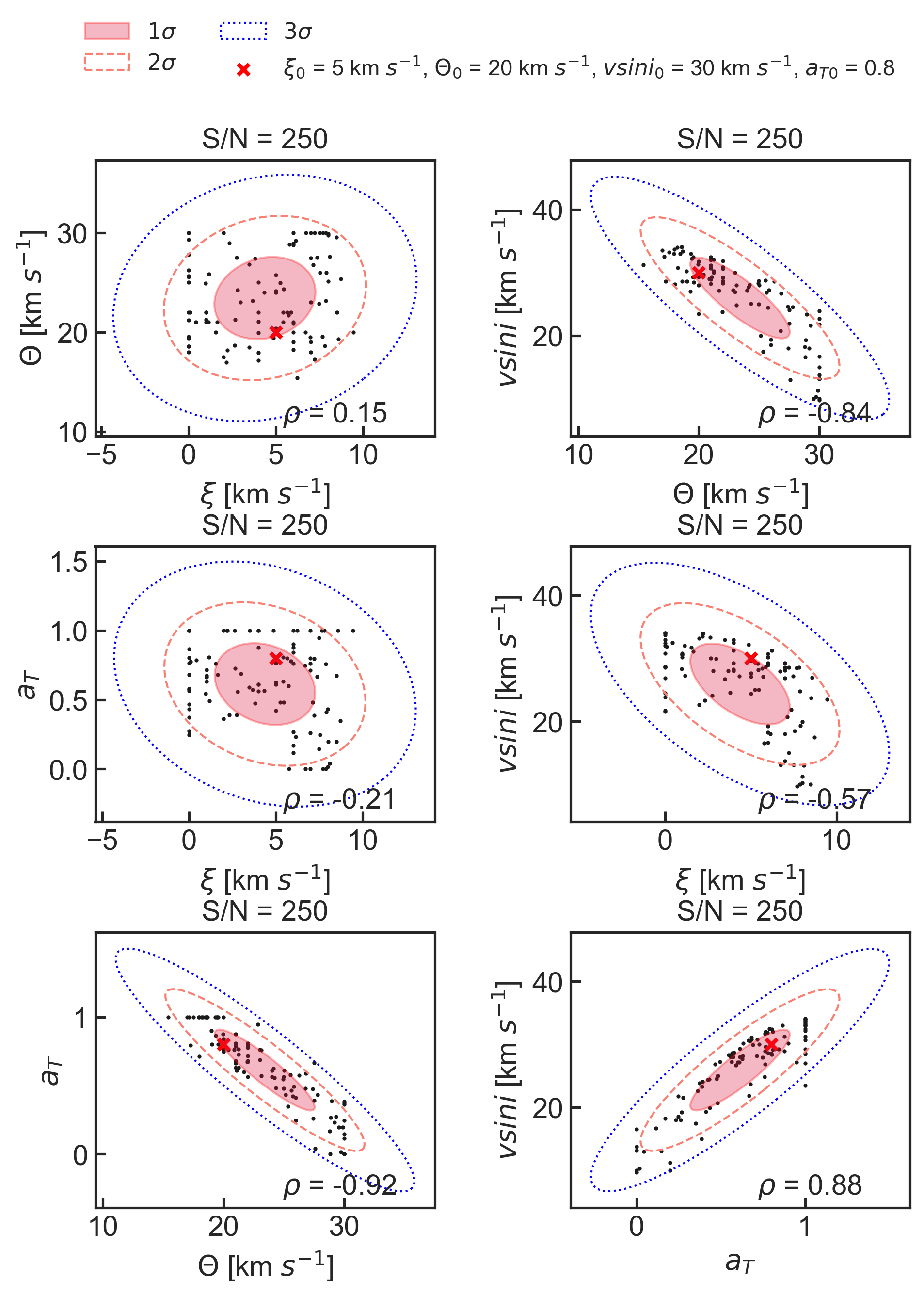}}
      \caption{ Same as Fig. \ref{cc2_mg_2_30_15_vmi} but \vsini\ and $a_{\rm T}$ are optimised as well as $\xi$ and $\Theta$, resulting in four free parameters. Only the results for S/N=250 are shown.  }
         \label{4D}
    \end{center}
 \end{figure*}  
 
 \clearpage


\end{document}